\documentclass[prb,aps,amsmath,amssymb,floatfix,preprint,showkeys]{revtex4-1}

\usepackage[export]{adjustbox}
\usepackage{amssymb}
\usepackage{graphicx}
\usepackage{dcolumn}
\newcolumntype{d}[1]{D{.}{.}{#1}}
\usepackage{bm}
\usepackage[utf8]{inputenc}
\usepackage{color}
\usepackage[caption=false]{subfig}

\usepackage{hyperref}
\hypersetup{
    colorlinks=true,
    citecolor=blue,
    urlcolor=black,
    pdfcreator={pdflatex},
}

\usepackage{natbib}
\usepackage{multirow}

\begin{document}
  
\title{Association and phase transitions in simple models for biological and soft matter condensates}

\author{Cecilia Bores$^1$, Antonio Diaz-Pozuelo$^2$, Enrique Lomba$^{2,3}$}
\email{Corresponding author: enrique.lomba@csic.es}

\affiliation{$^1$Department of Physics and Astronomy, Union College, 807 Union Street, Schenectady, 12308, New York, USA\\
$^2$Instituto de Qu\'{i}mica F\'{i}sica Blas Cabrera,
CSIC, Madrid, Spain\\
$^3$Grupo NAFOMAT, Facultade de Física, Universidade de Santiago de
Compostela, Santiago de Compostela, Spain}
\begin{abstract}
We investigate  a set of design principles that link specific features of
interparticle interactions to predictable structural and dynamic
outcomes in two-dimensional self-assembly, a framework relevant to
soft matter and biological condensates. Using extensive Molecular
Dynamics simulations of single- and two-component systems, we
systematically dissect how modifications to competing short-range
attraction and long-range repulsion (SALR) potentials (both isotropic
and anisotropic) serve as independent control parameters. In
particular, we have focused on tuning the repulsive barrier height,
decorating the attractive well with 
oscillatory components, and changing particle geometry. We  
demonstrate that these modifications dictate cluster size
distributions, the degree of intra-cluster ordering, the geometry of
the clusters, and the
propensity for inter-cluster crystallization. A key finding is the
decoupling of internal and global dynamics: oscillatory wells promote
solid-like order within clusters while maintaining liquid-like cluster
mobility. Furthermore, we show how asymmetric interactions in a binary
SALR mixture can be designed to induce internal phase segregation
within condensates. Complementing this, we observe that in anisotropic
models in which the short rage component of the interaction stems from
the presence of attractive patchy sites, stoichiometry and the
geometric distribution of the patches  are essential to control 
self-assembly and cluster morphology, whereas long-range repulsion can be
used to tune cluster size and polydispersity. The extracted
principles provide a causal road-map for engineering  self-assembled
materials and a set of basic physical concepts for interpreting the complex phase
behavior of biomolecular condensates.
\end{abstract}
\maketitle

\section{Introduction}

Self-assembly processes are essential for many key biological
processes such as the formation of  membraneless organelles (MLOs)
\cite{Hirose2023}, or the condensation of monoclonal
antibodies \cite{Yearley2014}, lysozime \cite{Liu2011} or intrinsically
disordered proteins \cite{Argudo2021}. In fact, self-assembly is the
fundamental principle governing how soft matter 
organizes itself from the nanoscale to the macroscopic world. It is
characterized by the spontaneous formation of   ordered structures
from its constituents through mostly
weak, non-covalent interactions. The process itself
is driven by the system's tendency to maximize entropy or minimize
free energy, rather than by the action of external agents. In soft and
biological matter, the relevant forces are typically weak—comparable to thermal
energy—allowing for fluidity and reorganization. The pioneering work
of Whitesides and Grzybowski \cite{Whitesides2002} highlighted
self-assembly as a universal phenomenon, whose presence is revealed
over a wide range of scales, from molecular clusters to living tissues. 

The specific structures formed are dictated by the nature of the
building blocks and thermodynamic conditions. Moreover, as Glotzer
and Solomon have shown\cite{Glotzer2007}, the
introduction of anisotropy—e.g. the location and number of associative
patches in  patchy particles—
opens an avenue to astonishing complexity, enabling the formation of
metamaterials. Competing interactions, such as the short-range
attraction and long-range repulsion (SALR), are known to create
equilibrium mesophases like clusters and gels, preventing full phase
separation. 

Perhaps most profoundly, self-assembly is the bridge between simple
physics and biological complexity.  In modern biology, as
detailed by Hyman and Rosen \cite{Hyman2014}, liquid-liquid phase separation—a form of
self-assembly—is seen as one of the key mechanisms for organizing the cell's
interior into membraneless organelles. This direct link to biological
function, and its misregulation in diseases, shows how soft matter
principles are essential for understanding life itself. In this
connection Sweatman and Lue\cite{Sweatman2019} have argued that
competing interactions might also be involved in the stabilization of
the liquid droplets within the process of liquid-liquid phase separation.  Finally, in
active matter, self-assembly takes on a non-equilibrium character,
creating dynamic patterns that define collective behavior in systems
from bacterial colonies to synthetic swimmers\cite{Sanchez2012}. 

In the particular case of molecular condensates, there is a striking
difference between those condensates that appear in the cytoplasm,
which exhibit an ample size polydispersity, and those occurring inside
the nuclei of eukaryotic cells, with a much smaller degree of size
polydispersity\cite{brangwynne2013phase}.  In the former case,
these organelles are the cell's response to external conditions (e.g.,
P-bodies, stress organelles) whose formation is governed by stochastic
nucleation events. The energy barrier for spontaneous nucleation is
low. The cytoplasm is a crowded environment exhibiting local concentration
fluctuations. When a stress signal occurs, numerous small,
unstable clusters form simultaneously across the cell. They then
coarsen and coalesce with smaller clusters. This process is often
interrupted by the dynamic 
nature of the cell, leaving a wide distribution of sizes which
actually fits well into the picture of a frozen metastable state in a liquid-liquid
separation\cite{brangwynne2013phase,banani2017molecular}.
On the other hand, size polydispersity is reduced when the condensate
exhibits a high nucleation barrier and surface tension. This is
typically explained in terms of the scaffold-client
model\cite{Orti2021}. In this case we will have a  limited number of
specific, high-affinity 
scaffold molecules (e.g., specific proteins with highly active sites) which drive the phase 
separation. Their concentration is tightly controlled and they usually have
very low saturation concentration, being key for the condensation of
the clients, other proteins or nucleic acids. In a recent work,
Diaz-Pozuelo and   
coworkers \cite{DiazPozuelo2025} revisited a SALR model in which the
nucleation barrier is due to the presence of a repulsive maximum in
the interaction potential, that once the cluster grows up to a certain
size reaches values well above the thermal energy. In the
scaffold-client model the nucleation barrier is due to the fact that
the system prefers to add molecules to existing nucleation sites
instead of creating new ones since it minimizes interfacial area and
thus surface energy. The process continues until the sites of the
scaffold molecules are saturated. Interestingly, this picture is
similar to the one proposed by Palaia and \v{S}ari\'c
\cite{Palaia2022}  where a binary mixture of  molecules with
associative patchy sites (between unlike sites) forms condensates.
The growth of the condensates saturates when the concentration of  one
of the components exceeds that of the other by a factor larger than
the number of associative sites per molecule.  The
resulting sample in the model of Palaia and
\v{S}ari\'c is nonetheless  widely polydisperse, by which this model fits more
properly into the picture of cytoplasmic aggregates. 

These biomolecular condensates can also  display internal phase
transitions. Such is the case of the  transactive
response DNA-binding Protein of 43 kDa (TDP-43)
\cite{Conicella2016}. Under certain conditions, TDP-43 droplets undergo internal
phase segregation at the droplet surface/solvent interface
\cite{PantojaUceda2021}. It is important to stress that the formation
of particular types of condensates as well as these internal phase
transitions have been related with the onset of neurodegenerative
diseases
\cite{brangwynne2013phase,banani2017molecular,PantojaUceda2021}. 

The formation of finite-sized clusters and modulated phases due to
competing short-range attraction and long-range repulsion is a
well-established phenomenon in three-dimensional systems, both
theoretically and experimentally \cite{Sciortino2004,
  Stradner2004}. Seminal work on 3D SALR fluids has rationalized the
stabilization of equilibrium cluster phases, the suppression of
macroscopic phase separation, and the rich interplay between
thermodynamics and structure \cite{Sweatman2019, Liu2019}. The
relevance of simple isotropic SALR interactions to 
the formation of biological condensates has been discussed in depth in
the literature (cf Ref.~\onlinecite{Godfrin2014} and references therein), being the lysozime condensates a
typical case example\cite{Chen2007b}. Our study
builds upon this foundation but focuses on the distinct physical
regime of two dimensions. The 2D geometry, relevant to membranes,
interfaces, and surface-associated biological assemblies, removes the
geometric frustration inherent in 3D packing
allowing for crystalline ordering both within and among clusters
\cite{Nelson2002}.  In two dimensions, the phase diagram of
such system includes the emergence of patterns ranging from clusters,
stripes or bubbles as has been shown in the literature using both lattice
\cite{Pekalski2014,Almarza2014} and continuum
\cite{Chacko2015,Bordin2018,Schwanzer2010} models. Here, following  the recent
work of Ref.~\onlinecite{DiazPozuelo2025} we will use the double
exponential interaction
proposed by Sear et al. \cite{Sear1999}, but in a 2D context. Previous 
studies under these conditions  have been performed,
among others, by Imperio and
Reatto\cite{Imperio2004,Imperio2006,Imperio2007}. For the same two
dimensional system, Schwanzer and  Kahl\cite{Schwanzer2010}  studied the competition between
clustering and vapor liquid condensation, and its cluster/particle
dynamics\cite{Schwanzer2016}. Confinement effects in disordered porous
media were also considered for this system by Bores et
al.~\cite{Lomba2014a,Bores2015}. 

In addition to the well known isotropic SALR
potential mentioned above\cite{DiazPozuelo2025}, we will extend our
study to a class of anisotropic SALR
interactions. In this instance the attractive component is
modeled via a patchy model as introduced by Palaia and
coworkers\cite{Palaia2022}, adding a long range repulsion that stems from
screened charges. This is in fact nothing but a 2D version variation
of the charged patchy particle model (CPPM) investigated in detail by
Yigit and coworkers \cite{Yigit2015} to model the self-assembly of globular
proteins. It is worth stressing that despite  being coarse
grained potentials, these models represent a higher degree of
approximation to molecular condensates, since the presence of active
sites and its role in self-assembly is accounted
for\cite{Yigit2015,Joseph2021}. We will illustrate the new features that anisotropy adds to
the SALR model.

In order to complement previous work, a substantial part of this work
will address  the
conditions that control polydispersity, namely, the height of the
potential maximum (the nucleation barrier), or the presence of
decorations in the attractive well, for the isotropic SALR
interactions, or, in CPPM
systems, stoichiometry, the number and geometry of attractive
sites, or the presence of repulsive long range forces. Polydispersity (or the lack of) is a characteristic 
feature of different types of biomolecular condensates, and has a
direct impact on geometric frustration. The growth of the repulsive
maximum of the SARL potential can have multiple physical origins, some
linked to the increase of the effective charge of the condensates due
to phosphorylation\cite{Monahan2017}, or changes in the pH
\cite{creighton_1993} or entropic/steric effects\cite{Schmidt2016}. In
this work we have chosen to model these complex process by the simple
addition of a Gaussian term on top of the maximum of the SALR
interaction, in order to illustrate the effects that this nucleation
barrier enhancement has on both the structure and dynamics of the clusters.

Concerning the decoration of the attractive component
  of the isotropic SALR interactions, it turns out that in a crowded,
  multi-component fluids, the organization 
  of water, ions, or small molecules around proteins/RNAs could lead
  to solvation forces that oscillate with distance. This is
  particularly relevant for charge-rich molecules such as nucleic
  acids, chromatin and nucleic-acid binding proteins. In this
  particular context the recent work of Tejedor et
  al. \cite{Tejedor2025} has shown that coarse-grained forces display
  clear oscillations within the attractive range of the
  interaction. The effect of these oscillations will be here
investigated through a simple model in which we
decorate the attractive part of the interaction with an oscillating
pair potential (OPP), whose intensity modulation  recalls the Friedel
oscillations of liquid metal potentials \cite{PettiforBOOK}. This
interaction was proposed by 
Mihalkovi\v{c} and  Henley  \cite{Mihalkovic2012} to model   amorphous
states in complex intermetallic compounds. Subsequently, Engel and coworkers showed
that an appropriate choice of parameters can lead to the formation of
icosahedral quasicrystals  in three dimensions
\cite{Engel2015}. However, in two dimensions we know that the
characteristic structural frustration 
that leads to amorphous states in three dimensions is absent. For a
radially symmetric potential like the SALR or SALR-OPP, the most
compact nucleation unit is the 
hexagon with a central particle, and this is a space filling
structure. This is not the case in three dimensions where the basic
unit, the tetrahedron, cannot fill the space
regularly\cite{Nelson2002}. As a consequence, decorating the attractive
part of the SALR potential with an OPP does not prevent the particles
to form  a triangular lattice when the clusters are cooled
down. In contrast, in
Ref.~\onlinecite{DiazPozuelo2025} it was shown that in 3D the
SALR system  is trapped into a glass-like state of clusters for
sufficiently low temperature and density. This is  a 
known result of the aforementioned frustration effects. Interestingly,
we will see that the formation of a triangular 
lattice of clusters, which readily occurs in the SALR case, is
hampered by the presence of oscillations in the attractive part of the
potential. We will also see that these oscillations severely impact
the dynamics of the system.

Binary mixtures of isotropic self-assembling SALR fluids will also be the subject of
investigation. Here our main interest dwells on the the conditions
that drive the system towards internal phase separation within
large bi-component condensates. We will assess which characteristics concerning range and
position of the repulsive maxima can lead  to a phase separation that
mimics the segregation experimentally observed in  condensates of
TDP-43\cite{PantojaUceda2021}.

Finally,  we will explore the behavior of two anisotropic SALR models, in
which the attractive component of the interaction is the one proposed
by Palaia and \v{S}ari\'c 
\cite{Palaia2022}, to which a screened Coulomb term (Yukawa) is added
in order to model the presence of charges and an implicit solvent with
counterions. These patchy SALR models (pSALR)  will be composed of
a dominant component A (the
linker) and a B-component. The resulting structure formed by
B-particles linked by A-particles would play the role of the
scaffold. Remaining non-binding A-particles either saturating B-sites
or attached by dispersive interactions, would be the clients of
the biomolecular condensate. In Ref.~\onlinecite{Palaia2022}
various topologies of patch distributions were studied, but in all
cases identical for A and B components. Here we will focus on 4-patch
B particles and 4- and 2-patch A linkers. Our pSALR
  models are endowed with 
screened charges either in the B component (BB-pSALR) or both in A and
B (AB-pSALR). The effects of this long
range repulsion and linker topology on cluster size, shape and size
polydispersity will be scrutinized and compared with the results of
the isotropic double exponential SALR. In addition, we have also
considered an uncharged model so as to better put our results in the
context of those of Palaia and \v{S}ari\'c\cite{Palaia2022}.

The rest of the paper is sketched as follows. The models and simulation
methods will be presented in the next Section. Results for  isotropic SALR
interactions are discussed in Section III. The most significant
results for our patchy SALR models will be presented in Section IV.
The paper is closed with a discussion of the potential implications of
our results, followed by a brief summary and future prospects.

\section{Model and Methods}\label{sec3}

\subsection{Model interactions}

\begin{figure}[t]
\centering
\includegraphics[width=14cm,clip]{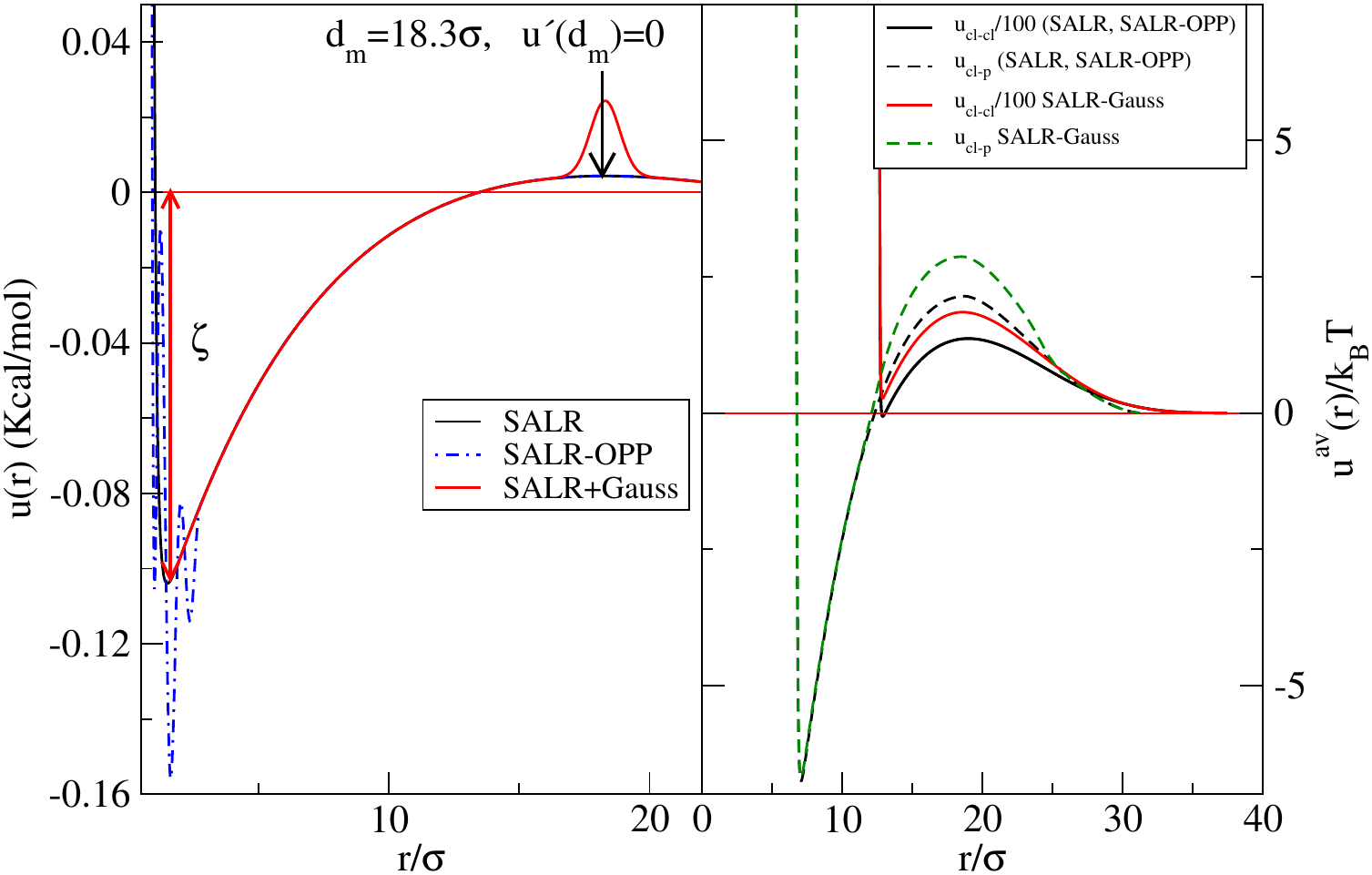}
   \caption{ (Left) SALR, SALR-Gauss, and SALR-OPP interaction potentials as described by
     Eqs.~(\ref{pot})m (\ref{gauss}), and (\ref{salr-opp}).  (Right) Average
     intercluster potential (solid  curves)  and cluster-particle potential
    (dashed curves), computed assuming a uniform average cluster
    density and spherical cluster radii using Eq.~(\ref{uav}).}
\label{ur}
\end{figure}

Our first isotropic SALR potential is identical to that used in
Ref.~\onlinecite{DiazPozuelo2025}, and it is composed of an analytic part 
\begin{equation}
\Phi(r)^{SALR}= \epsilon \left(K_re^{-\alpha_rr/\sigma}-K_ae^{-\alpha_ar/\sigma}\right),
 \label{salr}
 \end{equation}
to which for computational convenience a strongly repulsive inverse
potential is added. Both are truncated at $R_c$ to give 
\begin{equation}
\label{pot}
u(r) = \left\{\begin{array}{cc}
\phi^{SALR}(r) - \phi^{SALR}(R_c)+\epsilon\left(\frac{\sigma}{r}\right)^{10} & \mbox{if} \; r \le R_c \\
0 & \mbox{if} \; r > R_c 
\end{array} \right. .
\end{equation}
The potential parameters were  chosen to generate large
droplets: $K_r=1$, $K_a=1.5$, $\alpha_r=0.05$, $\alpha_a=0.12$,
$\epsilon=0.2$ kcal/mol and $\sigma=4$\r{A} and $R_c=100$ \r{A}. An
illustration of this interaction can be seen on
the left graph of Fig.~\ref{ur}. Note
that the size of the droplet is conditioned by the location of the
potential maximum at $d_m$. 
In order to explore how the potential barrier (the maximum at $d_m$)
affects the cluster distribution we have also considered an interaction
in which, in addition to the SALR potential, a Gaussian term of the form
\begin{equation}
  u_{gauss}(r) = \epsilon_g e^{-\kappa_g (r-d_m)^2}
  \label{gauss}
\end{equation}
  is added. Here we have considered $\epsilon_g/\epsilon=0.1$ and
  $\kappa_g/\sigma^2=1.6$. Again on the right panel of Fig.~\ref{ur} we
  can appreciate that the only effect of the Gaussian is to raise the
  height of the maximum, which will be shown to impact directly on
  the size of the aggregates. This interaction will be hereafter referred to
   as  SALR-Gauss. On the right panel we illustrate the
  cluster-cluster and cluster-particle effective interactions. The
  former quantity is   computed through
\begin{equation}
u_{cl-cl}^{av}(r;\rho_{cl}^{eff},R_{cl}^{ef}) = \left(\rho_{cl}^{eff}\right)^2\int_{S_{cl}} d{\bf
  s}_1\int_{V_{cl}} d{\bf s}_2 u(|{\bf r}-{\bf r}_1+{\bf r}_2|)
\label{uav}
\end{equation}
where the effective cluster radius, $R_{cl}$ is
  approximated as $R_{cl}\approx d_m$, the effective density within
  the cluster is $\rho_{cl}^{eff}=<N_{cl}>/(\pi d_m^2)$, and the
  average number of particles in the cluster is estimated from a
  cluster analysis for  $\rho\sigma^3=0.013$ at the lowest temperature
considered. $S_{cl}\sim \pi d_m^2$ is the average cluster area, and $d{\bf s}_i$
denotes the infinitesimal integration area. The cluster-particle
effective interaction is obtained by removing one of surface
integrals.  It can be
  appreciated that the presence of the Gaussian term increases both
  the cluster-cluster and cluster-particle potential maxima
  (cluster nucleation and coalescence barriers,  $\approx 3k_BT$ and
  $\approx 200k_BT$ respectively), but slightly reduces
  the extent of the interaction. 

Together with the two-exponential isotropic SALR interactions described
above,  we have also considered another isotropic SALR model with a
decorated attractive component, adding short period oscillations. In this
way  now the analytic part of the interaction reads
\begin{equation}
\phi(r)^{SALR-OPP}= \epsilon
\left(K_re^{-\alpha_rr/\sigma}-K_ae^{-\alpha_ar/\sigma}+
\sigma^3\frac{\cos (\kappa(\mu r - 5\sigma/4)-\psi)}{r^3}\right),
 \label{salr-opp}
 \end{equation}
with $\kappa = 8.5\sigma^{-1}$ and $\psi=0.47$, parameters taken from
Engel and coworkers \cite{Engel2015}. With this, one gets an
interaction of the form depicted by the dash-dotted curve in Fig.~\ref{ur}.
It is interesting to note that despite the considerably
deeper minimum, the average attractive energy integrated over the
attractive range is quite similar to that of the simple SALR
potential, due to the mutual cancellation of the strong
oscillations. 
In the Figure it can be seen that there is no apparent difference when comparing SALR and
SALR-OPP cluster-cluster effective potentials: the angular averaging
cancels any visible effect of the 
particle-particle interaction oscillations occurring below
$3\sigma$. 

\begin{figure}[t]
\includegraphics[width=10cm,clip]{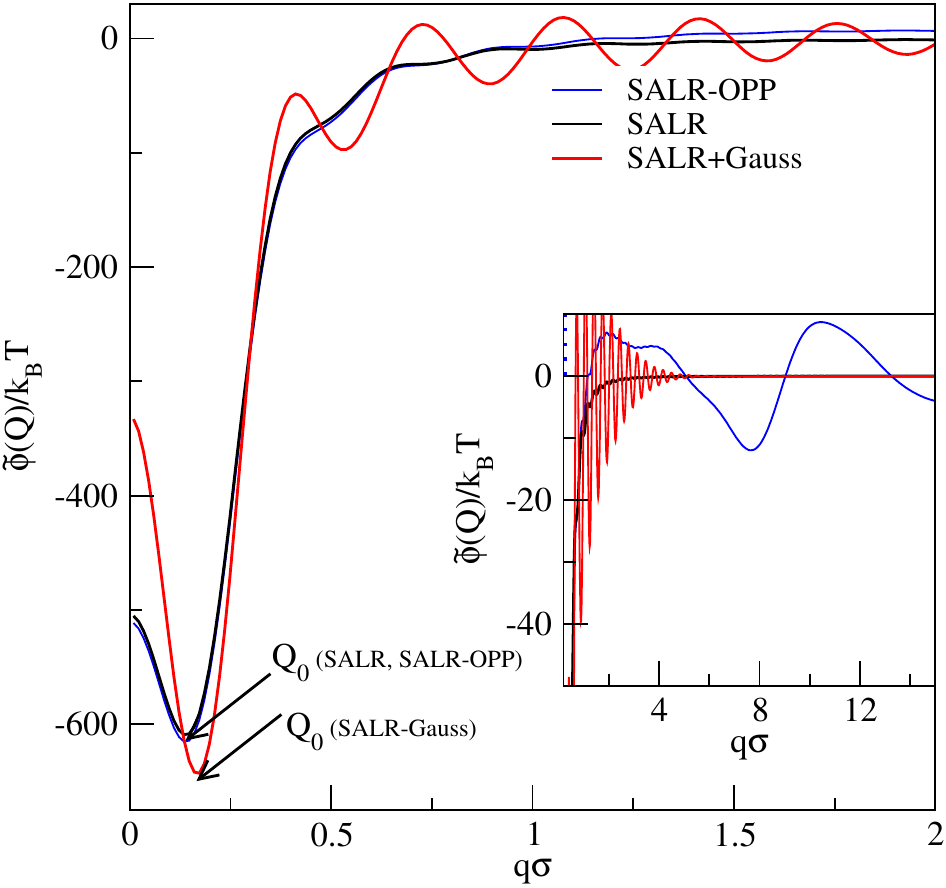}
   \caption{Short wavelength 2D Fourier transform of the SALR, SALR-Gauss
     and SALR-OPP interaction potentials. The characteristic wave
     lengths of the interactions, $Q_0$, are shown in the Figure. In
     the inset, the 
     long wavelength behavior of Fourier transform of the SALR, SALR-Gauss and
     SALR-OPP potentials is illustrated. The strong long range oscillations of
     $\tilde{\phi}^{SALR-OPP}(Q)$ stem from the marked narrow and deep
     oscillations of the potential below $3\sigma$. }
\label{phik}
\end{figure}

Now, as is  known (see the discussion below Eq.~(4) in
Ref.~\onlinecite{DiazPozuelo2025}), the 
characteristic wavelength of any modulation induced by this class of
potentials, $Q_0$, is given by the position of the minimum of the 
corresponding Fourier transform (in this case 2D) of the analytic
part of the potential. This quantity is  plotted in
Fig.~\ref{phik}. The location of $Q_0$ is determined  by the 
balance between attractive and repulsive forces. As seen in the right
graph of Figure
\ref{ur} the net effect of the OPP decoration on the effective
cluster-cluster and cluster-particle interactions is negligible. As a
consequence, the position of $Q_0$ for both SALR and SALR-OPP
potentials is identical, hence the corresponding
correlation lengths, $\lambda_0 = 2\pi/Q_0$ will be identical as well. In the inset one can
observe that the marked short range oscillations translate into long
range oscillations, which  reflect the positions of the narrow,
potential minima, the most intense occurring at $r/\sigma
\approx1.6$. On the other hand  for SALR-Gauss 
$Q_0$  is shifted to slightly larger values, consequently the
correlation length and the cluster size when dealing with globular
clusters will be smaller. This is consistent with the decrease in the
range of the effective cluster-cluster and cluster-particle
interactions that results from the presence of a higher maximum at
$d_m$.  This enhanced maximum also translates in the
appearance of marked medium range oscillations in $\tilde{\phi}(Q)$.
$\tilde{\phi}^{SALR}(Q)$ also displays 
traces of these oscillations but severely damped. These oscillations are just a
characteristic feature of the Fourier transform of an isolated intense 
maximum at medium range. 

\begin{figure}[t]
\subfloat[\label{4+4}]{\includegraphics[width=6cm,clip]{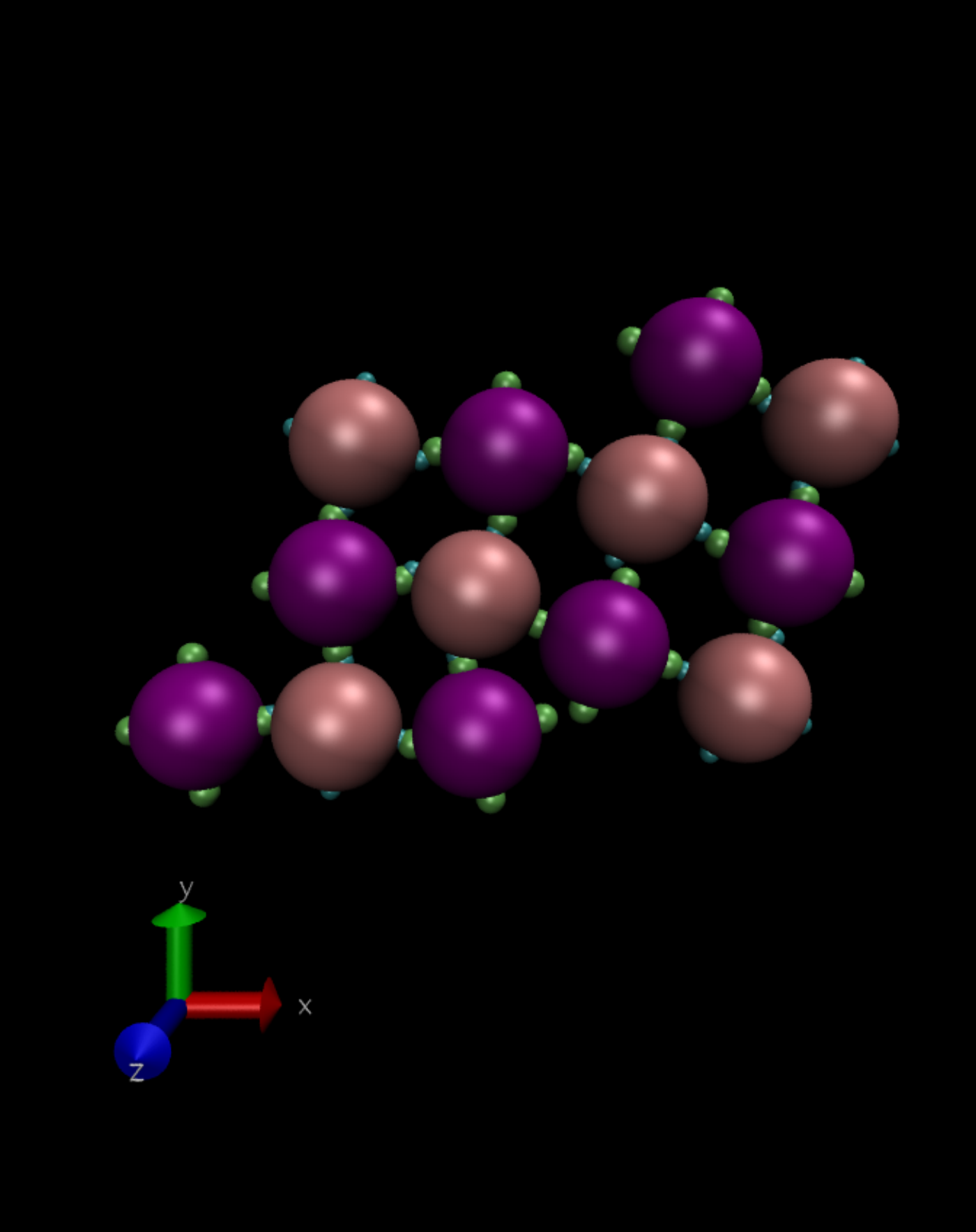}}
\hfill
\subfloat[]{\includegraphics[width=6cm,clip]{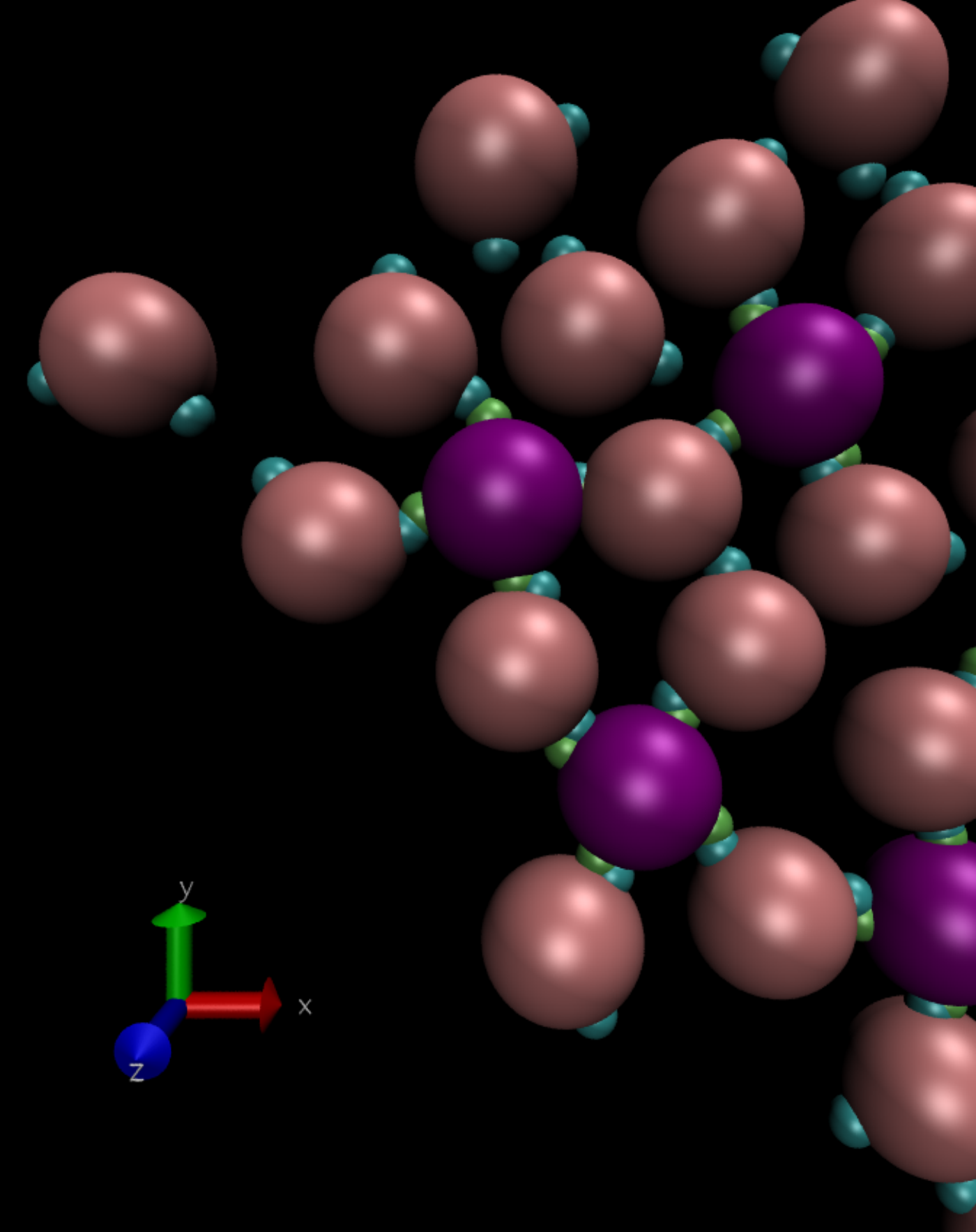}}
\caption{a) Configuration of a cluster of interacting 4-patch A-B
  particles b) The same for 2-patch A-particles and  4-patch
  B-particles. These are low temperature states in which inner
  particles of the clusters have all patches saturated. On the left
  particles display a 4-fold coordination corresponding to a body
  centered square lattice. On the right the two-site linkers prevent
  the presence of substitutional order and four-fold and five-fold
  coordinations occur along with vacancies.}
\label{2patch}
\end{figure}

As mentioned in the Introduction, in addition to these isotropic SALR
potentials we have also considered an associative model controlled by
specific interactions, as proposed by Palaia and \v{S}ari\'c
\cite{Palaia2022}, but with an added long range repulsion. Now we will
be dealing with a mixture of patchy 
particles in which one central site (A or B) is surrounded by patches
of type a (for A centers) or b (for B centers). The interaction
between patches is the source of the anisotropy. Centers do not
interact with patches of either type, and from the patch-patch
interactions only the a-b potential is
non negligible. This is attractive and with the form 
\begin{equation}
  u^{a-b}(r) = \left\{\begin{array}{cc} -\epsilon_{pp} & \mbox{if} \;
  r \le \sigma_{pp} \\
 -\epsilon_{pp}\cos^2\frac{\pi(r-\sigma_{pp})}{R_{pp}-\sigma_{pp}} &
 \mbox{if} \; \sigma_{pp} < r \le R_{pp} \\
0 & \mbox{if} \; r > R_{pp} 
  \end{array} \right.
  \label{pab}
\end{equation}
where $R_{pp} = 1.06\sigma$, $\sigma_{pp}=0.35\sigma$ and
$\epsilon_{pp} = 10\epsilon$. This latter value guarantees that
clusters will form when $k_BT/\epsilon\approx 1$ ($k_B$ being
Boltzmann's constant) and
below\cite{Palaia2022}. Like patches do not interact, and the 
central sites interact via a purely repulsive shifted and truncated Lennard-Jones
interaction of the form
\begin{equation}
  u_0(r) = \left\{\begin{array}{cc} -\epsilon_{cc}\left[\left(\frac{\sigma_{cc}}{r}\right)^{12}-2\left(\frac{\sigma_{cc}}{r}\right)^6+1\right] & \mbox{if} \;
  r \le \sigma_{cc} \\
0 & \mbox{if} \; r > \sigma_{cc} 
  \end{array} \right.
  \label{pcc}
\end{equation}
with $\sigma_{cc}=3.5\sigma$ and $\epsilon_{cc}=\epsilon_{pp}$ for all
interactions between A and B sites. In addition to these terms, in
order to make fluid clusters, as in Ref.~\onlinecite{Palaia2022} a
longer range dispersive interaction of the form
\begin{equation}
  u^{d}(r) = \left\{\begin{array}{cc} -\epsilon_{a} & \mbox{if} \;
  r \le \sigma_{cc} \\
 -\epsilon_{a}\cos^2\frac{\pi(r-\sigma_{cc})}{\sigma_{cc}} &
 \mbox{if} \; \sigma_{cc} < r \le 2\sigma_{cc} \\
0 & \mbox{if} \; r > 2\sigma_{cc} 
  \end{array} \right.
  \label{pcc-d}
\end{equation}
is added. Here $\epsilon_{a}=2\epsilon$. Now the center-center
interaction reads
\begin{equation}
  u_{cc}(r) = u_0(r) + u^d(r).
\end{equation}
We will be considering 4-patch B particles, with the patches placed on
the vertices of a square, and both similar 4-patch and 
2-patch A-particles. In the latter case the patches are placed at the same a-A
distance and the angle $\widehat{aAa}=120^\circ$.  In
Fig.~\ref{2patch} we present a close-up of a cluster of 4-patch A 
and B particles (left) and 2-patch A plus 4-patch B particles (right).

\begin{figure}[t]
\centering
\includegraphics[width=12cm,clip]{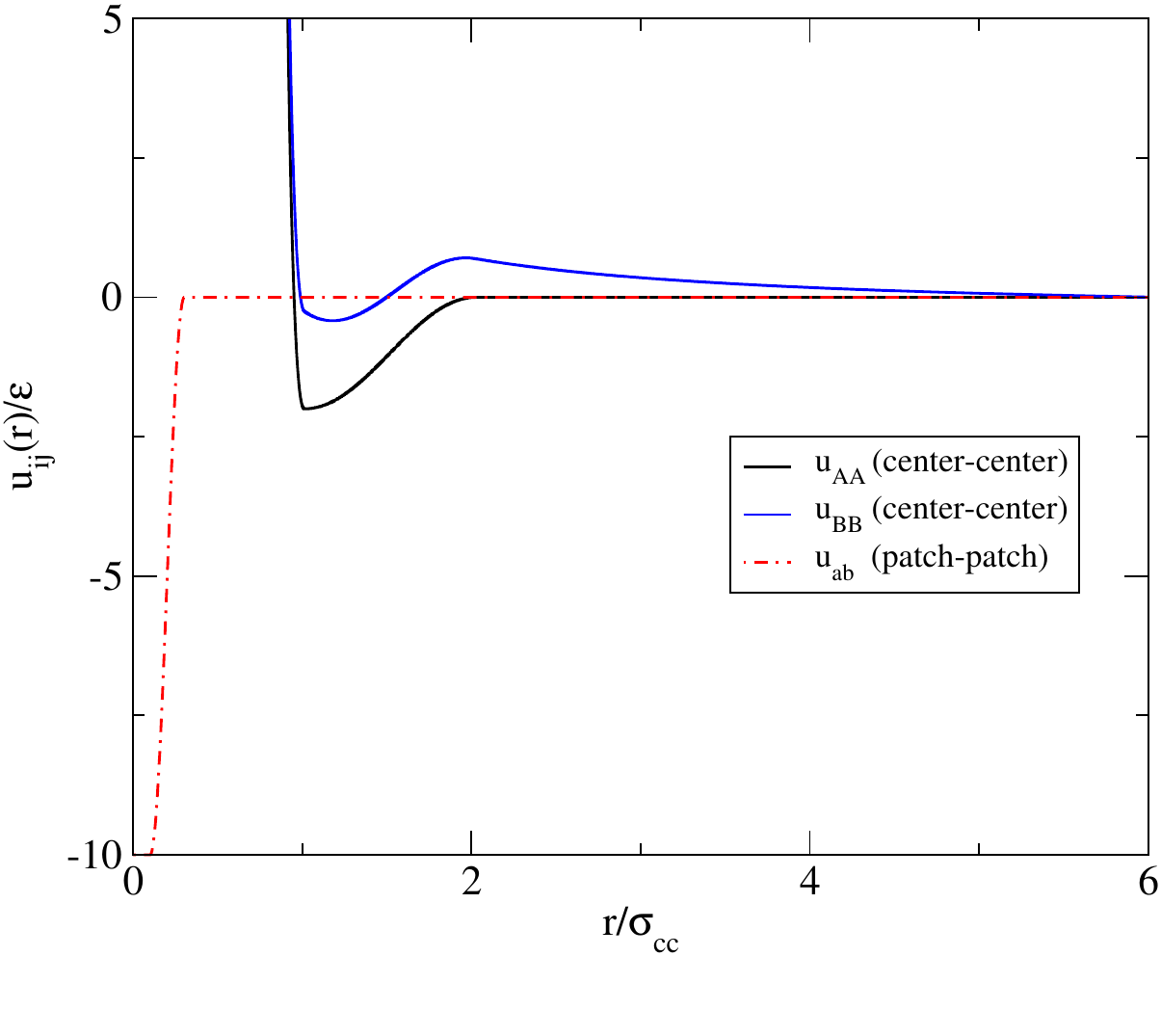}
   \caption{Center-center (black and blue curves) and unlike
     patch (dashed red curve) interactions
   of the BB-pSALR model. For the AB-pSALR the AA interaction is also
   long ranged and equal to BB. in the  model of Palaia and
  \v{S}ari\'c \cite{Palaia2022} the BB interaction is short ranged and
  equal to AA. All other interactions are set to zero.}
\label{upatch}
\end{figure}

From the results of Palaia and \v{S}ari\'c\cite{Palaia2022}, we know that the
model without long range repulsions yields  relatively wide cluster
size distributions, particularly when 
approaching equimolar mixtures, even if these cluster states are
thermodynamically arrested. This fits with the relatively large size
polydispersity seen in cytoplasmic MLOs, as mentioned in the
Introduction. On the other hand, the main building blocks of
biomolecular condensates are proteins - carrying a net positive or
negative charge that depends on the solution pH\cite{creighton_1993} - and nucleic acids -
always carrying a negative charge in solution\cite{watson_2014}.
For this reason, we have extended the model of Palaia and \v{S}ari\'c
by  adding a repulsive screened 
Coulomb (i.e. Yukawa-like) interaction to our associative patchy
particles, first just between B central 
sites --BB-pSARL model--, secondly between all A and B sites
(AB-pSARL model). This screened Coulomb interaction
accounts for the presence of same sign charges in the biomolecules
that are screened by the counterions present in the surrounding
medium.  Thus we will have
\begin{equation}
  u_{cc}(r) = u_0(r) +
  u^d(r)+H(R_y-r)K_p\left(\frac{e^{-\kappa_pr}}{r}-\frac{e^{-\kappa_pR_y}}{R_y}\right),
\label{yuk}
\end{equation}
where we have taken $\kappa_p\sigma_{cc}=0.05$,
$K_p=0.75\epsilon_{cc}$, $H(x)$ is a Heaviside function and the cutoff
$R_y=6\sigma_{cc}$. This will be just the B-B
interaction for the BB-pSARL model. All A-A, A-B and B-B interactions in
the AB-pSARL model follow Eq.~(\ref{yuk}). Since cluster formation can
also be driven by the 
presence of purely repulsive long range forces \cite{Hoffmann2006}, we
expect the effect of these repulsive long range interactions to stabilize finite
size clusters particularly in the case of uncharged 1:1 4-patch A+B particles, a
system whose ground state is a single cluster forming a body-centered square
lattice. An illustration of the relevant interactions  that describe
the BB-pSALR model can be found in Figure \ref{upatch}. The SALR
character of the complete BB interaction is readily visible in the
Figure. Differences
with the AB pSALR and the short range patchy model of Palaia and \v{S}ari\'c
\cite{Palaia2022} are described in the figure caption. 

\subsection{Simulation conditions}

In the case of isotropic SALR interactions we are
mostly interested in the 
micellar-like phases. Therefore, we have chosen a relatively low 
density, namely $\rho\sigma^3=0.013$. We have used the LAMMPS Molecular
Dynamics package 
\cite{LAMMPS_Thompson2022} to perform NVT simulations with a time step
set to 0.03$\tau_0$, using a reduced time unit
$\tau_0=(m\sigma^2/\zeta)^{1/2}$, where $m$ is the particle mass (set
to 1 for convenience), and  $\zeta$ is the depth of the interaction
(cf. Fig.~\ref{ur}). Simulations were 
started from 2D lattice structures of  100000-160000 
particles at reduced temperature ($T^*=k_B T / \zeta$) of 
$T^*=17.5$ and cooled the system down to $T^*=12, 9, 6, 3, 1$, along a
ramp 10 million step long. Systems were equilibrated for another
10 million steps and then averages were performed over 2000
configurations collected during a 20 million step production run.

\begin{figure}[t]
 \subfloat[SALR]{\includegraphics[width=0.30\linewidth]{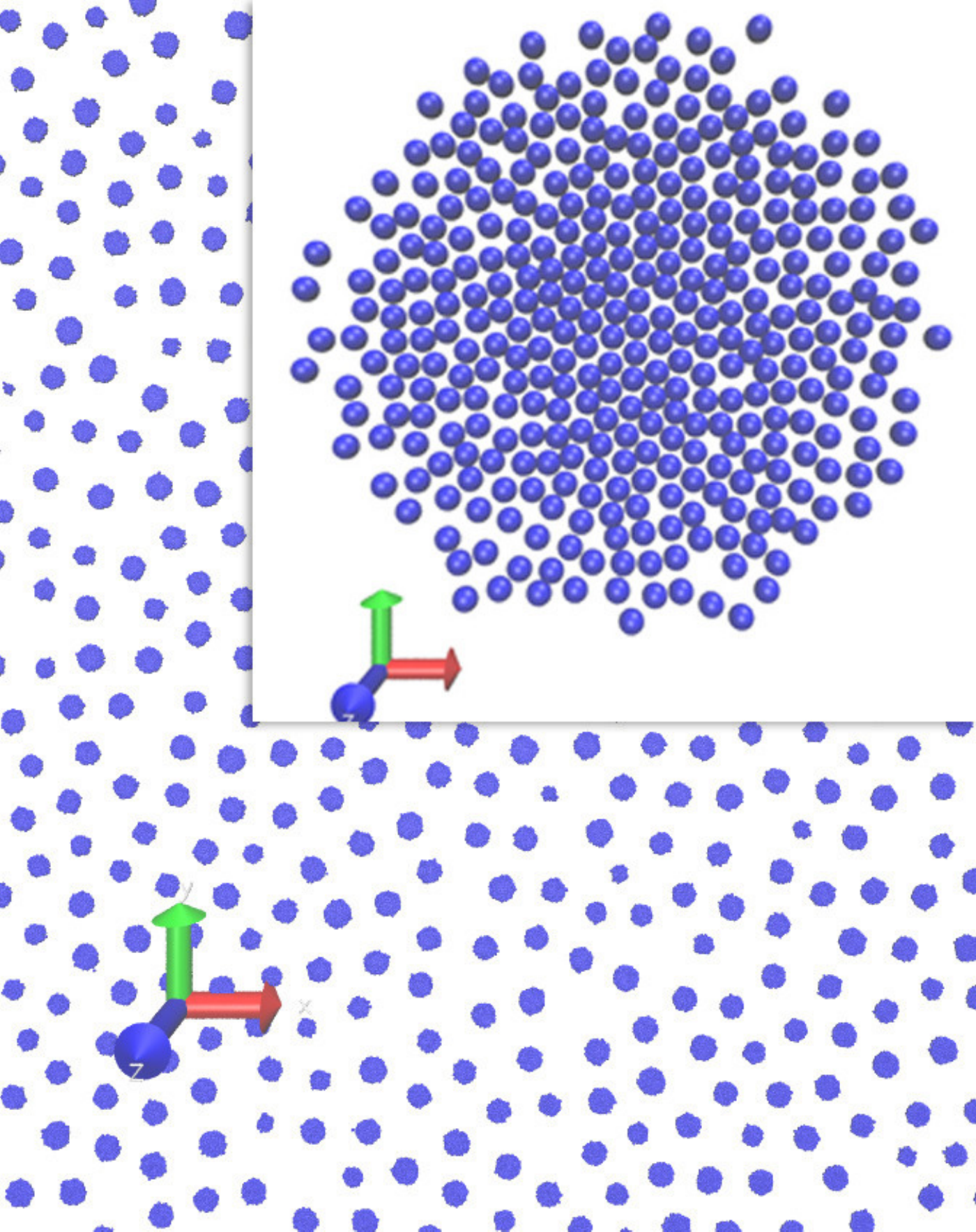}}
\hfill
 \subfloat[SALR-Gauss]{\includegraphics[width=0.30\linewidth]{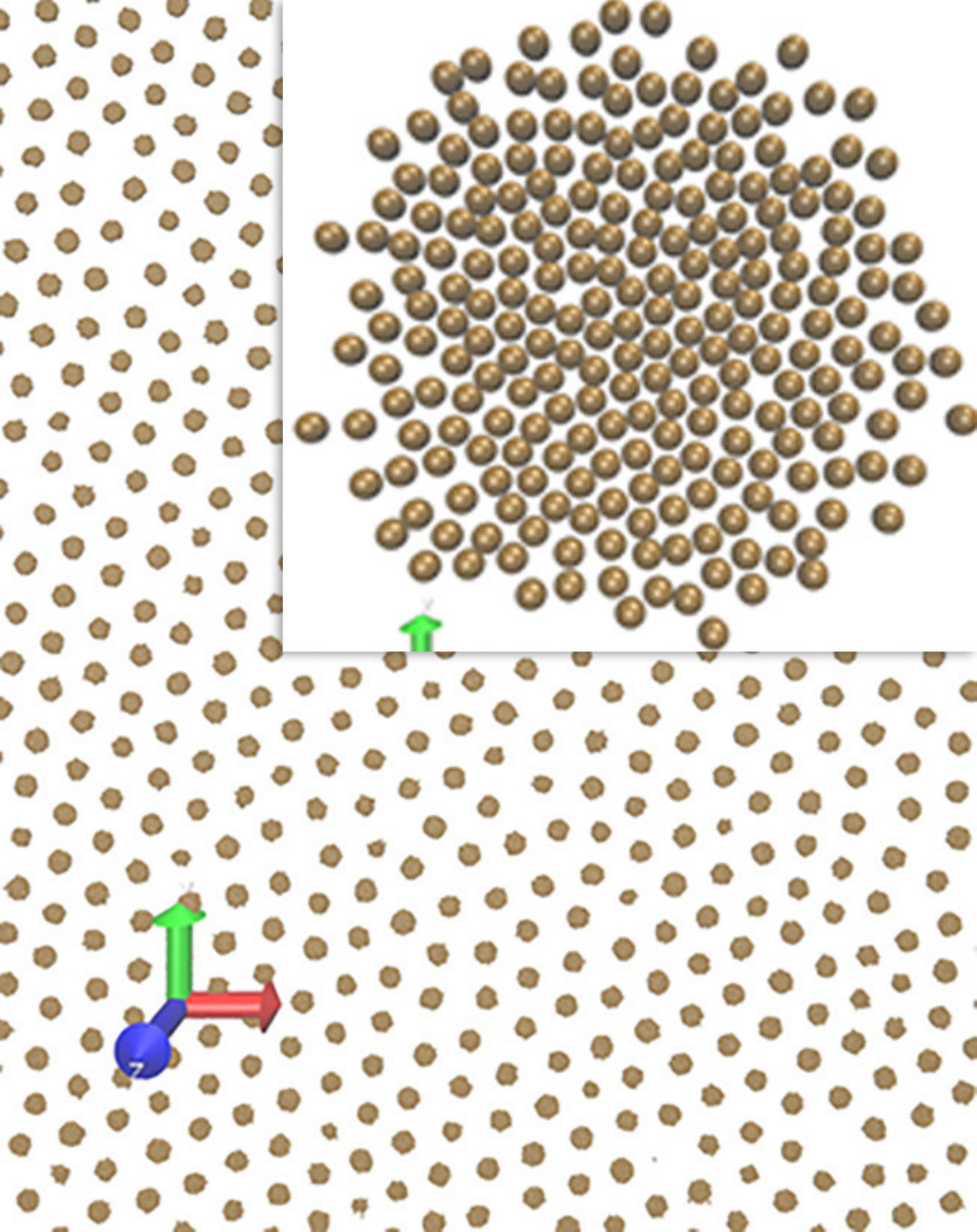}}
 \hfill
 \subfloat[SALR-OPP]{\includegraphics[width=0.30\linewidth]{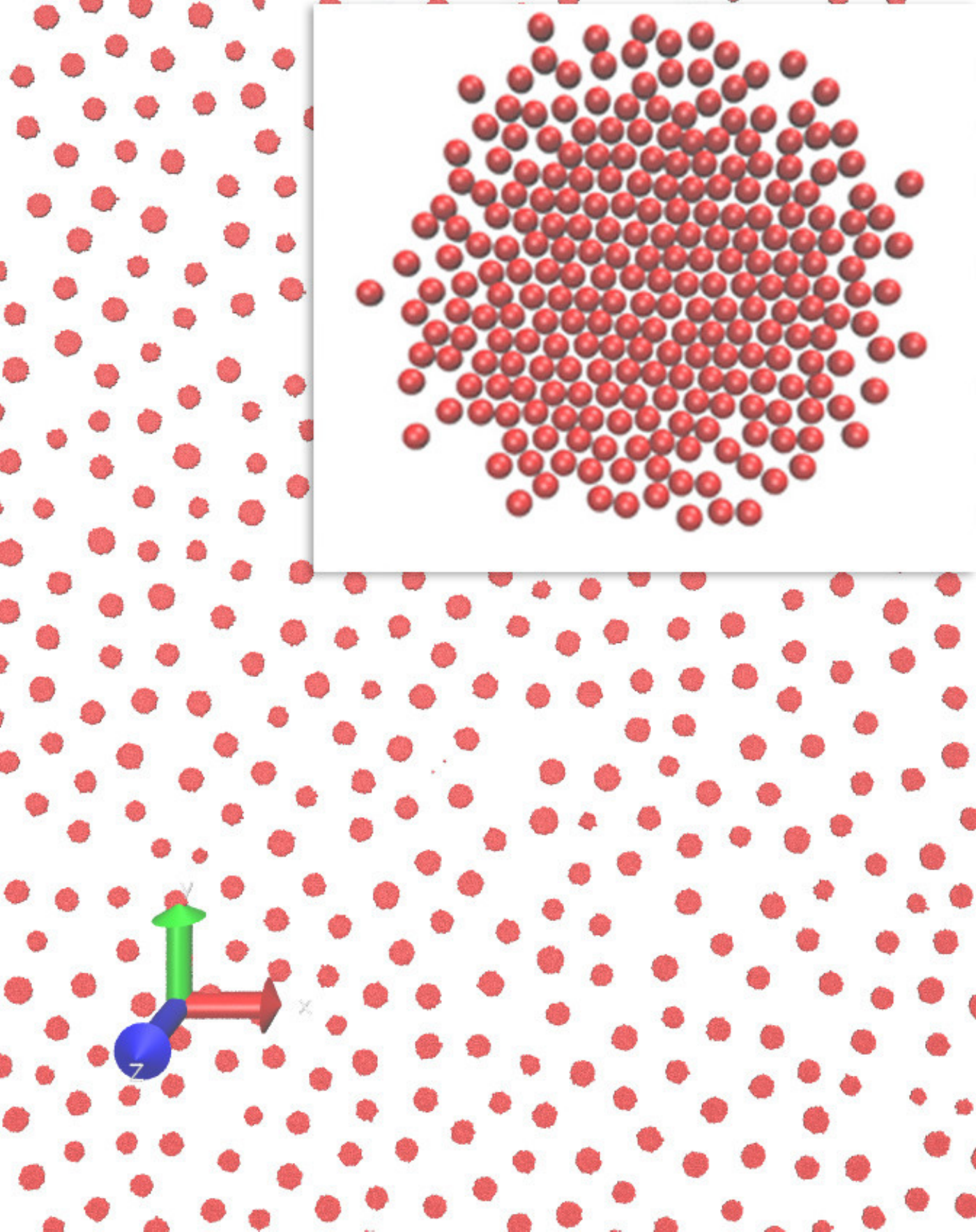}}

   \caption{Snapshots of the globular phase of: a) the SALR fluid b)
    SALR-Gauss c) SALR-OPP. The insets zoom in the configurations to display the internal
    structure of the clusters. All computed at $T^*=3$
    and $\rho\sigma^2=0.013$.}
\label{fig:clf}
\end{figure}

As to the patchy systems, density was set to
$\rho_A\sigma_{cc}^2=0.03$ as in Ref.~\onlinecite{Palaia2022}. Reduced
temperature is in this case defined as $k_BT/\epsilon_{cc}$ and all
simulations have been run at $k_BT/\epsilon_{cc}=1$. Again as in Palaia et
al.\cite{Palaia2022} the time step was set to $\tau = 0.01\tau_0^p$ 
with $\tau_0^p=(M_A\sigma_{cc}^2/\epsilon)^{1/2}$ with $M_B=5+n_p$,
being $n_p=4$ the number of patches in B particles. As to composition,
we have chosen to focus on equimolar systems ($n_A/n_B=1$) and systems
in which the number of A-particles can saturate all B-patches,
i.e. $n_A/n_B=4$. Simulations started from a square lattice
configuration at $k_BT/\epsilon_{cc}=25$ and cooled down to
$k_BT/\epsilon_{cc}=1$ along a 5 million step long ramp, further
equilibrated for another 20 million steps. Production runs were
50--60 
million steps long and averages calculated for 1000 equally spaced
configurations.

\section{Isotropic SALR systems}
\label{secRes1}
For simple SALR double exponential systems in 2D, Imperio and Reatto
\cite{Imperio2004,Imperio2006,Imperio2006a,Imperio2007} fully explored
the conditions of temperature and density that lead to the different
modulated phases that dominate the structural landscape. Moreover,
Archer and Wilding\cite{Archer2007a} thermodynamically identified two
first order transitions: a phase change at high dilution between a
vapor and a fluid of liquid-like spherical clusters (micelle-like
phase), and a transition between a liquid and a bubble
cluster phase. In Figure S1 in the supplementary information one can
appreciate the signatures of these two transitions in the two jumps
occurring both in pressure and in energy for low and high
densities. The high density transition is not exactly the one
mentioned by Archer and Wilding, since here it corresponds to a
transition from the lamellar onto the bubble phase. Given the fact that
these are plain NVT simulations it is not straightforward to assess
whether in this instance one has a first order transition, or just a
continuous structural transition. Figure S2 in the SI illustrates the
different modulated phases that occur as density is varied, even at a
relatively high temperature.

\subsection{Interaction potential and crystallization}
\begin{figure}[t]
 \includegraphics[width=0.75\linewidth]{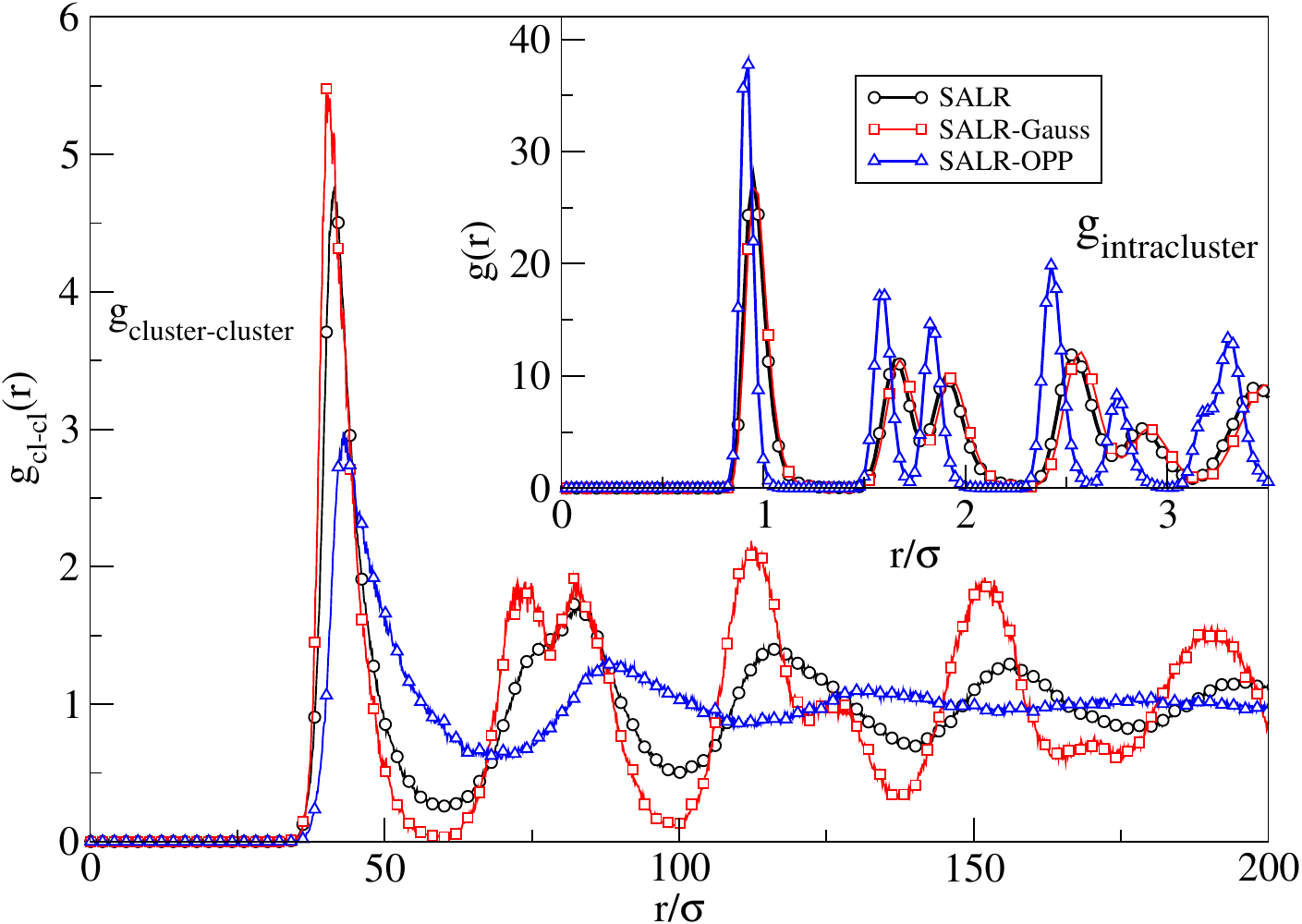}\\
 \caption{Cluster-cluster and intracluster (inset) pair distribution functions
   for our isotropic SALR interactions computed at $T^*=1$.\label{gcomp}}
 \end{figure}

\begin{figure}[h]
 \centering
 \includegraphics[width=0.75\linewidth]{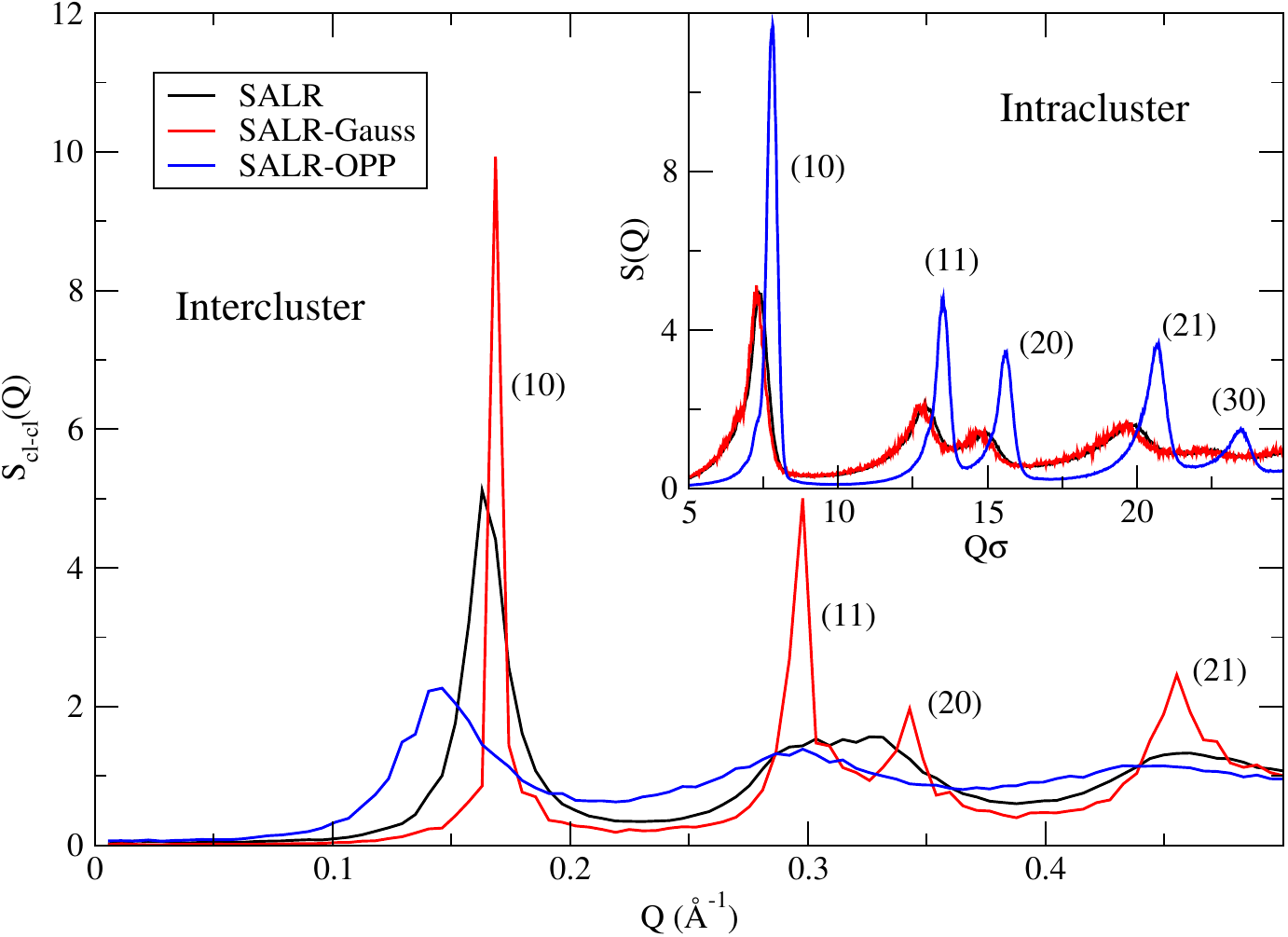}\\
 \caption{Intercluster and intracluster (inset) structure factors computed
   for isotropic SALR potentials at $T^*=1$. The numbers in parenthesis
   are the corresponding Miller indices of the fundamental
   diffractions of the triangular lattice.\label{sqtot}}
\end{figure}

Figure \ref{fig:clf} illustrates the formation of the
globular phases for  the three models of isotropic SALR interactions
considered in this work. From these simple snapshots taken at
relatively high temperature there is a couple of salient features to
mention. First, looking at the insets that display the internal
cluster structure, one sees that the inner structure of SALR-OPP clusters is
arranged in a triangular lattice. In contrast, both the SALR and
SALR-Gauss clusters are internally liquid like, even
  if some residual sixfold coordinations are still visible. On
the other hand, the cluster-cluster arrangement 
of the SALR model shows remnants of a triangular lattice, although with
multiple defects. The SALR-Gauss clusters form an almost perfect
triangular lattice of liquid clusters. In
contrast,  the SALR-OPP system displays a liquid like structure
of otherwise crystalline clusters. This
qualitative visual result can be quantified using the $\phi_m$ order
parameter\cite{Nelson1979}, namely
\begin{equation}
\phi_m = \frac{1}{N_n}\left<\sum_i^{N_n}e^{in\theta({\bf r}_{ij}}\right>
\label{phim}
\end{equation}
where the average is performed over all atoms (or cluster centers) and
configurations and the sum 
runs over $N_n$ nearest neighbors. The angle $\theta({\bf r}_{ij})$ is
formed by the vector joining the central atom, $i$, with its nearest
neighbor, $j$, with the $x$-axis. The brackets denote the ensemble
average. Here we add the restriction that $N_n$ must be equal to $n$.
This means that in order to calculate the relevant order parameter 
that monitors the build-up of  triangular lattice arrangements, $\phi_6$,
 only atoms with 6 nearest neighbors will be taken into
account. We have calculated this quantity for all particles, and for
cluster configurations ($\phi_6^{cl}$), where now clusters' geometric
center positions are used to evaluate the order parameter. In addition, we
have also calculated the average orientational order parameter profile
of the clusters, $\phi_6(r)$, as well as the cumulative order
parameter profile
\begin{equation}
\phi_6^c(r) = \frac{2}{r^2}\int_0^r r'\phi_6(r') dr'.
\label{phi6c}
\end{equation}
 This
latter quantity is plotted for the three systems of Figure 
\ref{fig:clf}  at $T^*=3$ in Figure S3 in the SI. From the cumulative  order
parameter we have chosen two values, namely,  $\phi_6^{c(1)}$, and
$\phi_6^{c(2)}$, which correspond to the average order parameter profile
calculated at the first and second particle layers around the
cluster's geometric center.  The cumulative order parameter
$\phi_6^{c}(r)$ typically displays a maximum in the first layer to
then decrease due to the lower particle coordination of the particles at
the cluster outer boundary (cf Figure S3 in the SI). In Table \ref{tab:order} we can see both
$\phi_6^{cl}$, $\phi_6^{c(1)}$, and $\phi_6^{c(2)}$ for three different
temperatures.  The results from Table \ref{tab:order} are a clear indication that
adding a Gaussian term to raise the repulsive barrier  of the SALR
potential favors the formation of a crystal-like triangular lattice
phase of clusters. Again, this is not the case when the SALR is decorated
with the OPP interaction. We see that the cluster-cluster order is
destroyed, but at the same time a triangular lattice local ordering
occurs at the intracluster level when the temperature is lowered. These
effects are also illustrated by looking at the cluster-cluster pair
distribution functions and the intracluster $g(r)$, both depicted in
Fig.~\ref{gcomp} for $T^*=1$, and  very specially the corresponding structure factors
presented in  Figure \ref{sqtot}.

\setlength{\tabcolsep}{15pt}
\begin{table*}[b]
 \caption{Cluster-cluster orientational order parameter ( $\phi_6^{cl}$) and cumulative  orientational
   order averages computed in the first  ($\bar{\phi}_6^{c(1)}$ and
   second layers ($\bar{\phi}_6^{c(2)}$\label{tab:order}}
 \begin{tabular}{ccccccc}
 \toprule

 $T^*$ & \multicolumn{2}{c}{SALR}
 &\multicolumn{2}{c}{SALR-Gauss}&\multicolumn{2}{c}{SALR-OPP} \\ \hline
          & $\phi_6^{cl}$ & $\bar{\phi}_6^{c(1)}/\bar{\phi}_6^{c(2)}$ &
 $\phi_6^{cl}$ & $\bar{\phi}_6^{c(1)}/\bar{\phi}_6^{c(2)}$&
 $\bar{\phi}_6^{cl}$ & $\bar{\phi}_6^{(1)}/\bar{\phi}_6^{(2)}$  \\ \hline
 1.0 & 0.137 & 0.71/0.09  & 0.745 & 0.59/0.07  &  0.018  &  0.76/0.12 \\
 3.0 & 0.024 & 0.39/0.10  & 0.627 & 0.27/0.07 &  0.020  & 0.53/0.11  \\
 6.0 & 0.037  & 0.15/0.02  & 0.001 & 0.06/0.01  & 0.019  & 0.37/0.06  \\
 \end{tabular}
 \end{table*}

\begin{figure*}
  \subfloat[\label{msd}]{\includegraphics[width=0.5\linewidth]{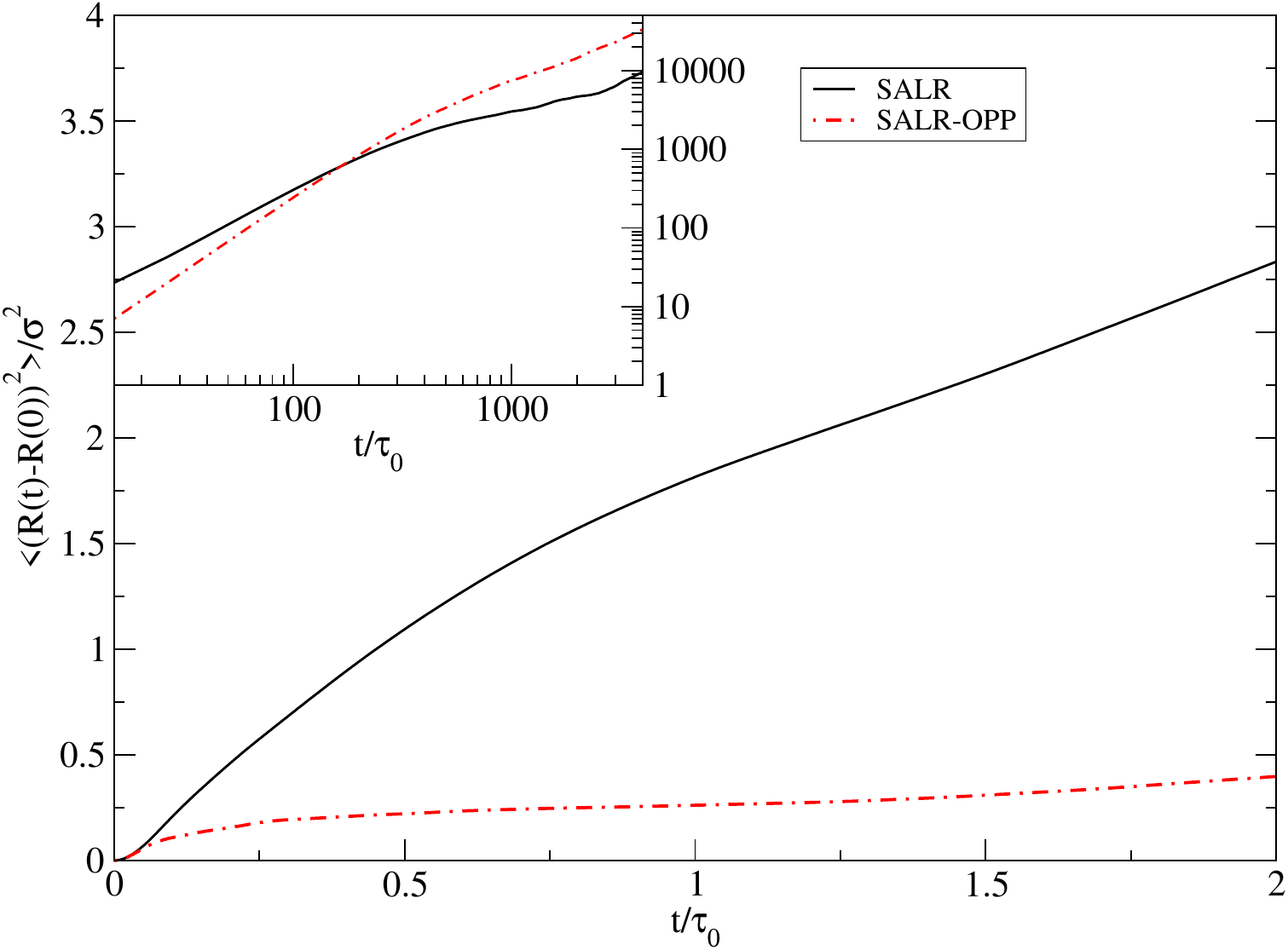}}
  \subfloat[\label{zw}]{\includegraphics[width=0.5\linewidth]{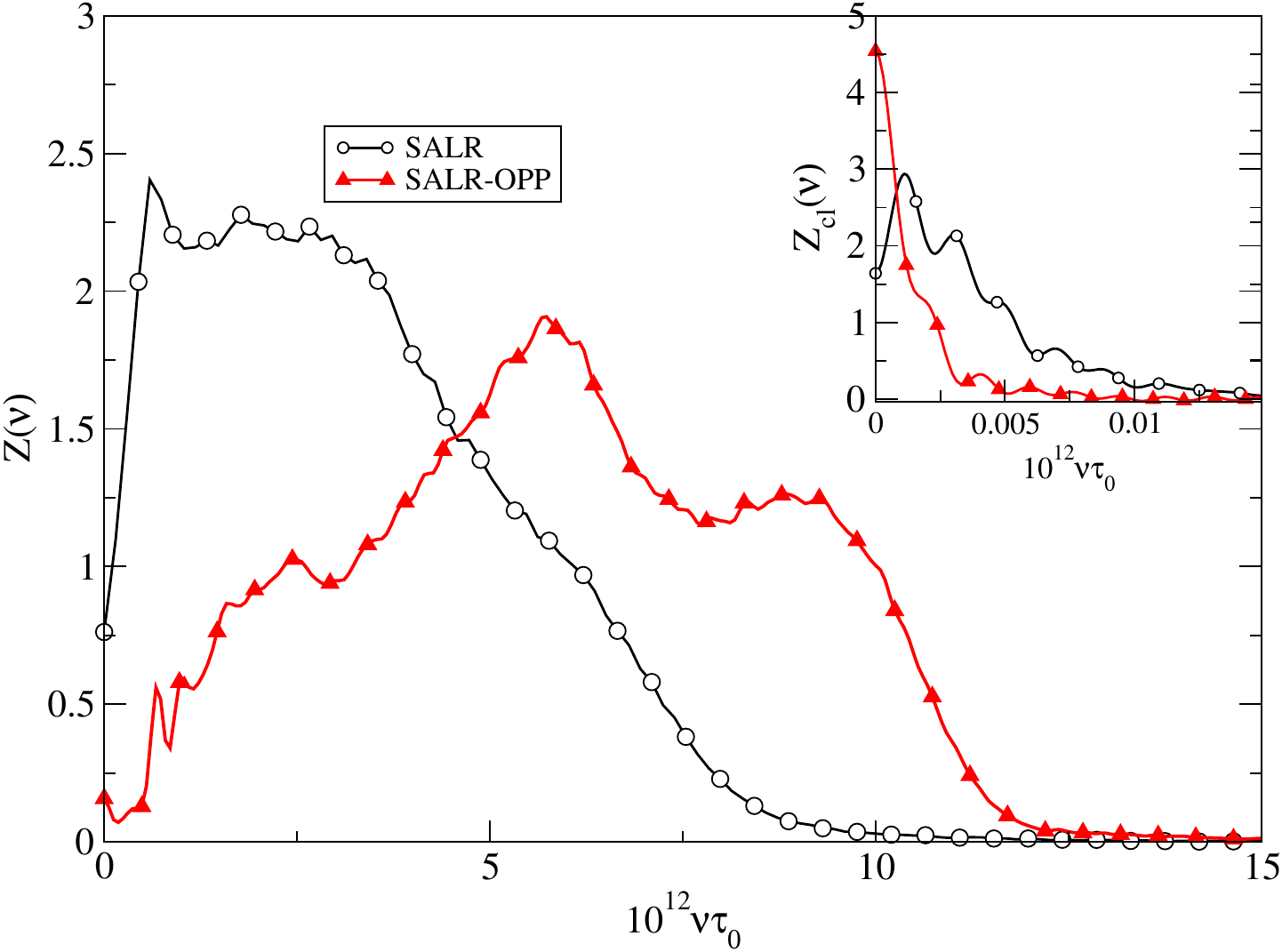}}
\caption{(a) Mean square displacements for the SALR and SALR-OPP
  fluids at $T^*=3$. The inset depicts the long time behavior. (b)
  Frequency spectra of the SALR and SALR-OPP
  fluids at $T^*=3$. The inset depicts the clusters center of mass
  frequency spectra.\label{dyn}}
\end{figure*}

We immediately see  that the effect of adding a Gaussian term to the
maximum of the SALR potential is precisely to facilitate the
crystallization of the intercluster lattice, as shown by the splitting
of the peaks in Fig.~\ref{gcomp} and very specially by the narrow
and high intercluster diffraction peaks shown in Figure \ref{sqtot}. These
correspond precisely to the main diffractions of a triangular lattice
powder diffractogram. On the other hand the OPP
decoration of the SALR minimum facilitates intracluster
crystallization, as evidenced by the huge and narrow peaks of the
$g(r)$ displayed in the inset of Fig.~\ref{gcomp} and the
corresponding structure factor in the inset of Figure~\ref{sqtot}. At the same time, the
coordination shells become narrower, most likely due to the fact that
particles get trapped in the deep and narrow minima of the OPP
potential (cf. Fig.~\ref{ur}). Even if the integrated averaged
attraction is the same for SALR and SALR-OPP, these deep minima have a
strong impact on intracluster particle mobilities, easing the
formation of local crystal-like structures. These effects on the
particle mobilities can be appreciated in Figs.~\ref{msd} and
\ref{zw} where we have plotted the mean square displacement and
frequency spectra for SALR and SALR-OPP fluids respectively. One
immediately sees that the diffusion of the SALR fluid is liquid-like
both for short times (intracluster) and long times (intercluster),
although it lags behind that of the SALR-OPP in the latter
instance. SALR-OPP clusters have a large mobility, whereas diffusion
at short times is hampered since particles get trapped in the deep
potential minima. This is more obvious when looking at the frequency
spectra (cf. Fig.~\ref{zw}). The vibrational density
  of states, $Z(\nu)$, quantity computed from the velocity autocorrelation
  function, in the case of   the SALR-OPP is clearly solid-like
($Z(\nu\rightarrow 0) \sim 0$, i.e., no diffusion) for
the range of intra-cluster movements. It displays pronounced maxima stemming
from particle vibrations in the different minima of the potential. In
contrast the SALR $Z(\nu)$ presents a single flat maximum with $Z(\nu=0)
\ne 0$, which indicates the presence of diffusivity within the
clusters. At much lower frequencies (cf inset) we have the frequency
spectra stemming from cluster movements. Here that of the SALR fluid
presents clear maxima, which correspond to phonon-like excitations of
the severely distorted triangular lattice formed by the clusters. 
SALR-OPP clusters behave liquid-like and with a large value of $Z_{cl}(0)$,
i.e. a large diffusion constant. Despite  of this liquid-like
behavior of the clusters, the presence of peaks particularly in the
plain SALR model, is the results of local vibrations in particle
cages. Recall that the SALR model retains a certain amount of
intercluster triangular order, although with multiple defects, and this
quasi-hexagonal cages that hamper particle movement induce the
presence of multiple maxima in $Z_{cl}(\nu)$, although much less intense
than those seen in the intracluster SALR-OPP $Z(\nu)$. 

\subsection{The role of cluster size}
\begin{figure}
  \includegraphics[width=0.7\linewidth]{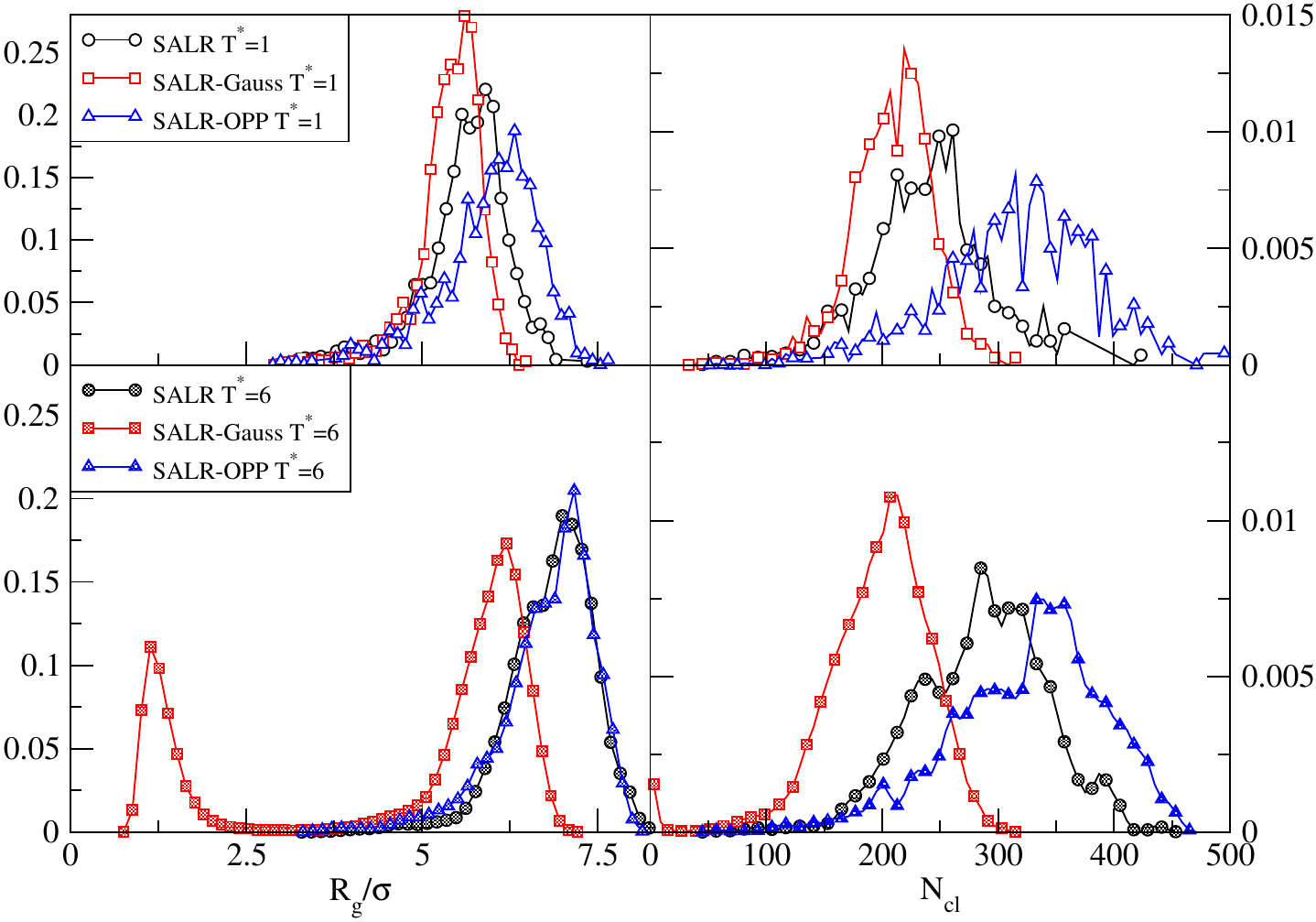}
  \caption{Gyration radius (left panels) and cluster
size (right panels) distributions for SALR, SALR-OPP and SALR-Gauss
fluids at two temperatures of interest.\label{clustRd}}
\end{figure}

The question to be answered is what might be the root cause of this
marked differences between the three systems under consideration. A look at the gyration radius and cluster
size distributions (Fig.~\ref{clustRd}) throws some light onto the problem. 
One can appreciate, that despite the fact that the characteristic
wavelength of the potentials, $Q_0$ is identical for SALR and
SALR-OPP, both the gyration radii and cluster size distributions are
wider and have a larger mean in the latter instance. A wider
distribution means a larger degree of polydispersity, which inhibits
the crystallization of the clusters and favors liquid-like
behavior. An extreme case is the SALR-Gauss which has even narrower
and more symmetric  distributions. Interestingly, for the highest
temperature an anomaly of small radii clusters occurs with a $R_g$
distribution that is bimodal. Nonetheless, the statistical weight of
these small clusters is minimal as reflected in the tiny maxima
occurring for small $N_{cl}$ in the cluster-size distribution. The
decrease of polydispersity favors the crystallization of the
SALR-Gauss fluid, as evidenced by the $g(r)$, $S(Q)$ and bond
orientational order parameters discussed above. As mentioned, the
intracluster peculiar  behavior of the SALR-OPP fluid is basically
conditioned by the presence of 
multiple narrow minima within the broad attractive valley of its
interaction potential. On one hand, at a temperature where other SALR
systems retain the features of fluid clusters, the fact that
particles are trapped in these narrow minima favors intra-cluster
crystallization. On the other, the presence of multiple minima favors
the stabilization of clusters of different sizes. In fact, the
cluster size distribution of the SALR-OPP shows multiple maxima and
minima in contrast with the rather smooth distributions present in
the SALR and SALR-Gauss systems. This is reflecting the fact that the presence
of these multiple minima in the interaction tends to favor certain
sizes. A general conclusion can be drawn: on one hand increasing the
maximum of the SALR potential (i.e. the nucleation barrier) reduces
cluster size and polydispersity, and the presence of multiple
attraction basins within the attractive range of an effective SALR
interaction induces the opposite effect, increasing intercluster
mobility but reducing intracluster diffusivity. 

\subsection{Long wavelength behavior and hyperuniformity}
Hyperuniformity  is a rather peculiar property in disordered systems
first discovered by Torquato and
Stillinger\cite{Torquato2003,Torquato2018a}, by which long wavelength
density fluctuations are heavily damped. In the last two decades
 it has been shown that  hyperuniform materials 
 display particularly interesting
 optical\cite{Froufe-Perez2017,Zhou2019,Milosevic2019} and
 acoustic properties\cite{RomeroGarcia2019,Cheron2022}. Recently
  its importance for the description of hidden order in biological
 systems has grown, in particular since Jiao and
 coworkers\cite{Jiao2014}  discovered the presence of disordered
 hyperuniformity in the spatial distribution of photoreceptors in
 avian retina. Later, in Ref.~\onlinecite{Lomba2020} it was shown that
 a simple effective model with long range interactions could reproduce
 qualitatively the spatial distribution found in avian retina. Klatt
 and coworkers\cite{Klatt2019} pointed out the existence of this type
 of universal hidden order in amorphous cellular geometries. More
 recently also it was reported the presence of hyperuniform cell packings on
 a growing surface\cite{Ross2025}. Interestingly, these findings in the
 biological realm have also had an impact in optical materials design as shown
 by Li et al. \cite{Li2018}.  In 
 the recent work of  Diaz-Pozuelo et 
al.\cite{DiazPozuelo2025} it was evidenced that a 3D counterpart of
our simple SALR model exhibited a 
cluster-cluster structure factor whose  long wavelength behavior was
consistent with the presence of effective hyperuniformity. On the
other hand, that  was not the case for the total structure factor (cf. Figure 10 and 11 in
Ref.~\onlinecite{DiazPozuelo2025}). It was
 argued that the reason for this different behavior might well lie in
 the polydispersity of the sample, and also on the insufficiently
 small-Q range accessible to the simulation in 3D.

\begin{figure}
   \includegraphics[width=0.5\linewidth]{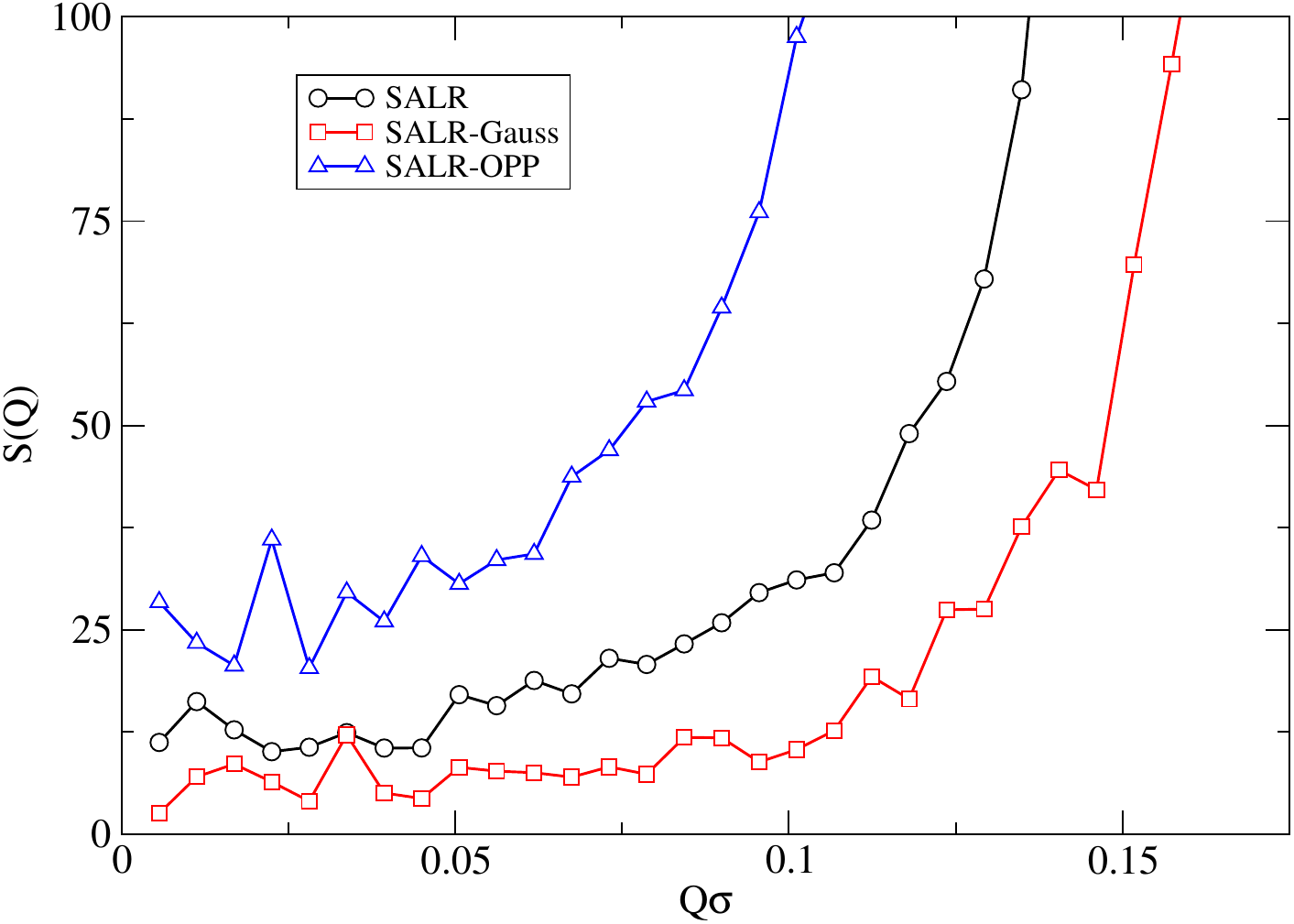}
   \caption{Low-Q behavior of the total (particle-particle) structure
     factor for the three isotropic SALR systems  for $T^*=1$.\label{sQlow}}
 \end{figure}

 As discussed above,  the potential
 implications of the presence of hyperuniformity in biological
 self-assembling systems are considerable. Since our samples
 are bidimensional one  can access Q-values almost one order of
 magnitude smaller than in Ref.~\onlinecite{DiazPozuelo2025} and
 therefore one can  better assess sample size effects. Additionally, our choice of
 interactions  facilitates an evaluation of the role played by
 polydispersity,  a feature that might also be responsible for the
 different behavior exhibited by the cluster-cluster and the total
 structure factor.  For this reason, we decided to explore to what
 extent our isotropic SALR systems are capable of displaying
 hyperuniformity. This is illustrated in Fig.~\ref{sQlow}, where we
 have plotted the low-Q total structure 
 factor of our three isotropic SALR systems for $T^*=1$. 
 One immediately sees that the trend is as expected. The smaller the
 polydispersity the more the low-Q behavior tends to hyperuniformity,
 i.e. $\lim_{Q\rightarrow 0}S(Q) \approx 0$. If one compares Fig.~\ref{sQlow} with the inset of Figure 11 in
 Ref.~\onlinecite{DiazPozuelo2025} one can appreciate that for all
 systems now the structure factor is one order of magnitude smaller,
 which is in part due to the sampling of smaller Q-values. But also, one
 appreciates a five-fold decrease in magnitude when comparing the
 results of SALR-OPP with those of SALR-Gauss (two-fold when comparing
 with SALR). This confirms our assumption that controlling
 polydispersity one can improve the hyperuniform behavior of the
 material. Note in passing, that for the SALR-Gauss the low-Q behavior
 is similar to that of the cluster-cluster $S(Q)$, namely, there is an
 entire region of $Q$ values where the value of $S(Q)$ seemingly falls below the
 effective hyperuniformity threshold\cite{Chen2018}
 (H=$S(Q_{max})/S(0))>10^3$), qualitatively recalling the behavior of 
 {\em stealthy hyperuniform} systems. 

 \subsection{Intra-cluster phase separation}\label{in_phase-sep}

\begin{figure}[b]
\includegraphics[width=0.7\linewidth]{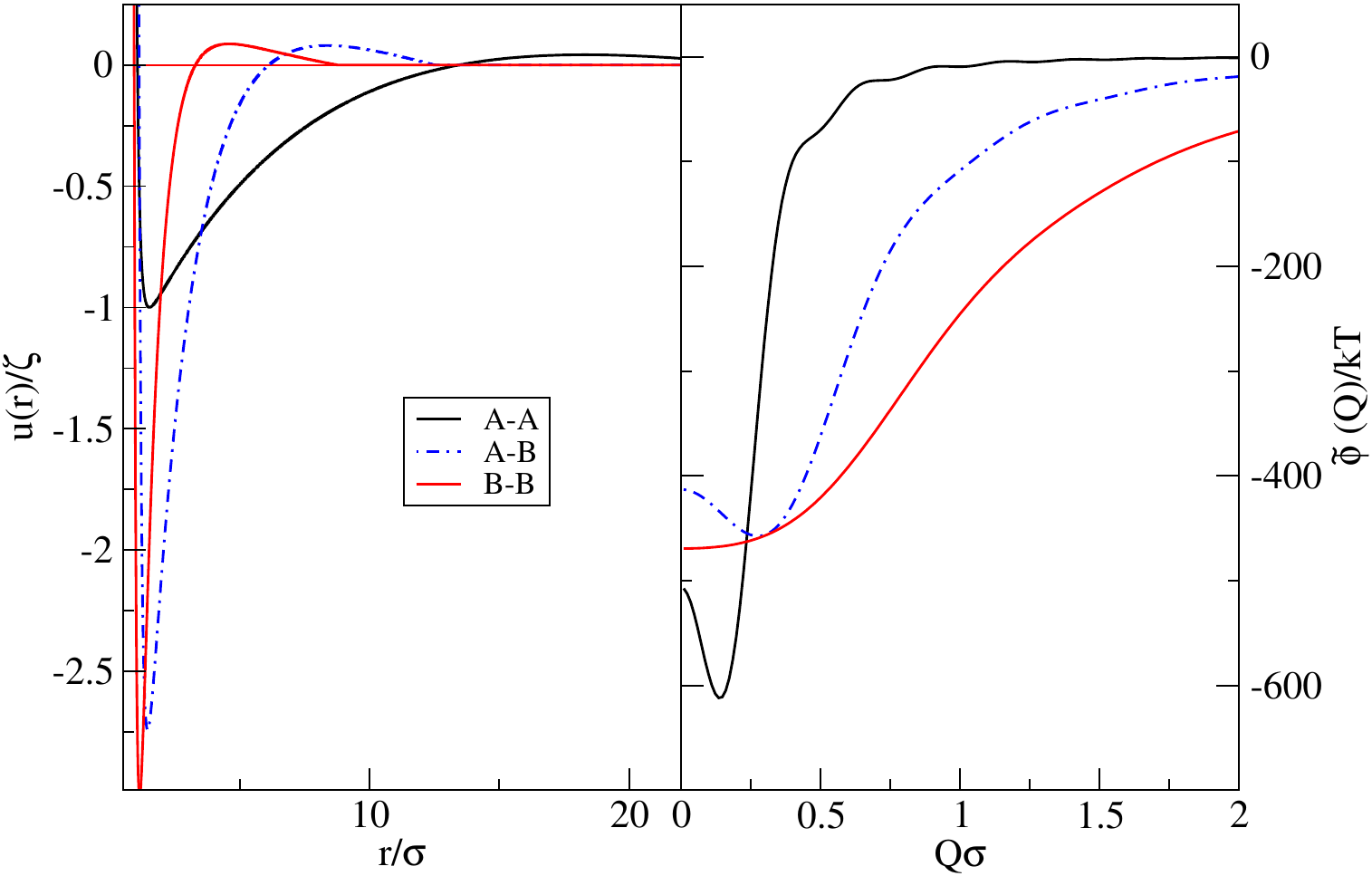}
\caption{Left) SALR mixture interactions giving rise to intracluster
  segregation. Right) Fourier transform of the analytic part of the
  interactions. The position of the minima correspond to the
  characteristic wavevectors defining the correlation lengths of the
  modulated phases.\label{umix}}
\end{figure}

In the previous Subsections we have seen how condensation into stable
aggregates can be controlled with simple isotropic SALR potentials,  and to
what extent one can manipulate the polydispersity of the samples and
the internal phase transitions of the clusters. Note that as discussed
before crystallization is not so straightforward in three dimensional
systems, which implies that when in this work we look at the formation
of crystal like structures, in 3D systems we would be mostly dealing with
amorphous or quasi-amorphous states \cite{DiazPozuelo2025}. The
formation of dynamically arrested states in aggregates is of extreme
relevance when studying biomolecular condensates given its
connection with well known pathologies\cite{Ranganathan2022}. On the
other hand, in some other instances like in the case of TDP-43 the
 condensate undergoes a demixing 
transition\cite{PantojaUceda2021}. Obviously this is the result of
changing thermodynamic 
conditions and interactions due to conformational changes. In many
instances a mere concentration change suffices to initiate the
transition, but one 
might also ask what changes should display the effective interactions
to induce such a change. In order to have a segregation of small
aggregates within the condensates as seen in
Ref.~\onlinecite{PantojaUceda2021}, a minimal condition is to have a
mixture of isotropic SALR particles: a dominant component (A) in which
the maximum of the interaction, $d_m$, appears at sufficiently long
distances, and a dilute component B that segregates at the surface of
large A-condensates. This implies a B-B interaction maximum at a much smaller
distance than $d_m$. Taking into account that B-particles will segregate within
A-clusters, if  the A-B interaction differs in range and/or intensity from
both B-B and A-A,  the net condensate-condensate
interaction will be anisotropic. This will hamper the formation of
an ordered lattice of clusters, an undesirable feature when modeling
biological systems.
\begin{table}[t]
 \caption{Potential parameters for the asymmetric interactions between the binary mixture components\label{tab:cl_sep_par}}
 \begin{tabular*}{\columnwidth}{@{\extracolsep\fill}llllllll@{\extracolsep\fill}}
 \toprule

  & $K_r^{ij}$ & $K_a^{ij}$ & $\alpha_r^{ij}$ & $\alpha_a^{ij}$ & $\epsilon_{ij}/\epsilon$ & $R_c^{ij}/\sigma$ \\ \hline

 \multicolumn{8}{l}{SALR} \\ \hline
 
 A-A & 1.0 & 1.5 & 0.05 & 0.12 & 1.0 & 25.0 \\
 A-B & 0.3 & 0.7 & 0.01 & 0.6 & 6.0 & 12.5 \\
 B-B & 0.2  & 9 & 0.12 & 1.15 & 1.25 &  8.75 \\
 \end{tabular*}
 \end{table}
With all this in mind, after some trials we arrived at the set of
parameters collected in Table \ref{tab:cl_sep_par}, using the functional form of
Eq.~(\ref{salr}). A representation of the corresponding interactions
as well as the Fourier transforms of their analytic components is
presented in Fig.~\ref{umix}

It is worth noticing that despite the presence of a clear maximum in
$u_{BB}(r)$, the minimum in $\tilde{\phi}_{BB}(Q)$ shrinks to
$Q_0\rightarrow 0$. This implies that no finite size modulation will
be directly induced by this interaction, but mostly by $u_{AB}$.

\begin{figure}
\includegraphics[width=0.7\linewidth]{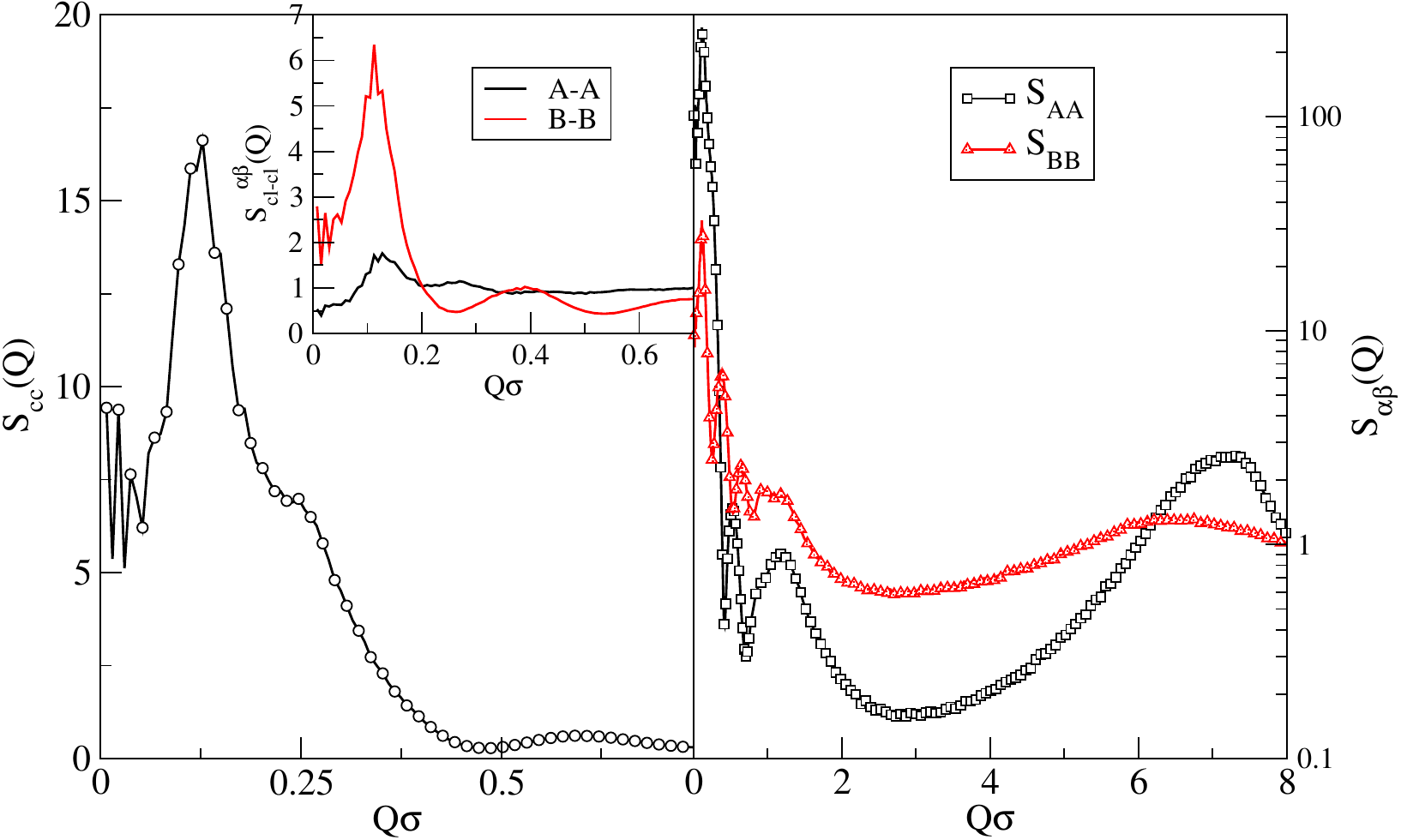}
\caption{Left) Concentration-concentration structure factor for the
  SALR mixture  for $x_B=1/2$ and $\rho\sigma^2=0.008$. The inset corresponds to the A and B
  cluster-cluster structure factors.  Right) AA and BB structure
  factors for the SALR mixture.\label{smix}}
\end{figure}

\begin{figure}[t]

\subfloat[$N_A=5N_B$, $\rho\sigma^2=0.0008$]{
\includegraphics[width=0.42\linewidth]{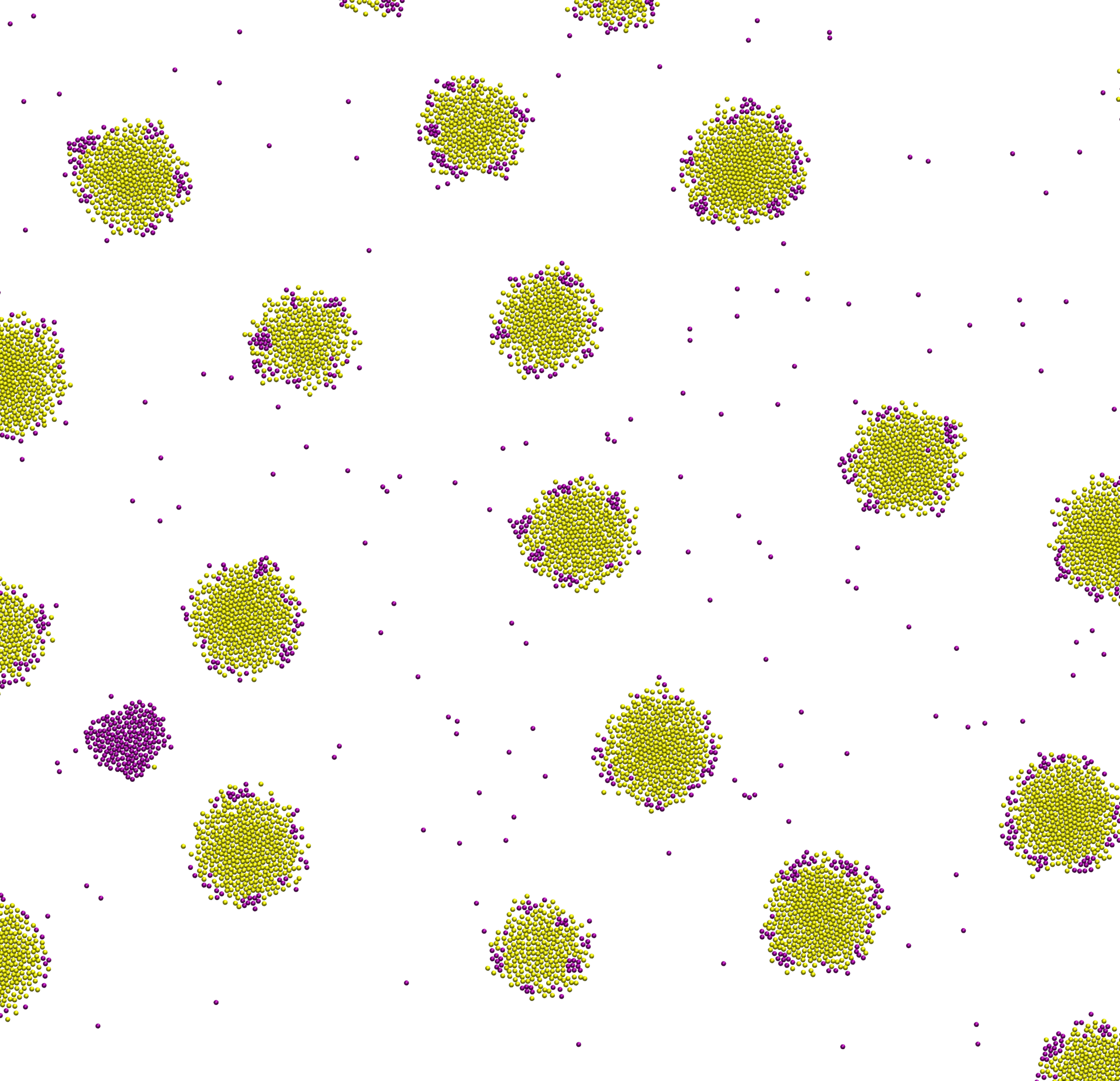}}
\hfill
\subfloat[$N_A=2N_B$, $\rho\sigma^2=0.0008$]{
\includegraphics[width=0.42\linewidth]{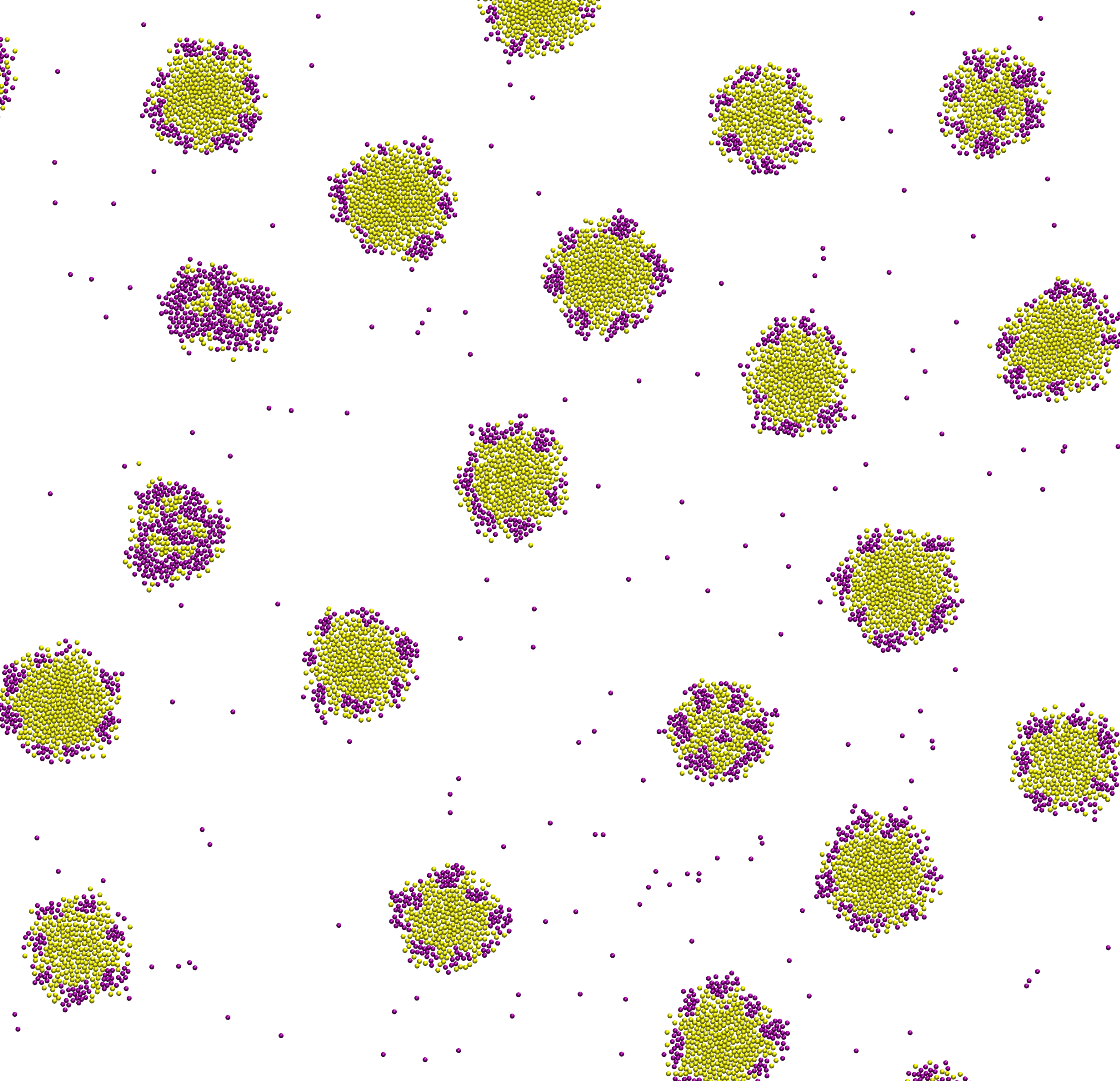}}
\hfill \\
\subfloat[$N_A=5N_B$, $\rho\sigma^2=0.02$]{
\includegraphics[width=0.42\linewidth]{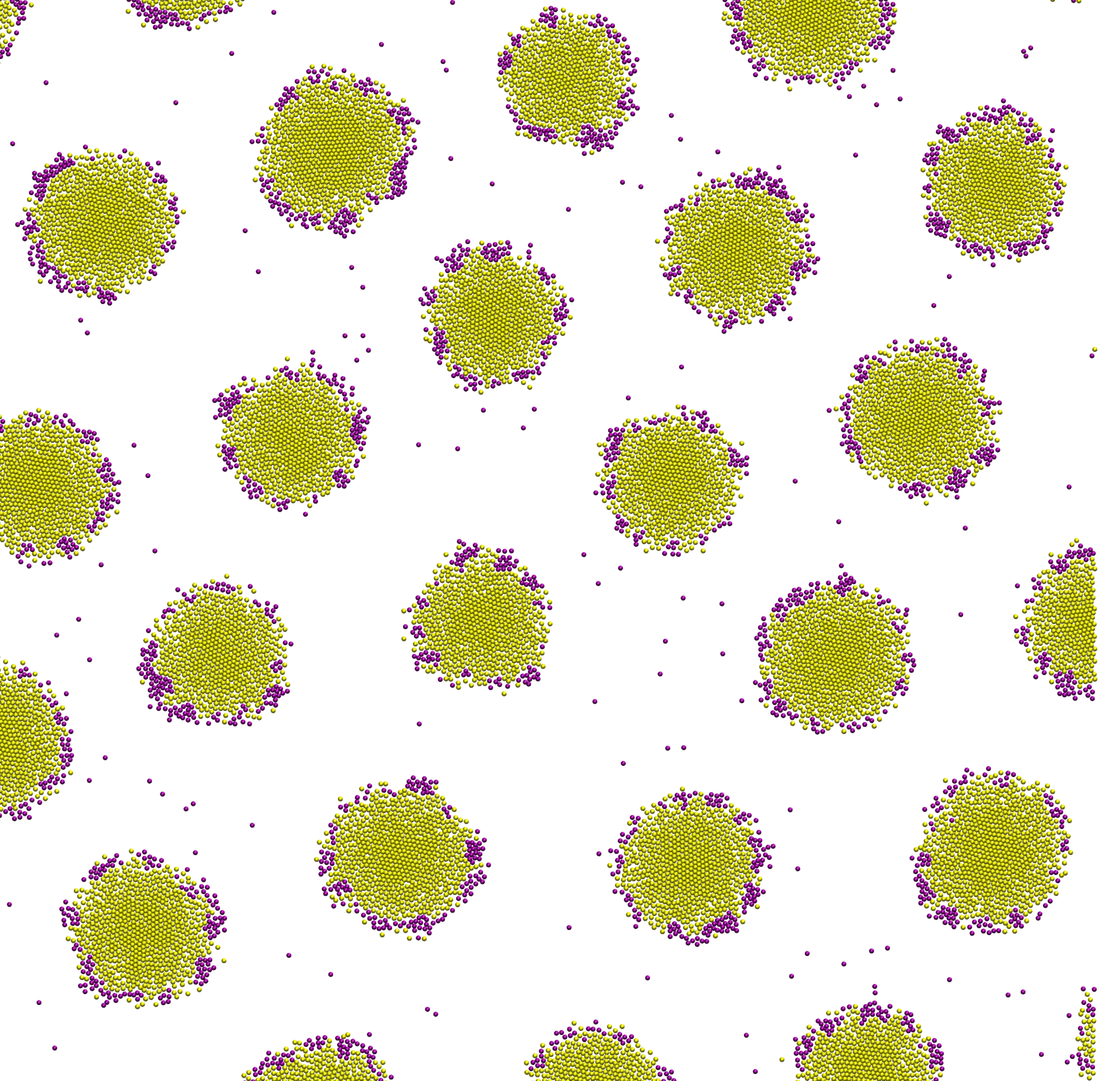}}
\hfill
\subfloat[$N_A=2N_B$,  $\rho\sigma^2=0.02$]{
\includegraphics[width=0.42\linewidth]{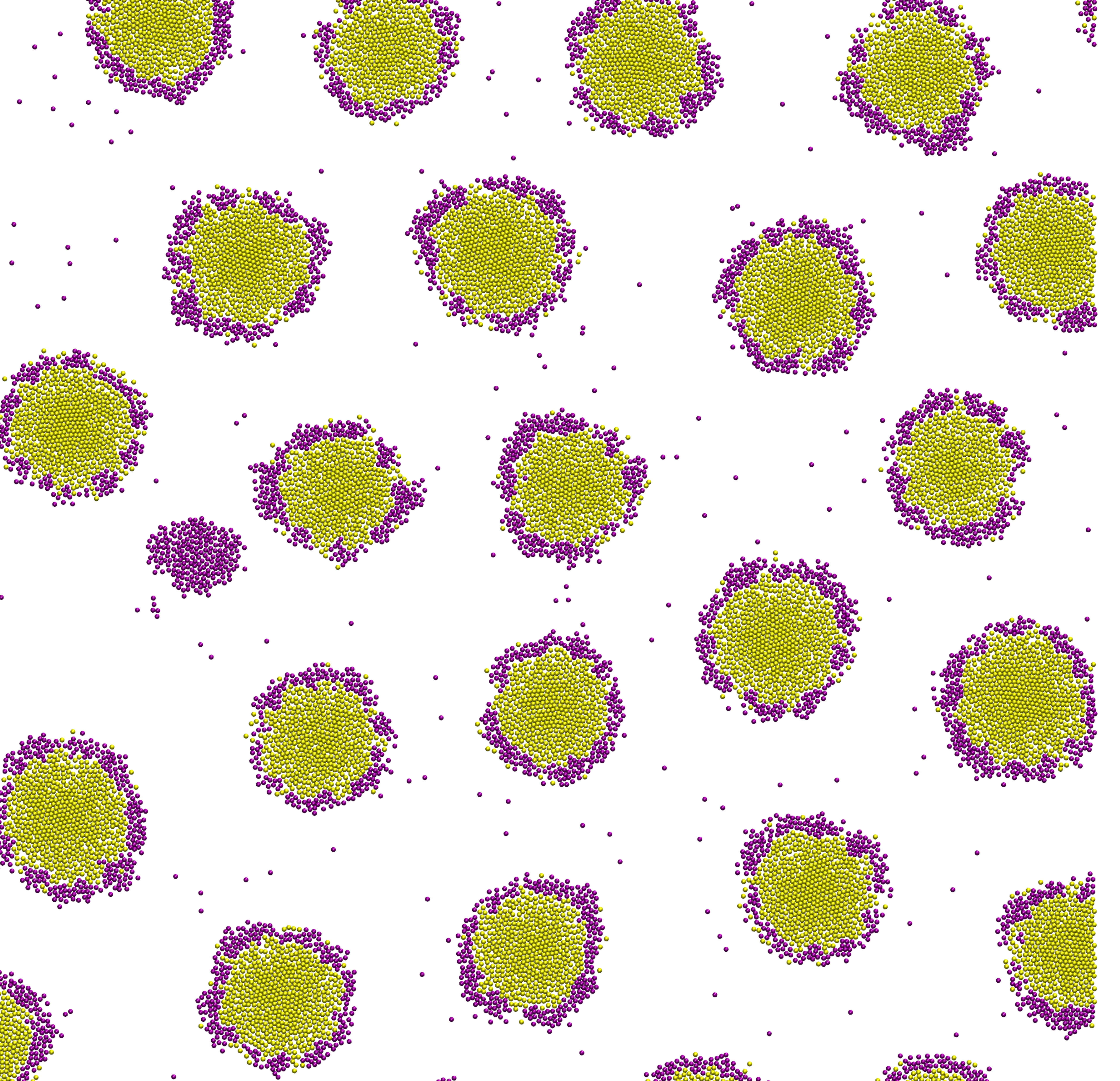}}
\hfill
\caption{Snapshots of the clustered system with asymmetric
  interactions promoting intra-cluster separation for two different
  concentrations of particles of type A (yellow) and B (violet).\label{fig:sep}}
\end{figure}

We have now considered systems at densities  $\rho\sigma^3
=$ 0.02, 0.008 and compositions $x_B=1/6, 1/3$ and $1/2$. The
presence of the phase separation is detected by the long wavelength
fluctuations of the concentration-concentration structure factor,
which is plotted in  the left panel of Fig.~\ref{smix} together the
corresponding partial AA and BB cluster-cluster structure factors
(see inset) and the total partial structure factors (left graph). The
concentration-concentration structure factor
is defined as 
\begin{equation}
S_{cc}(Q) = x_B^2S_{AA}(Q)+x_A^2S_{BB}(Q)-2x_Ax_BS_{AB}(Q).
\end{equation}
It  is known to take significantly large values at low wavevectors when concentration
fluctuations occur, and to tend to $x_Ax_B$ for random mixtures. The presence of a marked peak
around $0.15\sigma$ 
is an indication of modulated concentration fluctuations, not a simple
demixing. The relatively large values at $Q=0$ are the result of the
separation of B particles to form a gas in the intercluster
space. From the cluster-cluster 
structure factors depicted in the inset one can appreciate that the B
particles present clear modulations at intermediate Q, which is an
indication of correlations between relatively small clusters. On the
right panel, the AA and BB structure factors confirm these
conclusions. The peak of the cluster-cluster structure factors seen in
 the inset of the left panel of Figure \ref{smix} stem solely from correlations
 between  the geometric
 centers of the clusters. The large  values of the total partial
 structure factors on the right panel result from the convolution of
 the cluster-cluster structure factor with the inner cluster $S(Q)'s$.
These large peaks in  the total $S_{\alpha\beta}(Q)$ 
 near $Q_0$ indicate very strong short range correlations
 between the cluster 
 positions, i.e. a strong preference for a given interparticle
 distance.  In Figure S4 in the Supplementary Information one can
appreciate the difference in gyration radii of A (large) and B (small)
clusters. Some small radii A clusters appear, most likely due to the
presence of A particles inside B clusters. 
These effects can be visually analyzed  in Fig.~\ref{fig:sep} where we present a collection of snapshots that illustrate well
how B particles segregate within the large condensates, and even some
of them depart  to form B-rich domains. Interestingly, these
pictures resemble qualitatively those obtained for TDP-43 condensates
using fluorescence microscopy (cf Figure 3 in Ref.~\onlinecite{PantojaUceda2021}). 

Obviously, this simple model does not aim at acting as a realistic
representation of the transitions found in real biomolecular
condensates, but we believe it can provide some key elements to
understand  what changes should
be expected in the effective interactions between the constituent
particles during the process of internal segregation within the
condensates. For instance, the fact that the segregation leads mostly 
to clustering of B particles in the periphery of the condensates is
simply the result of the dominance of medium range repulsive
interactions between B particles. These tend to segregate the
B-clusters, whereas the presence of strong A-B attractions keeps the
B-aggregates within the condensates.

\begin{figure*}[t]
  \subfloat[]{\includegraphics[width=0.32\linewidth]{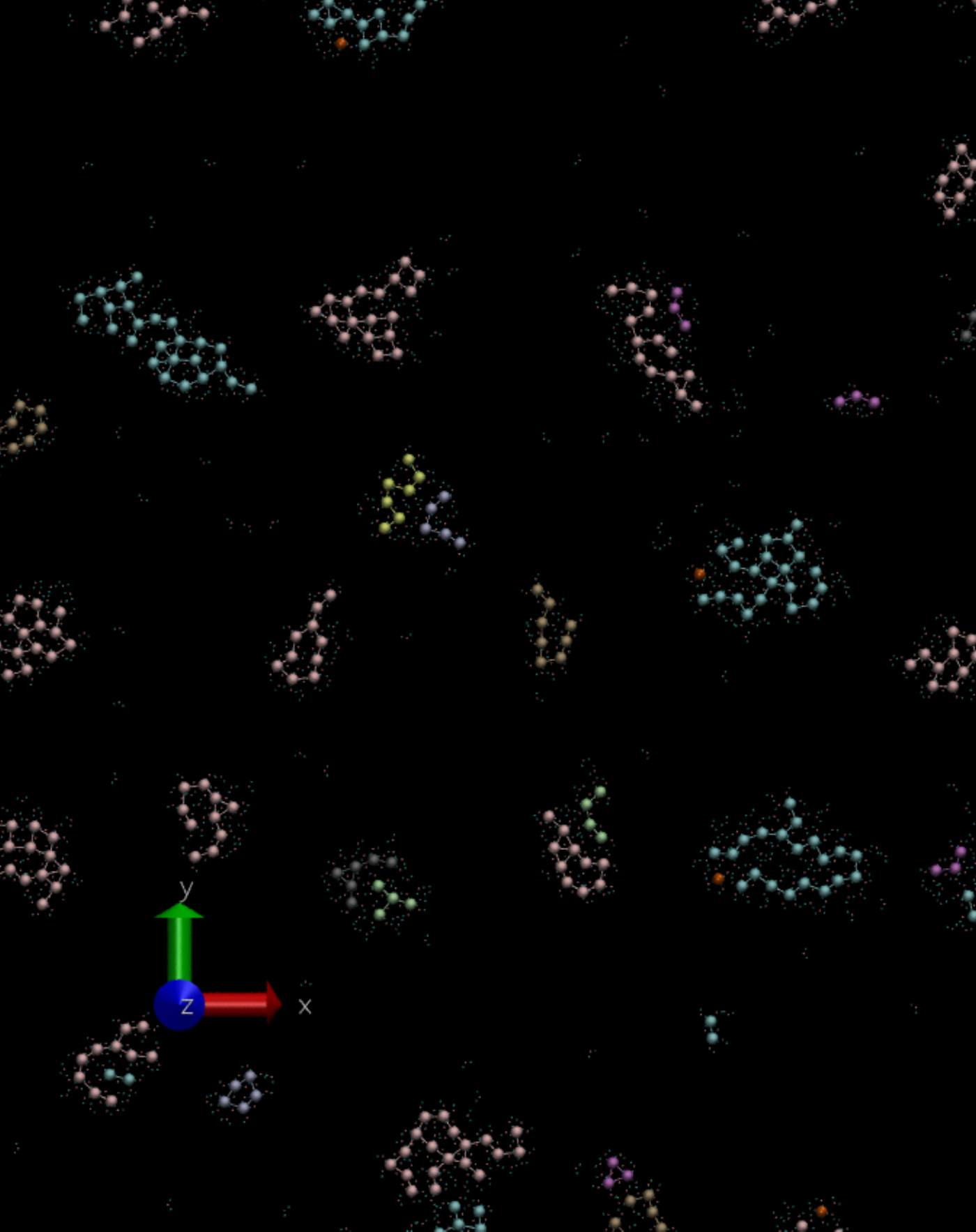}}
  \hspace{0.2cm}
  \subfloat[]{\includegraphics[width=0.32\linewidth]{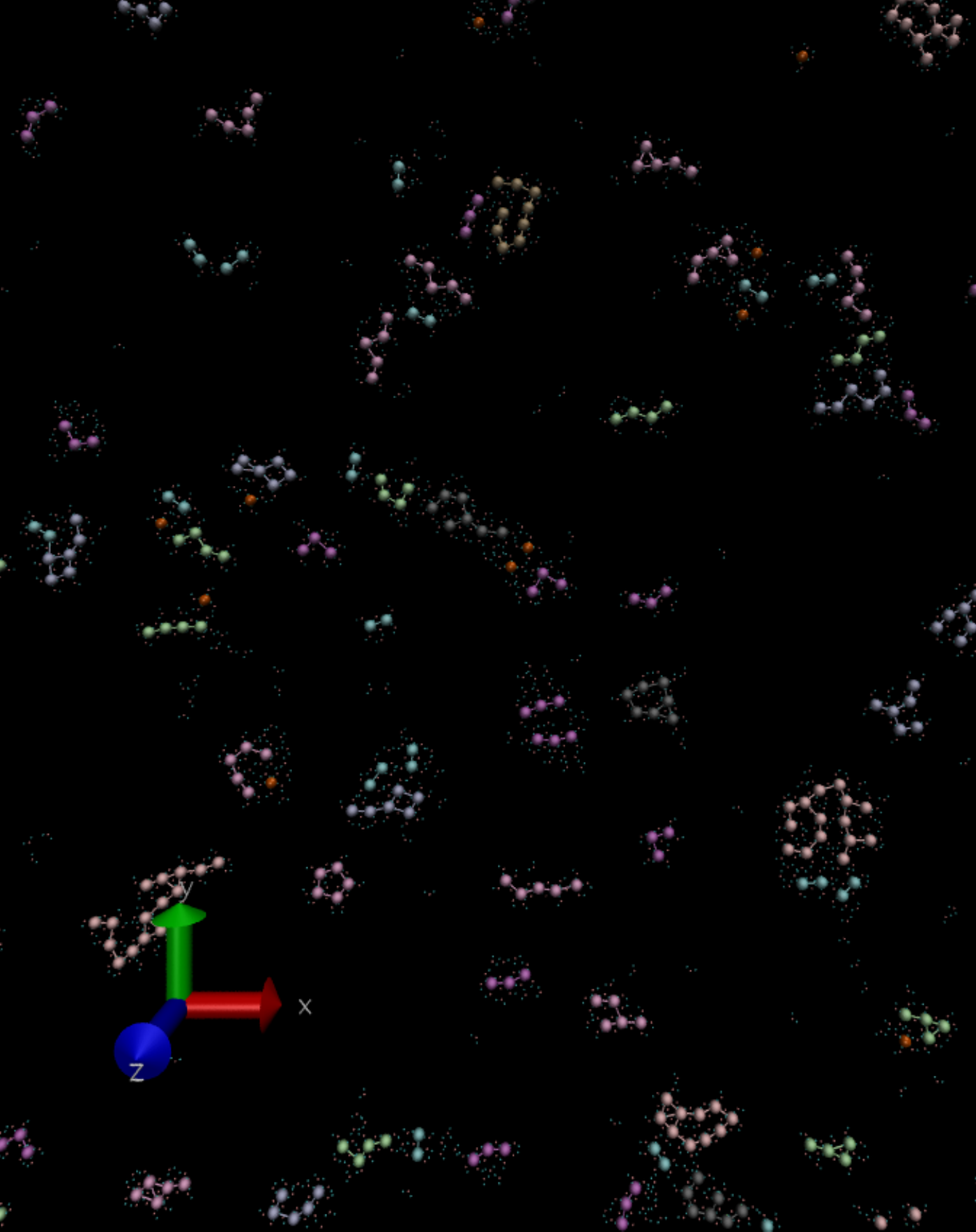}}\\
  \subfloat[]{\includegraphics[width=0.325\linewidth]{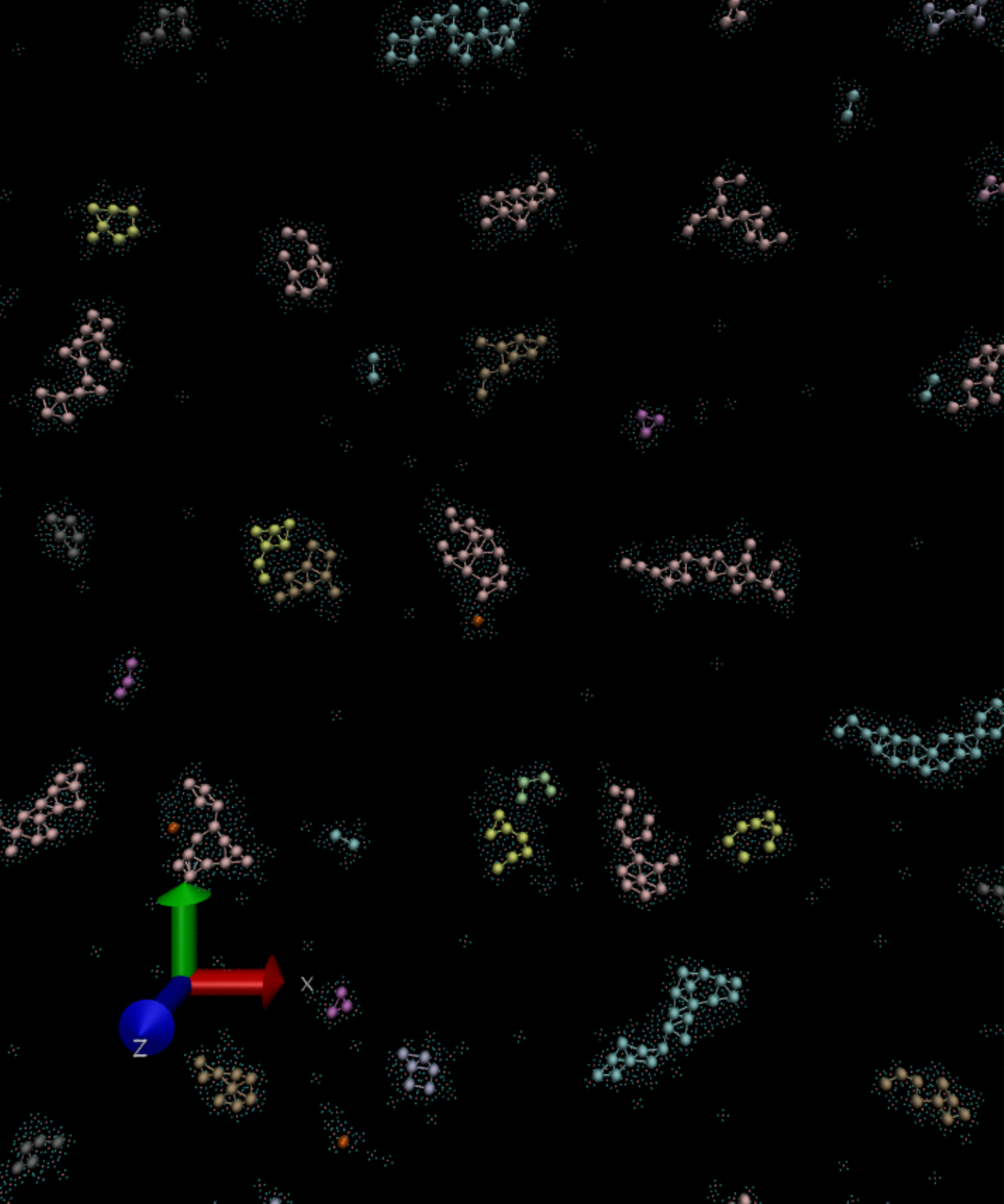}}
  \hspace{0.2cm}
 \subfloat[]{\includegraphics[width=0.31\linewidth]{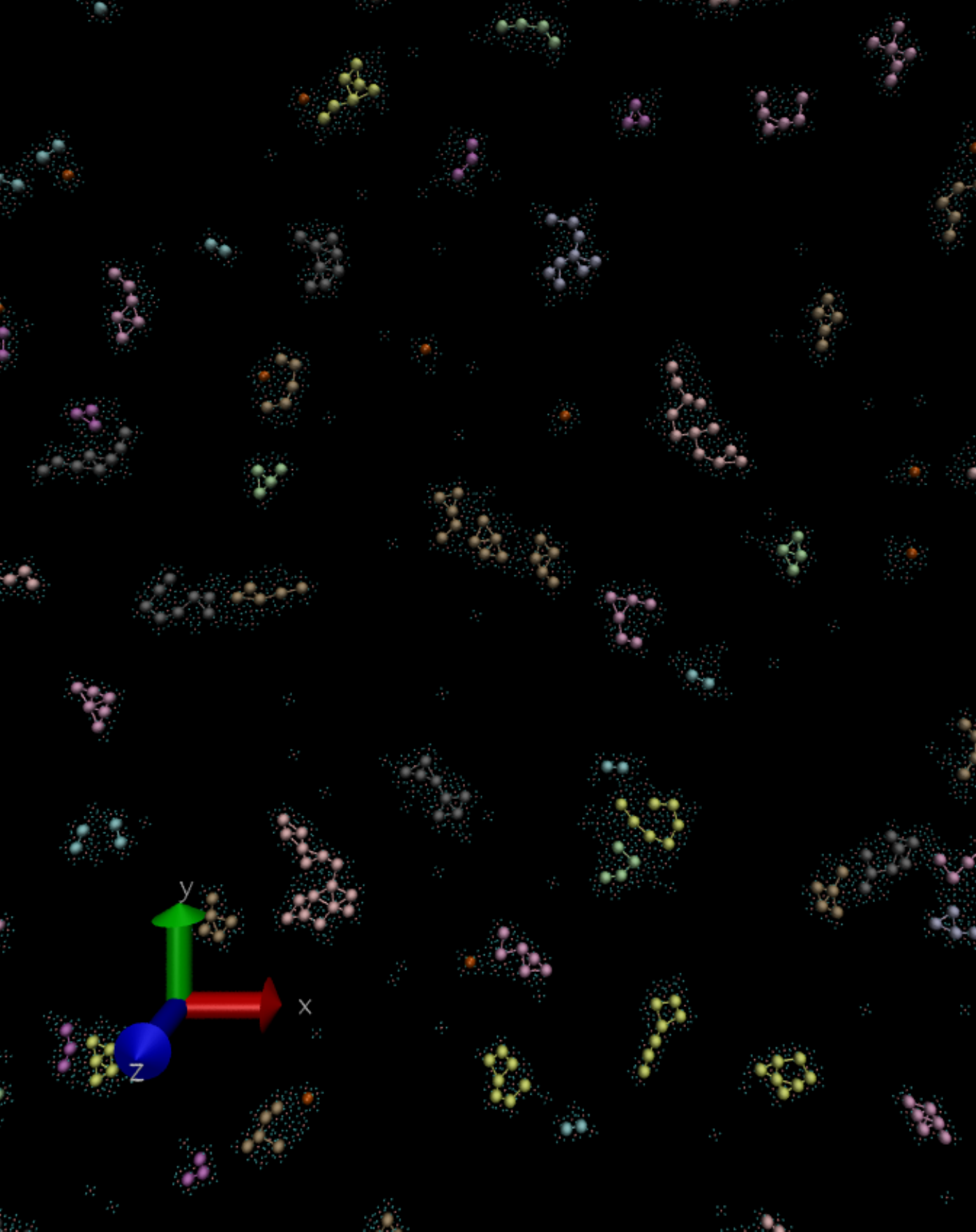}}
   \caption{Snapshots of clusters formed by 1:4  mixtures of 2- and 4-patch A
     particles (upper and lower pictures respectively) with 4-patch B
     particles. The systems  in the right figures include BB Yukawa
     interactions. Different colors designate different cluster
     sizes. Unconnected B monomers are depicted in
     orange.\label{fig:patch44} }

\end{figure*}

\section{Clustering via associating patchy particles}
In the previous section we have focused on systems with limited
polydispersity, with clear maxima in their cluster size distributions,
and basically with spherically-shaped (circularly-shaped in 2D)
condensates.  In the introduction it was mentioned
that many cytoplasmic membraneless organelles have very wide size
distributions, sometimes covering up to two orders of
magnitude. Palaia and \v{S}ari\'c \cite{Palaia2022} proposed in this
connection a patchy model in which clustering is driven by strong
associative forces in given sites, and cluster size is controlled by
stoichiometry, and to some extent also by kinetic barriers that prevent
arrested cluster states to fully condense into gel-like percolating
clusters. Now we will have strongly directional clustering
forces. A-particles have been considered both in 1:1 and 4:1
compositions. The cluster analysis has been
performed using B-particles as reference, and a cluster distance of
$2\sigma_{cc}$. Note that two B particles will always be linked by an
A-linker. As mentioned we have considered 4-patch and 2-patch linkers,
so, our results will illustrate how the geometry of the linker also
impacts the size and obviously the cluster
topology. More significantly, we will explore the
  morphology of the phases formed by the two anisotropic pSALR models
  introduced in Section III, under different conditions of linker topology and
  stoichiometry. In this connection, we have  paid special attention to the
role played by long range Yukawa repulsive forces 
acting between B-sites (BB-pSARL) or both A and B sites (AB-pSARL).

\begin{figure*}[t]
  \subfloat[\label{clustdp}]{\includegraphics[width=0.4\linewidth]{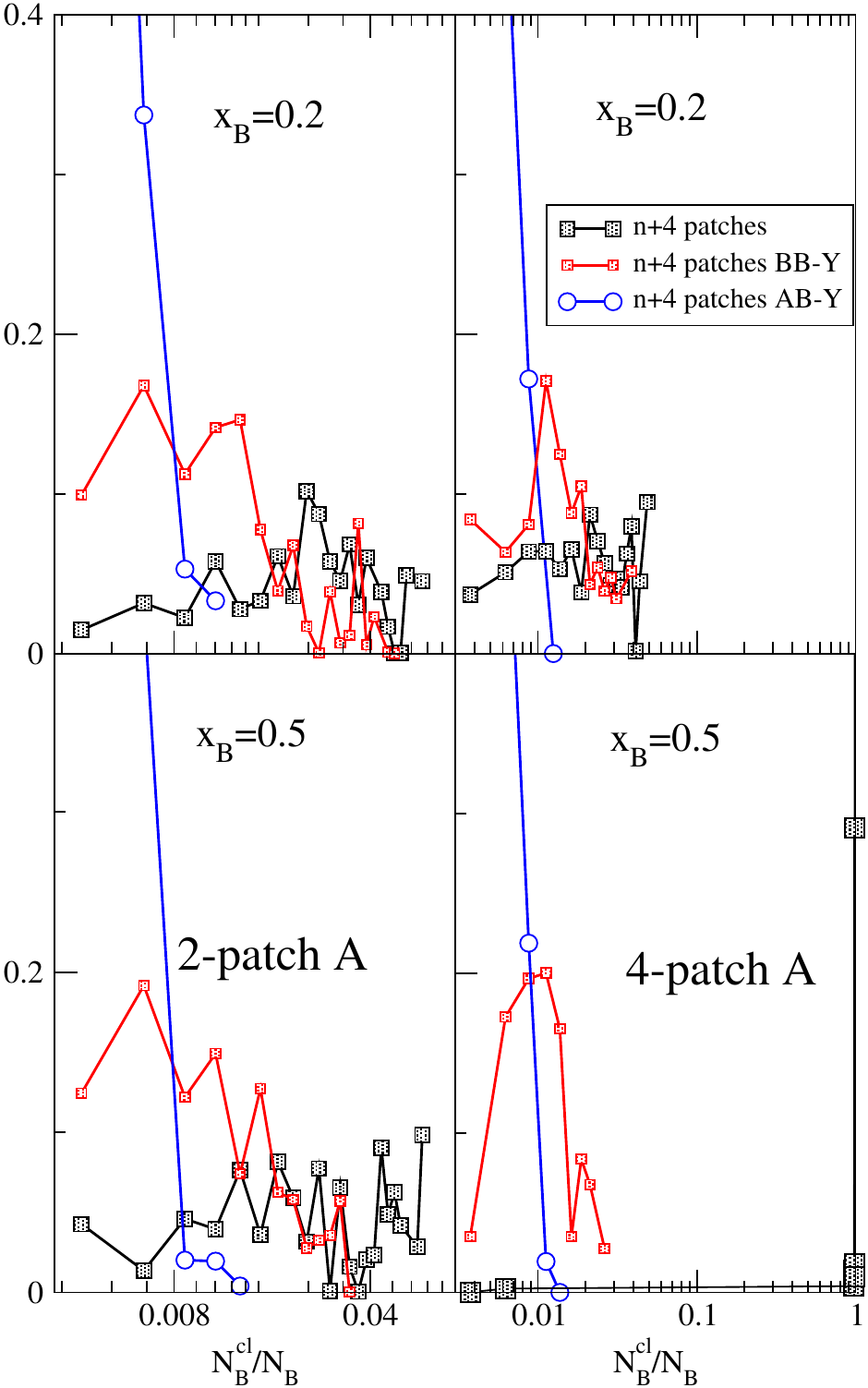}}
  \subfloat[\label{rhop}]{\includegraphics[width=0.44\linewidth]{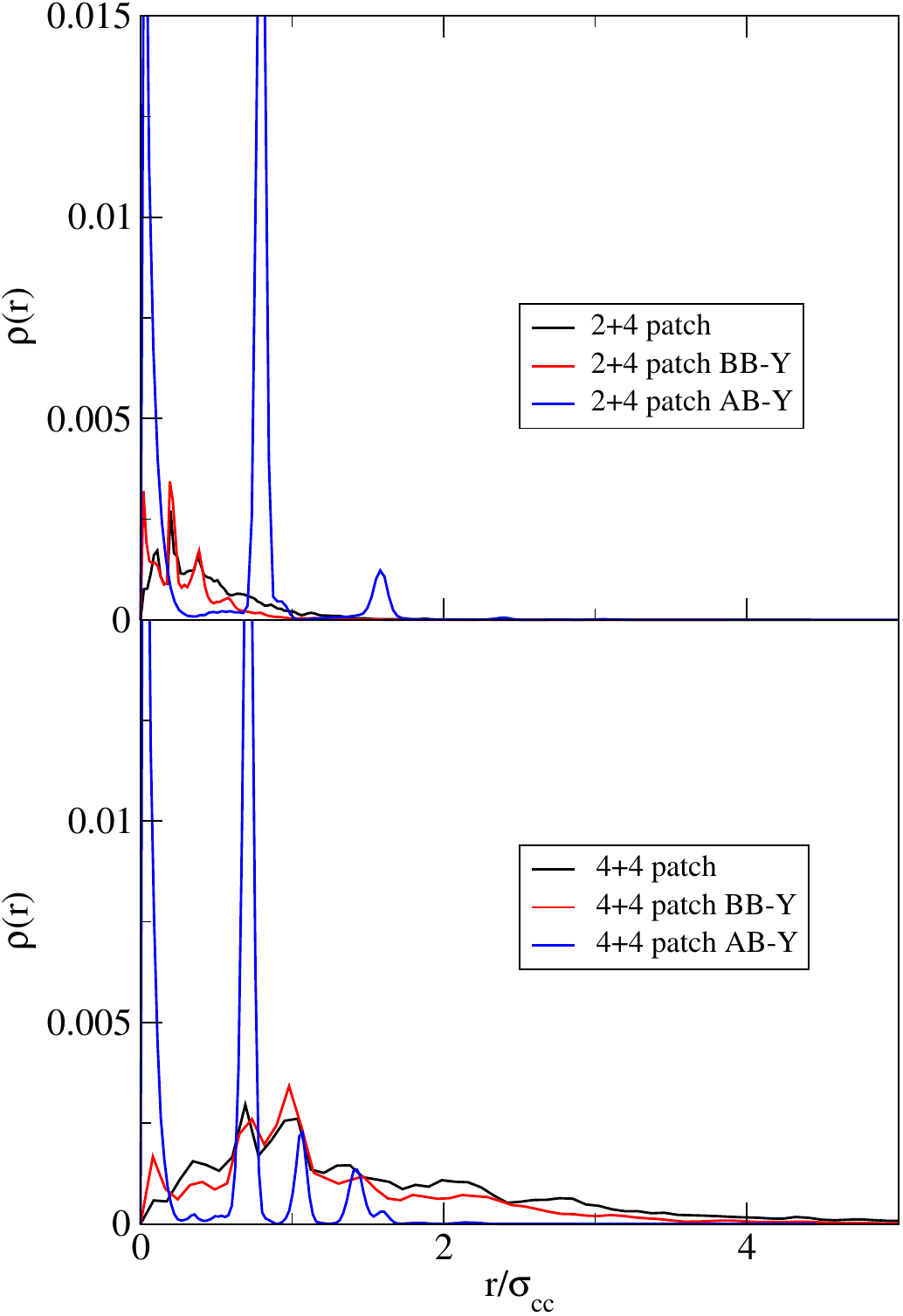}}
\caption{(a)Cluster size distributions for the two types of patchy models
  studied, with  2-patch A particles (left) and 4-patch A
  particles (right). Results correspond to $\rho_B\sigma_{cc}^2=0.03$
  and $T^*=1$. Compositions 1:1 (bottom) and 1:4 (upper graphs)  are
  considered. Note that for the equimolar mixture of 4-patch A and B
  particles the ground state corresponds to a single cluster spanning the
  whole sample. (b) Density profiles of 2-4 (top) and 4-4 patchy particle
  mixtures, with and without Yukawa interactions between BB and AB
  sites for $x_B=1/5$.}
\end{figure*}

\begin{figure*}[t]
  \subfloat[4p-A+4p-B]{\includegraphics[width=0.32\linewidth]{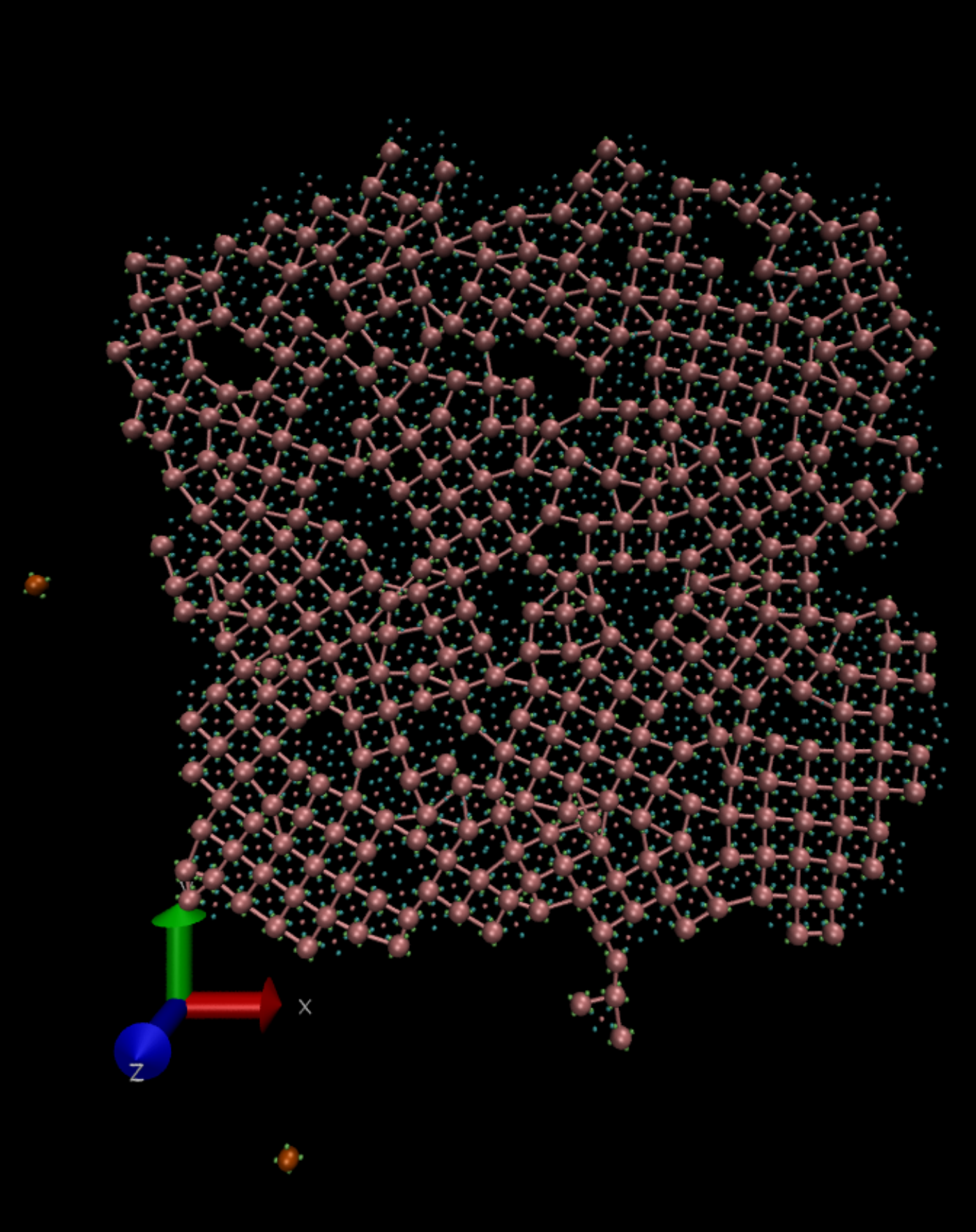}}
  \hfill
 \subfloat[4p-A+4p-B
   BB-pSARL]{\includegraphics[width=0.32\linewidth]{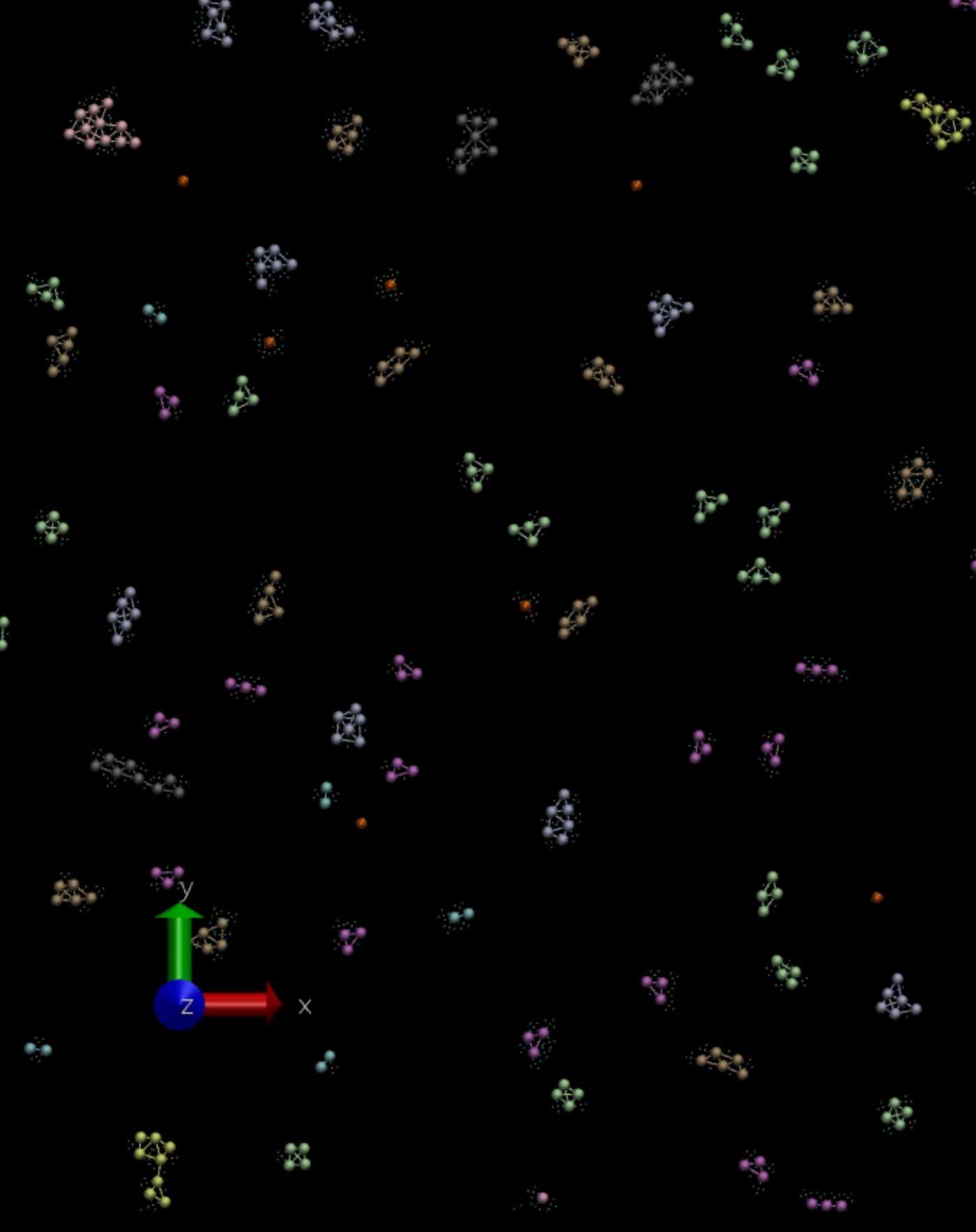}}
 \hfill
 \subfloat[4p-A+4p-B AB-pSARL]{\includegraphics[width=0.32\linewidth]{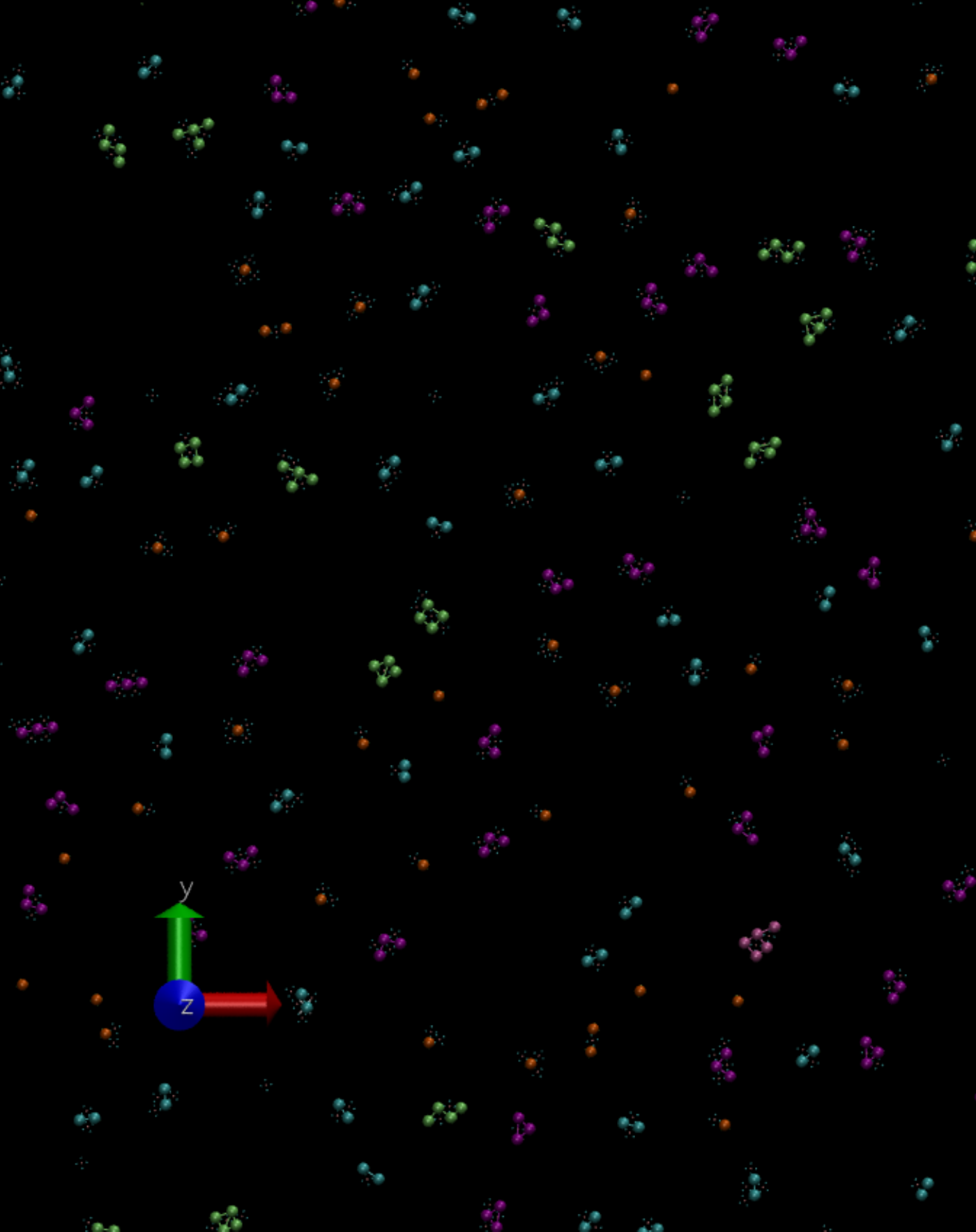}}\\

 \subfloat[2p-A+4p-B]{\includegraphics[width=0.32\linewidth]{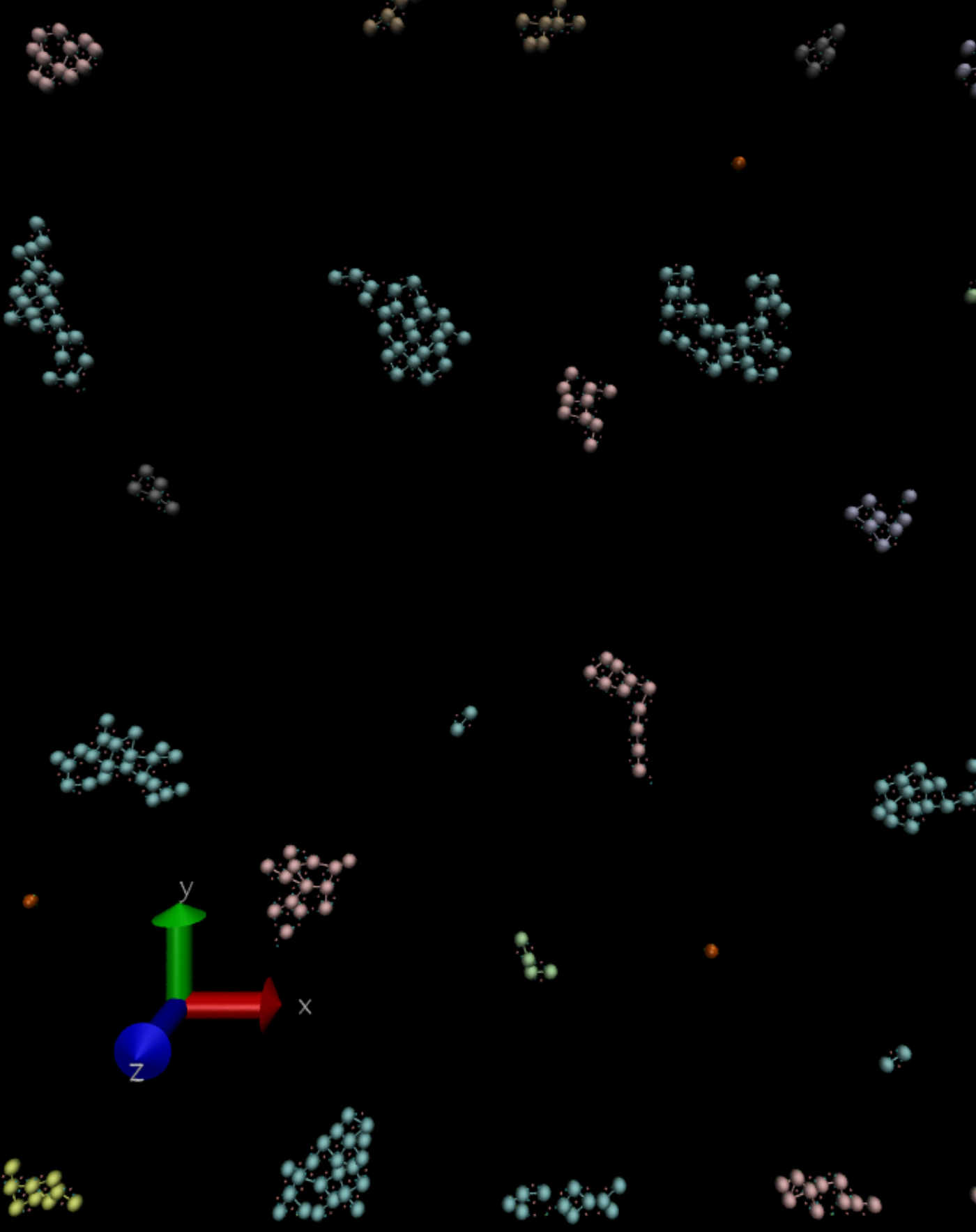}}
 \hfill
 \subfloat[2p-A+4p-B
   BB-pSARL]{\includegraphics[width=0.32\linewidth]{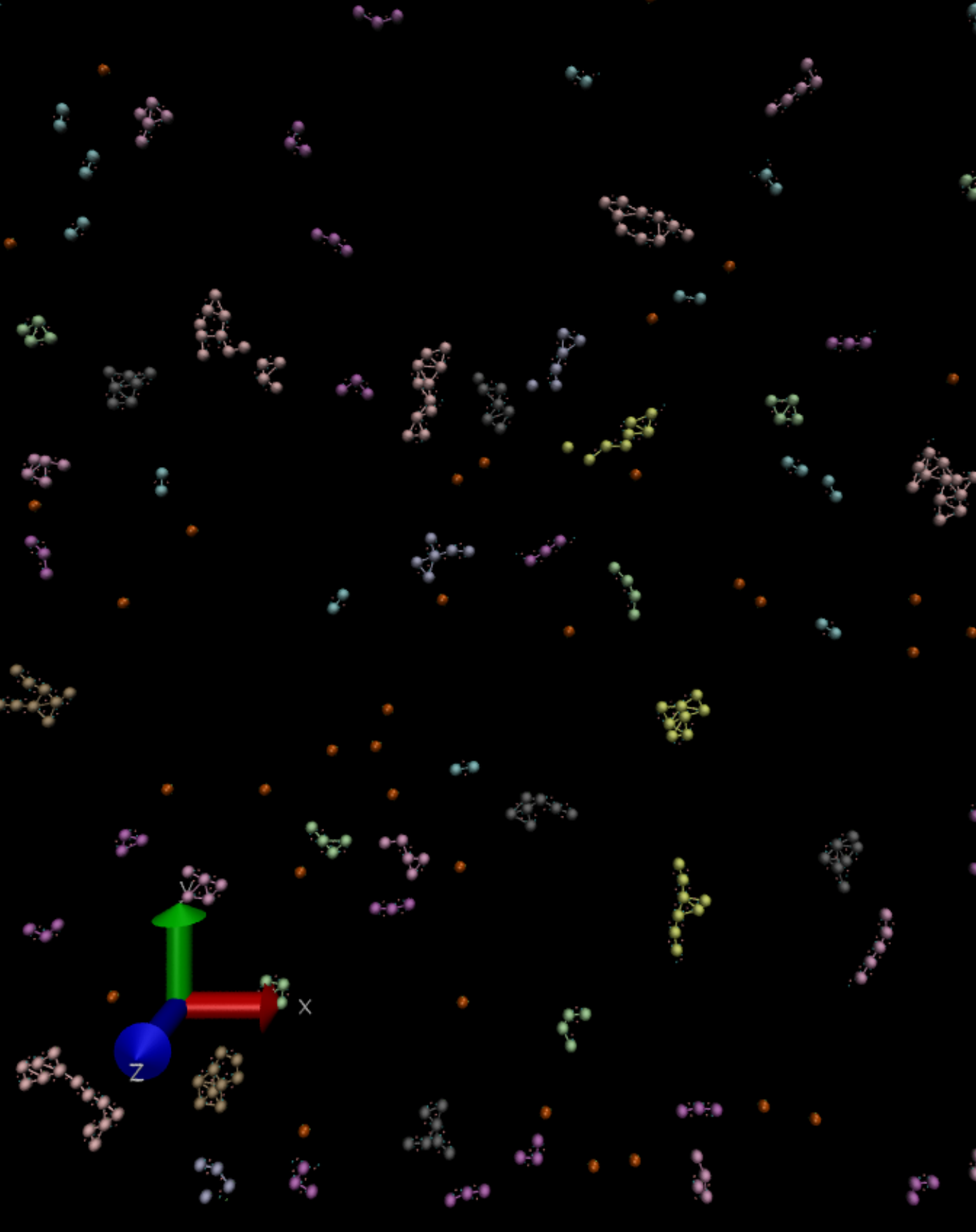}}
 \hfill
 \subfloat[2p-A+4p-B AB-pSARL]{\includegraphics[width=0.32\linewidth]{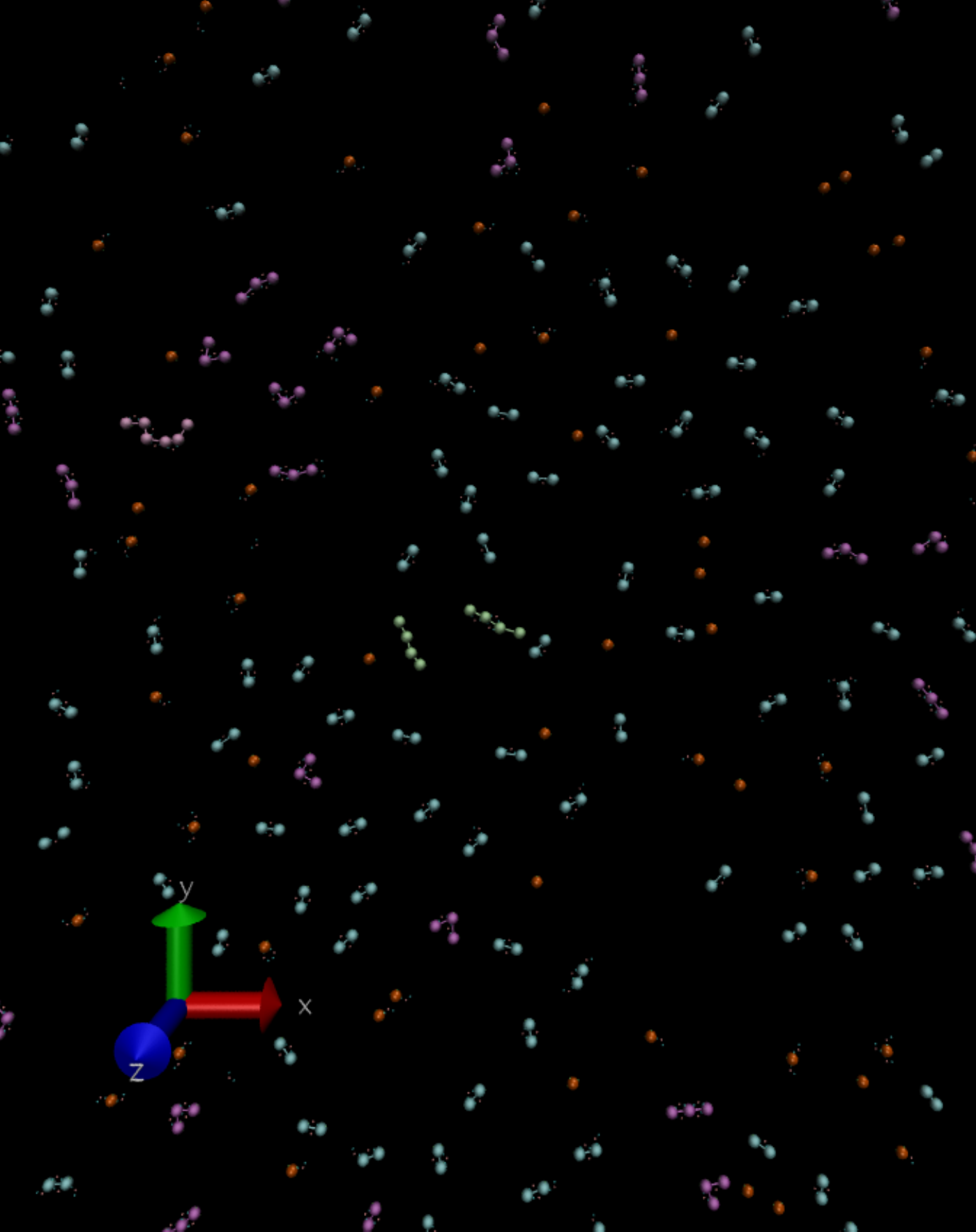}}

   \caption{Snapshots of clusters formed by equimolar mixtures of 4- and 2-patch A
     particles (upper and lower pictures respectively) with 4-patch B
     particles. The systems  in the central figures include BB Yukawa 
     interactions, and those in the right ones AA, AB Yukawa
     components as well. Larger spheres correspond to linked B
     particles. Different colors designate different cluster
     sizes. Unconnected B monomers are depicted in
     orange.\label{fig:patch11}} 

\end{figure*}

In Fig.~\ref{fig:patch44} we can see the snapshots of
1:4 mixtures of 2- and 4-patch A and B particles,  without (left) and with (right)
B-B Yukawa interactions (BB pSALR model). The presence of
2-patch linkers induces the formation of chains and reduces the size
of the clusters.  This is further illustrated in the upper graphs of
Fig.~\ref{clustdp} 
where the cluster size distributions are plotted as functions of the
ratio of B particles forming clusters vs the total number of
B-particles, $N_B^{cl}/N_B$. There are enough linkers to saturate
associative sites in B-particles, and therefore clusters are
finite. The effect of 2-patch linkers is to induce the build up of larger clusters with
the distribution displaying a maximum. In the case of 4-patch linkers
(equal number patches in  A and B particles), the
cluster size distribution is more uniform, as found by Palaia and \v{S}ari\'c
\cite{Palaia2022}, displaying fourfold coordination. 2-patch linkers favor a twofold coordination,
by which chains and rings are more common. The screened charges of the
BB pSALR model  enhance this tendency (chain-like conformations
minimize repulsion), and   somewhat decrease cluster size, inducing 
more pronounced maxima in the size distribution.  If charges are
present in both B and A sites, 
cluster size is considerably reduced, as can be seen both in
Fig.~\ref{clustdp} (blue curves with hollow circles) and in the snapshots of Fig.~\ref{fig:patch11} (left most column). 

Snapshots of equimolar compositions are shown in Figure \ref{fig:patch11},
and one can see there and in the bottom left panel of
  Fig.~\ref{clustdp}, --empty squares-- that without long range repulsion
 the (4-patch A) + (4-patch B) system collapses into a single
cluster (points gather together near $N_B^{cl}/N_B\sim 1$) in
equilibrium with a few free particles. The structure is a 
body-centered square lattice with vacancies and strong
dislocations. With 2-patch A linkers, clusters are again finite, due
to the imbalance of associative sites, and some chain-like structures
appear. Adding long range repulsion to the B-sites, reduces the cluster
size inducing a marked maximum, and with A and B long range repulsion,
5 to 2 particle clusters dominate, most of the structures being
linear. Recall that B-clusters include the 
corresponding A-particles acting as linkers (represented by small circles in the
picture to facilitate the visual identification of the clusters). This
means that the actual cluster size when accounting for A and B
particles is in this case double.

An interesting property that is worth inspecting is the average
cluster density profile. In the case of the isotropic SALR fluids these
profiles display roughly the shape of  a Fermi-Dirac distribution
(cf. Figure 6 in Ref.~\onlinecite{DiazPozuelo2025}), which evidences
the rather compact structure of the clusters. Patchy particle
clusters however, have shapes that considerably deviate from the
spherical (circular). This translates into rather different cluster
density profiles, as illustrated in Fig.~\ref{rhop}. The
effects on cluster size of both the type of linker and the presence of
long range repulsions are also apparent on the different extent of the
density profiles. But what is more relevant is the highly non
monotonous character of the profiles, particularly apparent when
screened charges are present both in  A and B sites. The {\em
  layered} profile structure in fact reflects the chain structure. The valleys
between peaks are filled by A-linkers. In all cases we have a clearly
more open structure. This is particularly relevant if we are modeling
biologically active condensates, since active sites would now be
accessible, whereas in the case of isotropic SALR condensates only the
surface is accessible. In any case, in this instance  small molecules
could in principle still diffuse through the cluster bulk due to differences
in chemical potential. Overall, however thinking in terms of nucleic
acid-protein complexes, the p-SALR mixture model seems more
promising.

\begin{figure*}
\subfloat[\label{2+4hd}]{\includegraphics[width=0.40\linewidth]{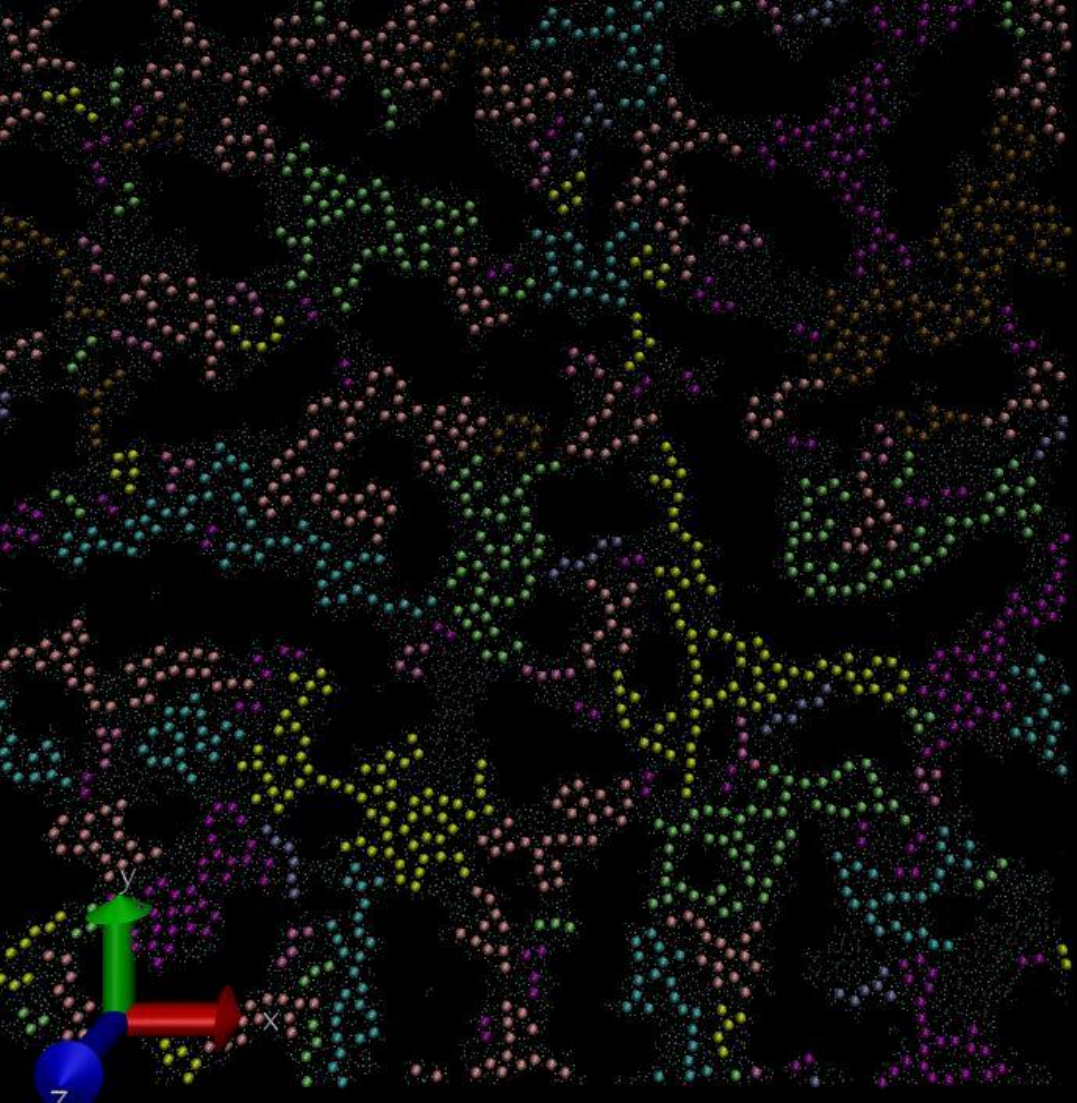}}
\hspace{0.25cm}
\subfloat[\label{lvsh}]{\includegraphics[width=0.55\linewidth]{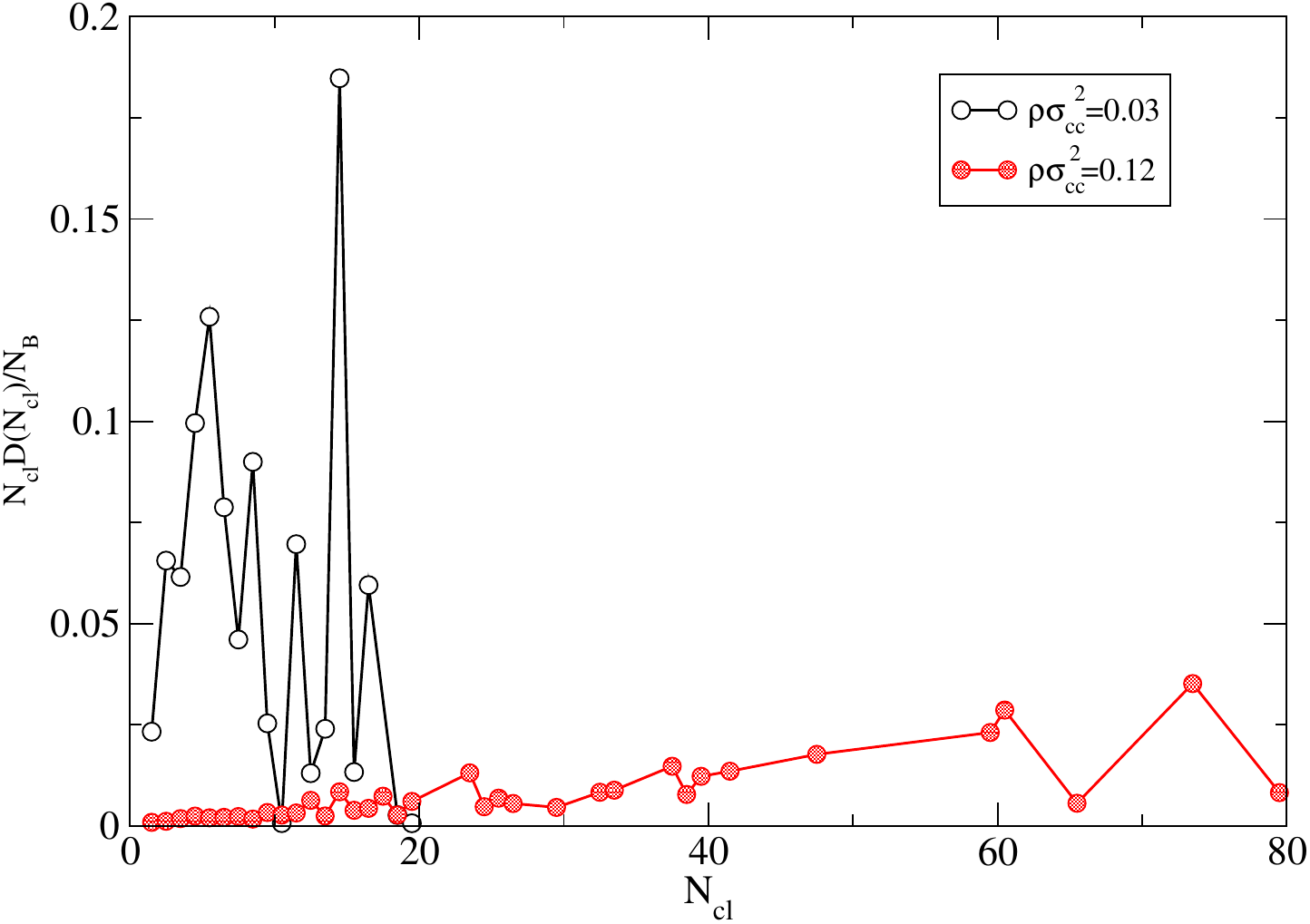}}
\caption{(a) Snapshot of a 2+4 BB-pSARL patchy fluid mixture at relatively
high density (b) Distribution of percentage of B-particles forming
part of clusters of size $N_{cl}$.}
\end{figure*}

Finally, it is worth to take a look at the effects of density on the
cluster topology.  In the case of isotropic SALR models we have seen how these
system undergo a transition from a globular phase towards a rather
regular lamellar phase that ends up in a well ordered bubble
phase. In the case of p-SALR models, the geometry of the clusters
forming the scaffold 
fully determines  the microscopic structure of the dense phase. Thus
for instance we see in Figure \ref{2+4hd} that a system of 2+4
patchy BB-pSARL particles with a composition ratio 4:1 and global density
$\rho\sigma_{cc}^2=0.12$ (four times that of the previous systems) ends
up in a disordered gel-like
structure. The 2-patch angular linkers form a percolating 
structure of entangled chains, far different from the ordered lamellar
phases displayed by 
isotropic SALR potentials.  Cluster size is also increased, and a larger
percentage of particles associate into larger clusters (cf. Figure \ref{lvsh}). The chain-like
character of the clusters is enhanced both by long range repulsion and
and  by the particular geometry of the two-site A-linkers. It is worth
mentioning that for these densities chains are frequent even when
A-particles have 4 associative sites (see Figure S7),  simply in order
to minimize intracluster repulsions. This effect is not so evident at
much lower densities 
(cf. Figure \ref{fig:patch44}).

\section{Discussion and Conclusions}

Our systematic exploration of SALR, decorated SALR, and patchy SARL
models reveals a set of general design principles that dictate the
structural and dynamic properties of two-dimensional self-assembled
aggregates. These principles establish a link between specific,
tunable features of the interparticle potential and the resulting
mesoscale architecture.

Analysis of the various models presented here, has evidenced that the
\textbf{repulsive barrier controls size and polydispersity} in SALR
models. Analysis of the effective cluster-cluster and cluster-particle
effective potentials shows  the repulsive maximum  acts as a
nucleation barrier. Increasing this barrier (as in the SALR-Gauss
model) directly reduces the equilibrium cluster size and, crucially,
suppresses size polydispersity. This narrow distribution enables the
emergence of long-range order, facilitating the crystallization of
clusters into a periodic lattice.

Additionally,    the presence \textbf{multiple attraction minima decouples internal and
global order.} Introducing oscillatory wells within the attractive
range (SALR-OPP) creates a landscape of deep, narrow minima. This
drastically reduces intra-cluster particle mobility, promoting
internal crystallization. Simultaneously, it stabilizes clusters of
varied sizes, and hence increases polydispersity. Thus suppresses
global lattice order. The result is a  composite state with solid-like
dynamics \emph{within} clusters that themselves display liquid-like
mobility. 
    
When dealing with binary mixtures, we have seen that
  appropriately tuned 
\textbf{asymmetric interactions drive internal demixing.} This
internal phase segregation within clusters---reminiscent of biological
processes like those in TDP-43 condensates---emerges from an interplay
of asymmetric interaction ranges. A scenario where a majority
component forms large condensates and a minority component has a much
shorter repulsive range naturally leads to the segregation of small
aggregates at the cluster interface, governed by a balance of specific
attractions and medium-range repulsions.  These small aggregates tend to concentrate on
the surface of the large condensates or fully segregate to the
intercluster space, our rough representation of a cell's cytoplasm.

Now, when dealing with anisotropic patchy SALR models, we find that
\textbf{anisotropy and stoichiometry are driving factors that
  condition  morphology and
  polydispersity.}  In contrast to isotropic SALR models, cluster 
 formation in associative patchy SALR systems is mostly governed by bonding
geometry and linker stoichiometry. This inherently leads to  broader
size distributions. The linker's valence is a primary determinant of
cluster shape: symmetric linkers (4-patch) promote compact domains,
while lower valence (2-patch) directs the formation of open,
chain-like or ring-like structures.

However, in all cases,  \textbf{long-range repulsion remains a universal tool to control
polydispersity, cluster size and to some extent morphology.} The
addition of a isotropic, 
long-range repulsive component (e.g., screened Coulomb potential) to any model---isotropic
or patchy---serves as a powerful, independent control parameter. It
universally counteracts coalescence, by which it limits cluster
growth, further modulating size distributions. In anisotropic systems,
can enhance the presence open morphologies by favoring linear arrangements that
minimize repulsive energy. 

It is worth stressing  that the models studied here are minimal by
design, intended to isolate the role of specific interaction
features. They are not quantitative models of any particular protein
or RNA. However, the principles derived from them provide a valuable
set of tools  for interpreting the complex
behavior of biomolecular condensates and for guiding the design of synthetic analogues.
In this connection, it might be of interest to recall that the principle
that a high repulsive barrier leads to monodisperse, ordered arrays
 offers a physical explanation for the regular spacing
and uniform size of certain nuclear bodies \cite{brangwynne2013phase},
where effective interactions might be tuned by chromatin geometry or
active processes. Conversely, the broad polydispersity inherent to
patchy models governed by stoichiometry  aligns naturally
with the heterogeneous size distribution of cytoplasmic condensates
like P-bodies, which form under kinetic control in a crowded
environment \cite{brangwynne2013phase,banani2017molecular}. 
Also, the SALR-OPP model demonstrates how specific interaction landscapes  can
generate phases that are internally rigid yet globally fluid. This
decoupling mirrors the behavior of stress granules, which can exhibit
solid-like internal properties while remaining mobile. Our results
suggest that the introduction of multiple, specific binding motifs
(e.g., via modular domains or linear motifs) could be a mechanism
cells use to tune condensate fluidity, with direct
implications for the liquid-to-solid transitions observed in
neurodegenerative disease \cite{Ranganathan2022}. 
Finally,  our binary SALR mixture  shows that internal phase
segregation, as seen in TDP-43 droplets \cite{PantojaUceda2021}, can
arise purely from asymmetric interaction ranges without invoking
conformational changes. This suggests that post-translational
modifications or changes in solution conditions that alter effective
interaction ranges could be sufficient to trigger such demixing.

In summary, through a systematic computational study of
two-dimensional self-assembling systems, we have demonstrated how
specific, tunable features of interparticle interactions serve as
independent design parameters to control cluster size, polydispersity,
morphology, and the coupling between internal and global dynamics. The
principles extracted here provide a road-map for engineering 2D
materials with desired mesoscale architecture and a framework for
interpreting the physical basis of condensate diversity in biology.

Looking forward, several directions emerge naturally from this
work. First, the principles we have established in 2D should be tested
in three dimensions, where geometric frustration and glassy arrest
play a more prominent role. Second, our models are equilibrium
systems, whereas biological condensates exist in a non-equilibrium,
active cellular environment. Introducing active forces---such as those
from ATP-dependent processes or molecular motors---will be crucial to
understand how activity modulates the phase behavior and dynamics of
condensates. Finally, the design rules we have explored can guide the
development of more detailed, component-specific models for particular
biomolecular systems, where sequence information and conformational
flexibility can be integrated into the effective interaction
potentials. Such efforts will bridge the gap between minimal physical
models and the complexity of living matter.

\section*{Supplementary material}
As Supplementary Information we include an illustration of snapshots
of the SALR model, in parallel with changes in pressure and
configurational energy as the system evolves along an isotherm with
increasing density. In addition, we also include the accumulated
averaged angular order parameter profiles for SALR, SALR-Gauss and
SALR-OPP. Gyration radii distributions for the
three isotropic SALR systems are also included. Finally, a variety of
snapshots of patchy systems illustrates the various topologies
generated for various densities and compositions.

\section*{Acknowledgments}
EL and ADP acknowledge  the  financial support from Grant No.
 PID2023-151751NB-I00 funded by MICIU/AEI/10.13039/501100011033 and by
 ``ERDF A way of making Europe.'' We would also like to acknowledge the Galicia
Supercomputing Center (CESGA) for the generous access to their
computer facilities. CB thanks Union College for their generous financial support.

\bibliography{refs}

@Book{PettiforBOOK,
  address   = {Oxford, England},
  author    = {D.G. Pettifor},
  publisher = {Clarendon Press},
  title     = {Bonding and structure of molecules and solids},
  year      = {1995},
}

@Book{Nelson2002,
  address   = {Cambridge, England},
  author    = {David R. Nelson},
  publisher = {Cambridge University Press},
  title     = {Deffects and Geometry in Condensed Matter},
  year      = {2002},
}

@article{Mihalkovic2012,
	author = {Mihalkovi{\v c}, Marek and Henley, C. L.},
	date = {2012/03/19/},
	day = {19},
	doi = {10.1103/PhysRevB.85.092102},
	id = {10.1103/PhysRevB.85.092102},
	j1 = {PRB},
	journal = {Physical Review B},
	journal1 = {Phys. Rev. B},
	month = {03},
	number = {9},
	pages = {092102--},
	publisher = {American Physical Society},
	title = {Empirical oscillating potentials for alloys from ab initio fits and the prediction of quasicrystal-related structures in the Al-Cu-Sc system},
	url = {https://link.aps.org/doi/10.1103/PhysRevB.85.092102},
	volume = {85},
	year = {2012}}

@article{Schwanzer2010,
	author = {Schwanzer, Dieter F and Kahl, Gerhard},
	doi = {10.1088/0953-8984/22/41/415103},
	journal = {Journal of Physics: Condensed Matter},
	month = {sep},
	number = {41},
	pages = {415103},
	title = {Two-dimensional systems with competing interactions: microphase formation versus liquid--vapour phase separation},
	url = {https://dx.doi.org/10.1088/0953-8984/22/41/415103},
	volume = {22},
	year = {2010}}

@article{Almarza2014,
	author = {Almarza, N. G. and P{\c e}kalski, J. and Ciach, A.},
	doi = {10.1063/1.4871901},
	isbn = {0021-9606},
	journal = {The Journal of Chemical Physics},
	journal1 = {J. Chem. Phys.},
	month = {8/6/2025},
	number = {16},
	pages = {164708},
	title = {Periodic ordering of clusters and stripes in a two-dimensional lattice model. II. Results of Monte Carlo simulation},
	url = {https://doi.org/10.1063/1.4871901},
	volume = {140},
	year = {2014},
	year1 = {2014/04/28}}

@article{Pekalski2014,
	author = {P{\c e}kalski, J. and Ciach, A. and Almarza, N. G.},
	doi = {10.1063/1.4868001},
	isbn = {0021-9606},
	journal = {The Journal of Chemical Physics},
	journal1 = {J. Chem. Phys.},
	month = {8/6/2025},
	number = {11},
	pages = {114701},
	title = {Periodic ordering of clusters and stripes in a two-dimensional lattice model. I. Ground state, mean-field phase diagram and structure of the disordered phases},
	url = {https://doi.org/10.1063/1.4868001},
	volume = {140},
	year = {2014},
	year1 = {2014/03/17}}

@article{Bordin2018,
	author = {Jos{\'e} Rafael Bordin},
	doi = {https://doi.org/10.1016/j.physa.2017.12.090},
	issn = {0378-4371},
	journal = {Physica A: Statistical Mechanics and its Applications},
	pages = {215-224},
	title = {Distinct aggregation patterns and fluid porous phase in a 2D model for colloids with competitive interactions},
	url = {https://www.sciencedirect.com/science/article/pii/S0378437117313420},
	volume = {495},
	year = {2018}}

@article{Chacko2015,
	author = {Chacko, Blesson and Chalmers, Christopher and Archer, Andrew J.},
	doi = {10.1063/1.4937941},
	isbn = {0021-9606},
	journal = {The Journal of Chemical Physics},
	journal1 = {J. Chem. Phys.},
	month = {8/6/2025},
	number = {24},
	pages = {244904},
	title = {Two-dimensional colloidal fluids exhibiting pattern formation},
	url = {https://doi.org/10.1063/1.4937941},
	volume = {143},
	year = {2015},
	year1 = {2015/12/24}}

@article{Conicella2016,
	author = {Conicella, Alexander E and Zerze, G{\"u}l H and Mittal, Jeetain and Fawzi, Nicolas L},
	cin = {Structure. 2016 Sep 6;24(9):1435-6. doi: 10.1016/j.str.2016.08.003. PMID: 27602988},
	copyright = {Copyright {\copyright}2016 Elsevier Ltd. All rights reserved.},
	crdt = {2016/08/23 06:00},
	date = {2016 Sep 6},
	dcom = {20171005},
	dep = {20160818},
	doi = {10.1016/j.str.2016.07.007},
	edat = {2016/08/23 06:00},
	gr = {T32 GM007601/GM/NIGMS NIH HHS/United States; P30 GM103410/GM/NIGMS NIH HHS/United States; P20 GM104937/GM/NIGMS NIH HHS/United States; R01 GM118530/GM/NIGMS NIH HHS/United States; P30 RR031153/RR/NCRR NIH HHS/United States; P20 RR018728/RR/NCRR NIH HHS/United States},
	issn = {1878-4186 (Electronic); 0969-2126 (Print); 0969-2126 (Linking)},
	jid = {101087697},
	journal = {Structure},
	jt = {Structure (London, England : 1993)},
	language = {eng},
	lid = {S0969-2126(16)30192-7 {$[$}pii{$]$}; 10.1016/j.str.2016.07.007 {$[$}doi{$]$}},
	lr = {20201109},
	mhda = {2017/10/06 06:00},
	mid = {NIHMS805438},
	month = {Sep},
	number = {9},
	own = {NLM},
	pages = {1537--1549},
	phst = {2016/06/20 00:00 {$[$}received{$]$}; 2016/06/20 00:00 {$[$}revised{$]$}; 2016/07/10 00:00 {$[$}accepted{$]$}; 2016/08/23 06:00 {$[$}entrez{$]$}; 2016/08/23 06:00 {$[$}pubmed{$]$}; 2017/10/06 06:00 {$[$}medline{$]$}; 2017/09/06 00:00 {$[$}pmc-release{$]$}},
	pii = {S0969-2126(16)30192-7},
	pl = {United States},
	pmc = {PMC5014597},
	pmcr = {2017/09/06},
	pmid = {27545621},
	pst = {ppublish},
	pt = {Journal Article},
	rn = {0 (DNA-Binding Proteins); 0 (Protein Aggregates); 0 (Recombinant Proteins); 0 (TARDBP protein, human)},
	sb = {IM},
	status = {MEDLINE},
	title = {ALS Mutations Disrupt Phase Separation Mediated by alpha-Helical Structure in the TDP-43 Low-Complexity C-Terminal Domain.},
	volume = {24},
	year = {2016}}

@article{Engel2015,
	author = {Engel, Michael and Damasceno, Pablo F. and Phillips, Carolyn L. and Glotzer, Sharon C.},
	date = {2015/01/01},
	doi = {10.1038/nmat4152},
	id = {Engel2015},
	isbn = {1476-4660},
	journal = {Nature Materials},
	number = {1},
	pages = {109--116},
	title = {Computational self-assembly of a one-component icosahedral quasicrystal},
	url = {https://doi.org/10.1038/nmat4152},
	volume = {14},
	year = {2015}}

@article{PantojaUceda2021,
	author = {Pantoja-Uceda, David AND Stuani, Cristiana AND Laurents, Douglas V. AND McDermott, Ann E. AND Buratti, Emanuele AND Mompe{\'a}n, Miguel},
	doi = {10.1371/journal.pbio.3001198},
	journal = {PLOS Biology},
	month = {04},
	number = {4},
	pages = {1-17},
	publisher = {Public Library of Science},
	title = {Phe-Gly motifs drive fibrillization of TDP-43's prion-like domain condensates},
	url = {https://doi.org/10.1371/journal.pbio.3001198},
	volume = {19},
	year = {2021}}

@article{Argudo2021,
	author = {Argudo, Pablo G. and Giner-Casares, Juan J.},
	doi = {10.1039/D0NA00941E},
	issue = {7},
	journal = {Nanoscale Adv.},
	pages = {1789-1812},
	publisher = {RSC},
	title = {Folding and self-assembly of short intrinsically disordered peptides and protein regions},
	url = {http://dx.doi.org/10.1039/D0NA00941E},
	volume = {3},
	year = {2021}}

@article{Liu2011,
	author = {Liu, Yun and Porcar, Lionel and Chen, Jinhong and Chen, Wei-Ren and Falus, Peter and Faraone, Antonio and Fratini, Emiliano and Hong, Kunlun and Baglioni, Piero},
	date = {2011/06/09},
	doi = {10.1021/jp109333c},
	isbn = {1520-6106},
	journal = {The Journal of Physical Chemistry B},
	journal1 = {The Journal of Physical Chemistry B},
	journal2 = {J. Phys. Chem. B},
	month = {06},
	number = {22},
	pages = {7238--7247},
	publisher = {American Chemical Society},
	title = {Lysozyme Protein Solution with an Intermediate Range Order Structure},
	type = {doi: 10.1021/jp109333c},
	url = {https://doi.org/10.1021/jp109333c},
	volume = {115},
	year = {2011},
	year1 = {2011}}

@article{Hirose2023,
	author = {Hirose, Tetsuro and Ninomiya, Kensuke and Nakagawa, Shinichi and Yamazaki, Tomohiro},
	copyright = {{\copyright}2022. Springer Nature Limited.},
	crdt = {2022/11/24 23:29},
	date = {2023 Apr},
	dcom = {20230328},
	dep = {20221123},
	doi = {10.1038/s41580-022-00558-8},
	edat = {2022/11/25 06:00},
	issn = {1471-0080 (Electronic); 1471-0072 (Linking)},
	jid = {100962782},
	journal = {Nat Rev Mol Cell Biol},
	jt = {Nature reviews. Molecular cell biology},
	language = {eng},
	lid = {10.1038/s41580-022-00558-8 {$[$}doi{$]$}},
	lr = {20230607},
	mh = {*Organelles/metabolism; *Biomolecular Condensates; RNA/metabolism; Gene Expression Regulation; Transcription Factors/metabolism},
	mhda = {2023/03/28 17:08},
	month = {Apr},
	number = {4},
	own = {NLM},
	pages = {288--304},
	phst = {2022/10/21 00:00 {$[$}accepted{$]$}; 2023/03/28 17:08 {$[$}medline{$]$}; 2022/11/25 06:00 {$[$}pubmed{$]$}; 2022/11/24 23:29 {$[$}entrez{$]$}},
	pii = {10.1038/s41580-022-00558-8},
	pl = {England},
	pmid = {36424481},
	pst = {ppublish},
	pt = {Journal Article; Research Support, Non-U.S. Gov't; Review},
	rn = {63231-63-0 (RNA); 0 (Transcription Factors)},
	sb = {IM},
	status = {MEDLINE},
	title = {A guide to membraneless organelles and their various roles in gene regulation.},
	volume = {24},
	year = {2023}}

@article{DiazPozuelo2025,
	author = {D{\'\i}az-Pozuelo, Antonio and Gonz{\'a}lez-Salgado, Diego and Lomba, Enrique},
	doi = {10.1063/5.0249688},
	eprint = {https://pubs.aip.org/aip/jcp/article-pdf/doi/10.1063/5.0249688/20406194/074903\_1\_5.0249688.pdf},
	issn = {0021-9606},
	journal = {The Journal of Chemical Physics},
	month = {02},
	number = {7},
	pages = {074903},
	title = {On the build-up of effective hyperuniformity from large globular colloidal aggregates},
	url = {https://doi.org/10.1063/5.0249688},
	volume = {162},
	year = {2025}}

@article{LAMMPS_Thompson2022,
	author = {Aidan P. Thompson and H. Metin Aktulga and Richard Berger and Dan S. Bolintineanu and W. Michael Brown and Paul S. Crozier and Pieter J. {in 't Veld} and Axel Kohlmeyer and Stan G. Moore and Trung Dac Nguyen and Ray Shan and Mark J. Stevens and Julien Tranchida and Christian Trott and Steven J. Plimpton},
	doi = {https://doi.org/10.1016/j.cpc.2021.108171},
	journal = {Computer Physics Communications},
	pages = {108171},
	title = {LAMMPS - a flexible simulation tool for particle-based materials modeling at the atomic, meso, and continuum scales},
	volume = {271},
	year = {2022}
}

@article{Whitesides2002,
    author = {Whitesides, George M. and Grzybowski, Bartosz},
    title = {Self-Assembly at All Scales},
    journal = {Science},
    volume = {295},
    number = {5564},
    pages = {2418--2421},
    year = {2002},
    doi = {10.1126/science.1070821}
}

@article{Glotzer2007,
    author = {Glotzer, Sharon C. and Solomon, Michael J.},
    title = {Anisotropy of building blocks and their assembly into complex structures},
    journal = {Nature Materials},
    volume = {6},
    number = {7},
    pages = {557--562},
    year = {2007},
    doi = {10.1038/nmat1949}
}

@article{Hyman2014,
    author = {Hyman, Anthony A. and Weber, Christoph A. and J\"{u}licher, Frank},
    title = {Liquid-Liquid Phase Separation in Biology},
    journal = {Annual Review of Cell and Developmental Biology},
    volume = {30},
    pages = {39--58},
    year = {2014},
    doi = {10.1146/annurev-cellbio-100913-013325}
}

@Article{Sweatman2019,
  author    = {Martin B. Sweatman and Leo Lue},
  journal   = {Advanced Theory and Simulations},
  title     = {The Giant {SALR} Cluster Fluid: A Review},
  year      = {2019},
  month     = {apr},
  number    = {7},
  pages     = {1900025},
  volume    = {2},
  doi       = {10.1002/adts.201900025},
  publisher = {Wiley},
}

@article{Sanchez2012,
    author = {Sánchez, Tim and Chen, Daniel T. N. and DeCamp, Stephen J. and Heymann, Michael and Dogic, Zvonimir},
    title = {Spontaneous motion in hierarchically assembled active matter},
    journal = {Nature},
    volume = {491},
    number = {7424},
    pages = {431--434},
    year = {2012},
    doi = {10.1038/nature11591}
}

@article{brangwynne2013phase,
  title={Phase transitions and size scaling of membrane-less organelles},
  author={Brangwynne, Clifford P and Hyman, Anthony A},
  journal={The Journal of Cell Biology},
  volume={203},
  number={6},
  pages={875--881},
  year={2013},
  doi={10.1083/jcb.201308087}
}

@article{banani2017molecular,
  title={The molecular language of membraneless organelles},
  author={Banani, Salman F and Lee, Richard A and Hyman, Anthony A and Rosen, Michael K},
journal={Journal of Biological Chemistry},
  volume={294},
  number={18},
  pages={7115--7127},
  year={2018},
  doi={10.1074/jbc.TM118.001192}
}

@Article{Orti2021,
  author    = {Orti, Fernando and Navarro, Alvaro M. and Rabinovich, Andres and Wodak, Shoshana J. and Marino-Buslje, Cristina},
  journal   = {Computational and Structural Biotechnology Journal},
  title     = {Insight into membraneless organelles and their associated proteins: Drivers, Clients and Regulators},
  year      = {2021},
  issn      = {2001-0370},
  pages     = {3964--3977},
  volume    = {19},
  doi       = {10.1016/j.csbj.2021.06.042},
  publisher = {Elsevier BV},
}

@Article{Sear1999,
  Title                    = {Spontaneous patterning of quantum dots at the air-water interface},
  Author                   = {Richard P. Sear and Sung-Wook Chung and Gil Markovich and William M. Gelbart and James R. Heath},
  Journal                  = {Phys. Rev. E},
  Year                     = {1999},
  Pages                    = {R6255},
  Volume                   = {59},
  DOI                      = {10.1103/PhysRevE.59.R6255}
}

@Article{Bores2015,
  author    = {C. Bores and N.G. Almarza and E. Lomba and G. Kahl},
  title     = {Adsorption of a two dimensional system with competing interactions in a disordered, porous matrix},
  journal   = {J. Phys. : Condens. Matter},
  year      = {2015},
  volume    = {27},
  pages     = {194127},
  doi       = {10.1088/0953-8984/27/19/194127},
}

@Article{Lomba2014a,
  author    = {Enrique Lomba and Cecilia Bores and Gerhard Kahl},
  journal   = {J. Chem. Phys.},
  title     = {Explicit spatial description of fluid inclusions in porous matrices in terms of an inhomogeneous integral equation},
  year      = {2014},
  pages     = {164704},
  volume    = {141},
  doi       = {10.1063/1.4898713},
}

@book{creighton_1993,
  title={Proteins: Structures and Molecular Properties},
  author={Creighton, Thomas E.},
  year={1993},
  publisher={W. H. Freeman},
  edition={2nd},
}

@book{watson_2014,
  title={Molecular Biology of the Gene},
  author={Watson, James D. and Baker, Tania A. and Bell, Stephen P. and Gann, Alexander and Levine, Michael and Losick, Richard},
  year={2014},
  publisher={Pearson},
  edition={7th},
  note={A foundational textbook of molecular biology. The early chapters on the structure of DNA and RNA clearly describe the polyanionic nature of the sugar-phosphate backbone.}
}

@Article{Hoffmann2006,
  author    = {Norman Hoffmann and Christos N. Likos and Hartmut L\"{o}wen},
  title     = {Microphase structuring in two-dimensional magnetic colloid mixtures},
  journal   = {J Phys : Condens Matter},
  year      = {2006},
  volume    = {18},
  pages     = {10193-10211},
  doi       = {10.1088/0953-8984/18/45/007},
}

@Article{Imperio2004,
  Title                    = {A bidimensional fluid system with competing interactions: spontaneous and induced pattern formation},
  Author                   = {A. Imperio and L. Reatto},
  Journal                  = {J. Phys.: Condens. Matter},
  Year                     = {2004},
  Pages                    = {S3769-S3789},
  Volume                   = {16},
  DOI                      = {10.1088/0953-8984/16/38/001}
}

@Article{Nelson1979,
  author    = {Nelson, David R. and Halperin, B. I.},
  journal   = {Physical Review B},
  title     = {Dislocation-mediated melting in two dimensions},
  year      = {1979},
  issn      = {0163-1829},
  month     = mar,
  number    = {5},
  pages     = {2457--2484},
  volume    = {19},
  doi       = {10.1103/physrevb.19.2457},
  publisher = {American Physical Society (APS)},
}

@Article{Ranganathan2022,
  author    = {Ranganathan, Srivastav and Liu, Junlang and Shakhnovich, Eugene},
  journal   = {Essays in Biochemistry},
  title     = {Different states and the associated fates of biomolecular condensates},
  year      = {2022},
  issn      = {1744-1358},
  month     = dec,
  number    = {7},
  pages     = {849--862},
  volume    = {66},
  doi       = {10.1042/ebc20220054},
  editor    = {Mukhopadhyay, Samrat},
  publisher = {Portland Press Ltd.},
}

@article{Sciortino2004,
  title = {Phase Diagram of Patchy Colloids: Towards Empty Liquids},
  author = {Sciortino, Francesco and Mossa, Stefano and Zaccarelli, Emanuela and Tartaglia, Piero},
  journal = {Phys. Rev. Lett.},
  volume = {93},
  issue = {5},
  pages = {055701},
  numpages = {4},
  year = {2004},
  month = {Jul},
  publisher = {American Physical Society},
  doi = {10.1103/PhysRevLett.93.055701},
  url = {https://link.aps.org/doi/10.1103/PhysRevLett.93.055701}
}

@Article{Imperio2006,
  Title                    = {Microphase separation in two-dimensional systems with competing interactions},
  Author                   = {A. Imperio and L. Reatto},
  Journal                  = {J. Chem. Phys.},
  Year                     = {2006},
  Pages                    = {164712},
  Volume                   = {124},
  DOI                      = {10.1063/1.2185618}
}

@Article{Chen2007b,
  author    = {Chen, Sow-Hsin and Broccio, Matteo and Liu, Yun and Fratini, Emiliano and Baglioni, Piero},
  journal   = {J. Appl. Crystallogr.},
  pages     = {s321--s326},
  title     = {The two-Yukawa model and its applications: the cases of charged proteins and copolymer micellar solutions},
  volume    = {40},
  year      = {2007},
  issn      = {0021-8898},
  month     = apr,
  number    = {s1},
  doi       = {10.1107/s0021889807006723},
  publisher = {International Union of Crystallography (IUCr)},
}

@Article{Monahan2017,
  author    = {Monahan, Zachary and Ryan, Veronica H and Janke, Abigail M and Burke, Kathleen A and Rhoads, Shannon N and Zerze, Gül H and O’Meally, Robert and Dignon, Gregory L and Conicella, Alexander E and Zheng, Wenwei and Best, Robert B and Cole, Robert N and Mittal, Jeetain and Shewmaker, Frank and Fawzi, Nicolas L},
  journal   = {EMBO J.},
  pages     = {2951--2967},
  title     = {Phosphorylation of the FUS low‐complexity domain disrupts phase separation, aggregation, and toxicity},
  volume    = {36},
  year      = {2017},
  issn      = {1460-2075},
  month     = aug,
  number    = {20},
  doi       = {10.15252/embj.201696394},
  publisher = {Springer Science and Business Media LLC},
}

@Article{Schmidt2016,
  author    = {Schmidt, H. Broder and Görlich, Dirk},
  journal   = {Trends Biochem. Sci.},
  pages     = {46--61},
  title     = {Transport Selectivity of Nuclear Pores, Phase Separation, and Membraneless Organelles},
  volume    = {41},
  year      = {2016},
  issn      = {0968-0004},
  month     = jan,
  number    = {1},
  doi       = {10.1016/j.tibs.2015.11.001},
  publisher = {Elsevier BV},
}

@Article{Tejedor2025,
  author    = {R. Tejedor, Andrés and Aguirre Gonzalez, Anne and Maristany, M. Julia and Chew, Pin Yu and Russell, Kieran and Ramirez, Jorge and Espinosa, Jorge R. and Collepardo-Guevara, Rosana},
  journal   = {ACS Cent. Sci.},
  pages     = {302--321},
  title     = {Chemically Informed Coarse-Graining of Electrostatic Forces in Charge-Rich Biomolecular Condensates},
  volume    = {11},
  year      = {2025},
  issn      = {2374-7951},
  month     = feb,
  number    = {2},
  doi       = {10.1021/acscentsci.4c01617},
  publisher = {American Chemical Society (ACS)},
}

@Article{Yigit2015,
  author    = {Yigit, Cemil and Heyda, Jan and Dzubiella, Joachim},
  journal   = {J. Chem. Phys.},
  title     = {Charged patchy particle models in explicit salt: Ion distributions, electrostatic potentials, and effective interactions},
  volume    = {143},
  year      = {2015},
  issn      = {1089-7690},
  month     = aug,
  number    = {6},
  doi       = {10.1063/1.4928077},
  publisher = {AIP Publishing},
}

@Article{Joseph2021,
  author  = {Jerelle A. Joseph and Aleks Reinhardt and Anne Aguirre and Pin Yu Chew and Kieran O. Russell and Jorge R. Espinosa and Adiran Garaizar and Rosana Collepardo-Guevara},
  journal = {Nature Computational Science},
  title   = {Physics-driven coarse-grained model for biomolecular phase separation with near-quantitative accuracy},
  year    = {2021},
  month   = {nov},
  number  = {11},
  pages   = {732--743},
  volume  = {1},
  doi     = {10.1038/s43588-021-00155-3},
}

@Article{Li2018,
  author    = {Li, Xinzhi and Das, Amit and Bi, Dapeng},
  journal   = {Proc. Natl. Acad. Sci.},
  pages     = {6650--6655},
  title     = {Biological tissue-inspired tunable photonic fluid},
  volume    = {115},
  year      = {2018},
  issn      = {1091-6490},
  month     = jun,
  number    = {26},
  doi       = {10.1073/pnas.1715810115},
  publisher = {Proceedings of the National Academy of Sciences},
}

@Article{Ross2025,
  author    = {Ross, R.J.H. and Masucci, G.D. and Lin, C.Y and Iglesias, T.L. and Reiter, S. and Pigolotti, S.},
  journal   = {Phys. Rev. X},
  pages     = {021064},
  title     = {Hyperdisordered Cell Packing on a Growing Surface},
  volume    = {15},
  year      = {2025},
  doi       = {10.1103/physrevx.15.021064},
}

@Article{Jiao2014,
  author    = {Yang Jiao and Timothy Lau and Haralampos Hatzikirou and Michael Meyer-Hermann and Joseph C. Corbo and Salvatore Torquato},
  title     = {Avian photoreceptor patterns represent a disordered hyperuniform solution to a multiscale packing problem},
  journal   = {Phys. Rev. E},
  year      = {2014},
  volume    = {89},
  pages     = {022721},
  doi       = {10.1103/PhysRevE.89.022721},
}

@Article{Lomba2020,
  author    = {Enrique Lomba and Jean-Jacques Weis and Leandro Guis{\'{a}}ndez and Salvatore Torquato},
  journal   = {Physical Review E},
  title     = {Minimal statistical-mechanical model for multihyperuniform patterns in avian retina},
  year      = {2020},
  pages     = {012134},
  volume    = {102},
  doi       = {10.1103/physreve.102.012134},
}

@Article{Klatt2019,
  author    = {Klatt, Michael A. and Lovrić, Jakov and Chen, Duyu and Kapfer, Sebastian C. and Schaller, Fabian M. and Schönhöfer, Philipp W. A. and Gardiner, Bruce S. and Smith, Ana-Sunčana and Schröder-Turk, Gerd E. and Torquato, Salvatore},
  journal   = {Nature Communications},
  title     = {Universal hidden order in amorphous cellular geometries},
  volume    = {10},
  year      = {2019},
  issn      = {2041-1723},
  month     = feb,
  number    = {1},
  doi       = {10.1038/s41467-019-08360-5},
  publisher = {Springer Science and Business Media LLC},
}

@Article{Liu2019,
  author    = {Liu, Yun and Xi, Yuyin},
  journal   = {Current Opinion in Colloid \& Interface Science},
  title     = {Colloidal systems with a short-range attraction and long-range repulsion: Phase diagrams, structures, and dynamics},
  year      = {2019},
  issn      = {1359-0294},
  month     = feb,
  pages     = {123--136},
  volume    = {39},
  doi       = {10.1016/j.cocis.2019.01.016},
  publisher = {Elsevier BV},
}

@InProceedings{Imperio2006a,
  Title                    = {Fluctuations and pattern formation in fluids with competing interactions},
  Author                   = {A. Imperio and D. Pini and L Reatto},
  Booktitle                = {International Workshop on Collective Phenomena in Macroscopic System},
  Year                     = {2006},
  Address                  = {Villa Olmo, Como, Italy},
  Month                    = {December},
  URL                      = {http://arxiv.org/abs/cond-mat/0703060}
}

@Article{Imperio2007,
  Title                    = {Microphase morphology in two-dimensional fluids under lateral confinement},
  Author                   = {Alessandra Imperio and Luciano Reatto},
  Journal                  = {Phys. Rev. E},
  Year                     = {2007},
  Pages                    = {040402(R)},
  Volume                   = {76},
  DOI                      = {10.1103/PhysRevE.76.040402}
}

@Article{Godfrin2014,
  author    = {P. Douglas Godfrin and Néstor E. Valadez-Pérez and Ramon Castañeda-Priego and Norman J. Wagner and Yun Liu},
  journal   = {Soft Matter},
  title     = {Generalized phase behavior of cluster formation in colloidal dispersions with competing interactions},
  year      = {2014},
  pages     = {5061 - 5071},
  volume    = {10},
  doi       = {10.1039/C3SM53220H},
}

@Article{Torquato2003,
  Title                    = {Local density fluctuations, hyperuniformity, and order metrics},
  Author                   = {Salvatore Torquato and Frank H. Stillinger},
  Journal                  = {Phys. Rev. E},
  Year                     = {2003},
  Pages                    = {041113},
  Volume                   = {68},
  DOI                      = {10.1103/PhysRevE.68.041113}
}

@Article{Chen2018,
  author    = {Duyu Chen and Enrique Lomba and Sal Torquato},
  journal   = {Phys. Chem. Chem. Phys.},
  title     = {Binary Mixtures of Charged Colloids: A Potential Route to Synthesize Disordered Hyperuniform Materials},
  year      = {2018},
  pages     = {17557 -1 7562},
  volume    = {20},
  doi       = {10.1039/c8cp02616e},
}

@Article{Yearley2014,
  author    = {Yearley, Eric J. and Godfrin, Paul D. and Perevozchikova, Tatiana and Zhang, Hailiang and Falus, Peter and Porcar, Lionel and Nagao, Michihiro and Curtis, Joseph E. and Gawande, Pradad and Taing, Rosalynn and Zarraga, Isidro E. and Wagner, Norman J. and Liu, Yun},
  journal   = {Biophys. J.},
  title     = {Observation of Small Cluster Formation in Concentrated Monoclonal Antibody Solutions and Its Implications to Solution Viscosity},
  year      = {2014},
  issn      = {0006-3495},
  month     = apr,
  number    = {8},
  pages     = {1763--1770},
  volume    = {106},
  doi       = {10.1016/j.bpj.2014.02.036},
  publisher = {Elsevier BV},
}

@Article{Archer2007a,
  Title                    = {Phase behavior of a fluid with competing attractive and repulsive interactions},
  Author                   = {Andrew J. Archer and Nigel B. Wilding},
  Journal                  = {Phys. Rev. E},
  Year                     = {2007},
  Pages                    = {031501},
  Volume                   = {76},
  DOI                      = {10.1103/PhysRevE.76.031501}
}

@Article{Stradner2004,
  author    = {Anna Stradner and Helen Sedgwick and Frederic Cardinaux and Wilson C. K. Poon and Stefan U. Egelhaaf and Peter Schurtenberger},
  journal   = {Nature},
  pages     = {492 - 495},
  title     = {Equilibrium cluster formation in concentrated protein solutions and colloids},
  volume    = {432},
  year      = {2004},
  doi       = {10.1038/nature03109},
}

@Article{Palaia2022,
  author    = {Palaia, Ivan and \v{S}ari\'c, Andela},
  journal   = {J. Chem. Phys.},
  pages     = {194902},
  title     = {Controlling cluster size in 2D phase-separating binary mixtures with specific interactions},
  volume    = {156},
  year      = {2022},
  issn      = {1089-7690},
  month     = may,
  number    = {19},
  doi       = {10.1063/5.0087769},
  publisher = {AIP Publishing},
}

@Article{Schwanzer2016,
  author    = {Schwanzer, Dieter F and Coslovich, Daniele and Kahl, Gerhard},
  journal   = {J. Phys. Condens. Matter},
  title     = {Two-dimensional systems with competing interactions: dynamic properties of single particles and of clusters},
  year      = {2016},
  issn      = {1361-648X},
  month     = aug,
  number    = {41},
  pages     = {414015},
  volume    = {28},
  doi       = {10.1088/0953-8984/28/41/414015},
  publisher = {IOP Publishing},
}

@Article{Torquato2018a,
  author    = {Salvatore Torquato},
  title     = {Hyperuniform states of matter},
  journal   = {Phys. Rep.},
  year      = {2018},
  volume    = {745},
  pages     = {1--95},
  month     = {jun},
  doi       = {10.1016/j.physrep.2018.03.001},
}

@Article{Cheron2022,
  author    = {Chéron, Elie and Groby, Jean-Philippe and Pagneux, Vincent and Félix, Simon and Romero-García, Vicent},
  journal   = {Physical Review B},
  title     = {Experimental characterization of rigid-scatterer hyperuniform distributions for audible acoustics},
  year      = {2022},
  issn      = {2469-9969},
  month     = aug,
  number    = {6},
  pages     = {064206},
  volume    = {106},
  doi       = {10.1103/physrevb.106.064206},
  publisher = {American Physical Society (APS)},
}

@Article{Froufe-Perez2017,
  author    = {Luis S. Froufe-P{\'{e}}rez and Michael Engel and Juan Jos{\'{e}} S{\'{a}}enz and Frank Scheffold},
  title     = {Band gap formation and Anderson localization in disordered photonic materials with structural correlations},
  journal   = {Proc. Natl. Acad. Sci. U.S.A.},
  year      = {2017},
  volume    = {114},
  number    = {36},
  pages     = {9570--9574},
  month     = {aug},
  doi       = {10.1073/pnas.1705130114},
}

@Article{RomeroGarcia2019,
  author    = {Romero-García, V. and Lamothe, N. and Theocharis, G. and Richoux, O. and García-Raffi, L.M.},
  journal   = {Physical Review Applied},
  title     = {Stealth Acoustic Materials},
  year      = {2019},
  issn      = {2331-7019},
  month     = may,
  number    = {5},
  pages     = {054076},
  volume    = {11},
  doi       = {10.1103/physrevapplied.11.054076},
  publisher = {American Physical Society (APS)},
}

@Article{Zhou2019,
  author    = {Zhou, Wen and Tong, Yeyu and Sun, Xiankai and Tsang, Hon Ki},
  title     = {Hyperuniform disordered photonic crystal polarizers},
  journal   = {ArXiv},
  year      = {2019},
  doi       = {https://arxiv.org/abs/1908.00759},
}

@Article{Milosevic2019,
  author    = {Milošević, Milan M. and Man, Weining and Nahal, Geev and Steinhardt, Paul J. and Torquato, Salvatore and Chaikin, Paul M. and Amoah, Timothy and Yu, Bowen and Mullen, Ruth Ann and Florescu, Marian},
  journal   = {Scientific Reports},
  title     = {Hyperuniform disordered waveguides and devices for near infrared silicon photonics},
  year      = {2019},
  issn      = {2045-2322},
  month     = dec,
  number    = {1},
  volume    = {9},
  doi       = {10.1038/s41598-019-56692-5},
  publisher = {Springer Science and Business Media LLC},
}

\end{document}